\documentclass[10pt,journal,compsoc]{IEEEtran}
\usepackage{chngcntr} %

\usepackage{multicol}

\usepackage{booktabs} 
\usepackage{enumitem}
\usepackage{multirow}
\usepackage{array}
\usepackage{graphicx}
\usepackage{grffile}
\usepackage{subfigure}
\usepackage{multirow}
\usepackage{tabularx}
\usepackage{bbm}
\usepackage[nocompress]{cite}
\usepackage{amsmath, amssymb, mathtools}
\usepackage[justification=centering]{caption}
\usepackage[numbers]{natbib}
\usepackage{threeparttable}
\usepackage{tabularx}
\usepackage{multirow}
\usepackage{booktabs}
\usepackage{hyperref} 
\usepackage{xcolor}
\usepackage{array, makecell}

\newtheorem{definition}{Definition}[section]%

\ifCLASSOPTIONcompsoc
  \usepackage[nocompress]{cite}
\else
  \usepackage{cite}
\fi

\begin{document}

\title{A Comprehensive Survey of Artificial Intelligence Techniques for Talent Analytics}

\author{Chuan Qin,~\IEEEmembership{Member,~IEEE,}
        Le Zhang,
        Yihang Cheng, 
        Rui Zha,
        Dazhong Shen,~\IEEEmembership{Member,~IEEE} \\
        Qi Zhang, 
        Xi Chen, 
        Ying Sun,~\IEEEmembership{Member,~IEEE,} 
        Chen Zhu,~\IEEEmembership{Member,~IEEE} \\
        Hengshu Zhu*,~\IEEEmembership{Senior Member,~IEEE,}
        Hui Xiong*,~\IEEEmembership{Fellow,~IEEE}

\IEEEcompsocitemizethanks{

\IEEEcompsocthanksitem C. Qin and H. Zhu are with the Computer Network Information Center, Chinese Academy of Sciences, Beijing, China and University of Chinese Academy of Sciences, Beijing, China. E-mail: chuanqin0426@gmail.com, zhuhengshu@gmail.com.

\IEEEcompsocthanksitem Y. Cheng is with the Computer Network Information Center, Chinese Academy of Sciences, Beijing, China. E-mail: chengyihang544@gmail.com

\IEEEcompsocthanksitem L. Zhang is with the Business Intelligence Lab, Baidu Inc., Beijing, China. E-mail: zhangle0202@gmail.com.

\IEEEcompsocthanksitem R. Zha and X. Chen are with the School of Computer Science and Technology, University of Science and Technology of China, Anhui, China. E-mail: crui0210@gmail.com, chenxi0401@mail.ustc.edu.cn.

\IEEEcompsocthanksitem C. Zhu is with the School of Management, University of Science and Technology of China, Anhui, China. E-mail: zc3930155@gmail.com.

\IEEEcompsocthanksitem D. Shen is with the College of Computer Science and Technology, Nanjing University of Aeronautics and Astronautics, Nanjing, China. E-mail: dazh.shen@gmail.com

\IEEEcompsocthanksitem Q. Zhang is with the Shanghai Artificial Intelligence Laboratory. E-mail: zhangqi.fqz@gmail.com.

\IEEEcompsocthanksitem Y. Sun is with the Thrust of Artificial Intelligence, The Hong Kong University of Science and Technology (Guangzhou), China. E-mail: yings@hkust-gz.edu.cn

\IEEEcompsocthanksitem H. Xiong is with the Thrust of Artificial Intelligence, The Hong Kong University of Science and Technology (Guangzhou), China and Department of Computer Science and Engineering, The Hong Kong University of Science and Technology Hong Kong SAR, China, E-mail: xionghui@ust.hk

\IEEEcompsocthanksitem H. Zhu and H. Xiong are the corresponding authors.

}}
\IEEEtitleabstractindextext{%

\begin{abstract}
In today’s competitive and fast-evolving business environment, it is {critical} for organizations to rethink how to make talent-related decisions in a quantitative manner.  Indeed, the recent development of Big Data and Artificial Intelligence (AI) techniques has revolutionized human resource management. The availability of large-scale talent and management-related data provides unparalleled opportunities for business leaders to comprehend organizational behaviors and gain tangible knowledge from a data science perspective, which in turn delivers intelligence for real-time decision-making and effective {talent management} for their organizations. In the last decade, talent analytics has emerged as a promising field in applied data science for human resource management, garnering significant attention from AI communities and inspiring numerous research efforts. To this end, we present an up-to-date and comprehensive survey on AI technologies used for talent analytics in the field of human resource management. Specifically, we first provide the background knowledge of talent analytics and categorize various pertinent data. Subsequently, we offer a comprehensive taxonomy of relevant research efforts, categorized based on three distinct application-driven scenarios at different levels: talent management, organization management, and labor market analysis. In conclusion, we summarize the open challenges and potential prospects for future research directions in the domain of AI-driven talent analytics.
\end{abstract}

}

\maketitle
\IEEEdisplaynontitleabstractindextext
\IEEEpeerreviewmaketitle

\IEEEraisesectionheading{\section{Introduction}\label{introduction}}

\begin{figure}[t!]
\centering
\vspace{-7mm}
\includegraphics[scale=0.3]{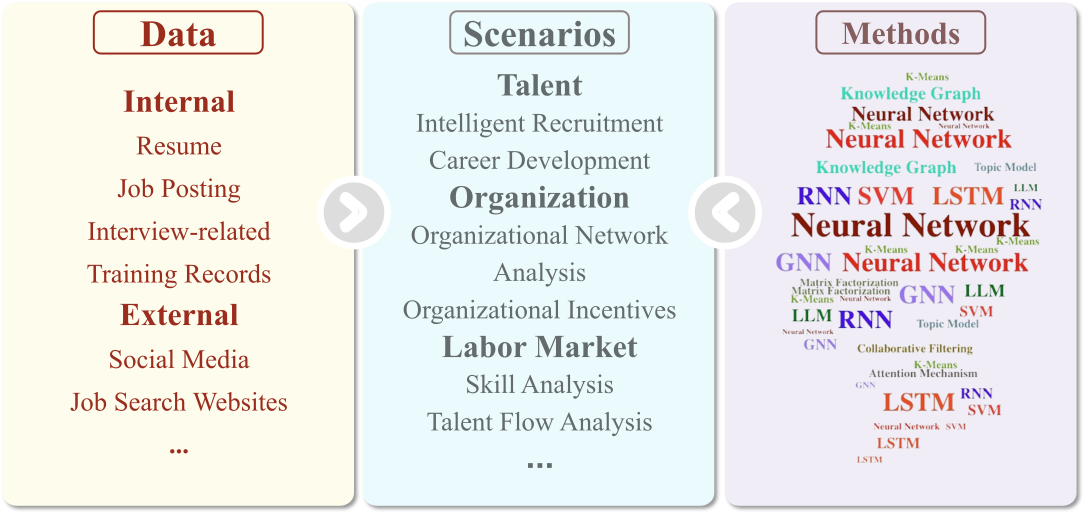}
\vspace{-2mm}
\caption{Graphical abstract of this survey from data to the proposed methods.}
\vspace{-6mm}
\label{fig:scenario} 
\end{figure}

\IEEEPARstart{I}n the world of volatility, uncertainty, complexity, and ambiguity (VUCA), talent are always precious treasures and play an important role in business success. To cope with the fast-evolving business environment and maintain competitive edges, it is critical for organizations to rethink how to make talent-related decisions in a quantitative manner. Thanks to the era of big data, the availability of large-scale talent data provides unparalleled opportunities for business leaders to understand the {patterns of} talent and management, which in turn deliver intelligence for effective decision-making and management for their organizations~\cite{sivathanu2019technology,sharma2017talent}. Along this line, as an emerging applied data science direction in human resource management, talent analytics has attracted a wide range of attention from both academic and industry circles. Specifically, talent analytics, also {known as} workforce analysis or people analytics, focuses on leveraging data science technologies to analyze extensive sets of talent-related data, empowering organizations with informed decision-making capabilities that enhance their organizational and operational effectiveness~\cite{kaur2017trends}. In practice, talent analytics plays a pivotal role in strategic human resource management (HRM), encompassing diverse applications such as talent acquisition, development, retention, as well as examining organizational behaviors and external labor market dynamics.
Generally, the research directions of talent analytics can be divided into three categories, as illustrated in Figure~\ref{fig:scenario}, including talent management, organization management, and labor market analysis. 

To be specific, first, talent management is a constant strategic process of attracting and {hiring high-potential employees, training them in needed skills}, motivating them to improve their performance, and retaining them to keep organizational competitiveness. In this particular scenario, talent analytics primarily {focuses} on individual-level analysis. For instance, it can help {human resource} managers find the right talents for different jobs in a practical way~\cite{qin2018enhancing,yao2022knowledge}, and it can reasonably {make employee performance or turnover predictions}~\cite{li2017prospecting}. Second, organization management is the art of fostering collaboration among talent and guiding organizations toward achieving success. In this scenario, talent analytics can diagnose the health of an organization and measure organizational performance by leveraging various relationship information among talents and organizations, such as organizational structure, communication patterns, and project collaborations~\cite{ye2022mane}. It can also assist the organization in effectively structuring and optimizing teams~\cite{wi2009team}. {Third, talent analytics can be leveraged from an external, macro-level perspective, particularly in labor market analysis scenarios.} It is crucial to devise talent and organizational strategies. For instance, by analyzing talent demands within the labor market, managers can effectively craft recruitment strategies~\cite{zhu2016recruitment,zhang2021talent}.

Historically, {talent analytics was proposed within the conception of HRM around the 1920s}~\cite{jackson1995understanding}. Talent analytics is usually manual in the early stages of HRM before 1970~\cite{mathys1982survey}. {At this stage, Human Resource Systems (HRS) or Human Resource Management Systems (HRMS) serve as the primary tools for talent analytics.}~\cite{boon2019systematic}. There are various types of these systems, {such as commitment or control systems}~\cite{arthur1994effects}. The assessment of the talent performance with these systems {is} conducted by the human resource managers~\cite{gerhart2000measurement}. At the same time, talent assessment-related theories have emerged, including human capital theory and resource-based theory~\cite{jackson1995understanding}.
{With the occurrence of mechanical automation, human resource information systems (HRIS) arose around the 1940s; however, before 1970, HRIS was based on sorting and tabulating equipment. At the same time, the main functions of HRIS are keeping employee records automatically, and there is no computer support}~\cite{desanctis1986human}. However, from this stage, talent analytics could be supported by aggregated information.
With the development of information systems motivated by more and more data generated during management, HRIS enhanced by computer system {has} been widely adopted in 1970s~\cite{broderick1992human}, in this stage, HRIS is a combination of the database, the computer application, the software and the hardware to record, manage, and operate human resource data. This development trend is verified by a survey, which confirms that 60\% of Fortune 500 companies use HRIS to support daily HRM operations~\cite{ball2001use}. Besides, in this stage, people analytics, {refer to} a novel, quantitative, evidence-based, and data-driven approach to manage the workforce, {was} proposed to raise the efficiency of core human resource functions such as recruiting~\cite{giermindl2022dark}. Some standard statistical methods {were adopted in this step, such as correlation analysis and simple regressions~\cite{leicht2022challenges}; this is generally referred to as descriptive analytics.} At the same time, due to the rich functions of HRIS, many evaluation studies on the specific functions of software have been carried out, such as organizational performance, turnover, and so on.

With the development of artificial intelligence (AI) algorithms, some advanced regression techniques, data mining, text mining, web mining, and forecast calculations {have been} used in talent analytics around the 2010s. 
On the one hand, talent analytics is facing digital disruption, which enables the availability of large-scale relevant data. For instance, a world-renowned job search site had 11.3 million active jobs as of January 2022~\cite{indeed202102}.
Meanwhile, LinkedIn, the largest online professional network, had 774 million members from around 200 countries as of March 2022~\cite{LinkedIn202203}, building up a wealth of labor market data. Moreover, numerous enterprises are setting up their Digital Human Resource Management Systems (Digital HRMS), enabling the collection, storage, and processing of a huge amount of talent and organizational information in a digital environment~\cite{vardarlier2020digital}. 
On the other hand, with the advent of talent-related big data, advanced AI techniques have rapidly revolutionized a series of {practices in this field} at an alarming rate, which in turn deliver intelligence for decision-making and management for their organizations. 
In this stage, {deep learning} methods have enabled the new paradigm in person-job fit~\cite{qin2018enhancing,bian2019domain} and person-organization fit~\cite{sun2019impact,sun2021modeling}, so as to achieve the efficient and accurate talent selection and development. Text mining methods have been adopted in the employer brand analysis based on the large-scale labor market data~\cite{lin2017collaborative,lin2020enhancing}, which enables the forward-looking strategic plans created for the business. At the same time, several high-tech companies are gradually incorporating AI technologies into their HRMS. 
With the strong automation capability of {large language models (LLMs)}, autonomous people analytics {was introduced around the 2020s}~\cite{papachristou2024network}. 

Accordingly, AI in talent analytics in this paper includes supervised learning, unsupervised learning, deep learning, reinforcement learning, knowledge representation, natural language processing, and so on~\cite{benbya2021special}. These techniques construct different business capabilities in people analytics, which are automation of structured (or semistructured) work processes, engagement with employees and managers, decision-making through extensive analysis of a large amount of data, {and} creation of novel outcomes~\cite{benbya2021special}.
This survey attempts to provide a comprehensive review of the rapidly evolving AI techniques for talent analytics. {Based on our investigation, we first provide a detailed taxonomy of relevant data, which lays a foundation for better leveraging AI techniques to understand talent, organizations, and their management.}
Generally, talent behaviors {are reflected} in three levels, including {the} individual level, organizational level, and market level.
Accordingly, the research efforts of the AI techniques for talent analytics from the corresponding three aspects, including talent management, organization management, and labor market analysis. Finally, we identify challenges for future AI-based talent analytics and suggest potential research directions.


\vspace{-4mm}
\section{Data for Talent Analytics}\label{data}
\vspace{-1mm}
{As enterprises undergo accelerated digital transformation, a large amount of data related to talent analytics has been accumulated. In this section, we introduce the data collected from various scenarios to provide readers with a foundational understanding of the research data and the motivations behind model design. Generally, the data can be classified into \textit{internal data} (collected from enterprise management systems) and \textit{external data} (collected from the external labor market).}


\vspace{-2mm}
\subsection{Internal Data}
Based on the described objects, internal data can be broadly divided into three categories: recruitment data, employee data, and organizational data.

\subsubsection{Recruitment Data}
\begin{figure}[t!]
	\centering
	\begin{minipage}[t]{0.95\linewidth}
		\centering
		\includegraphics[width=1.0\textwidth]{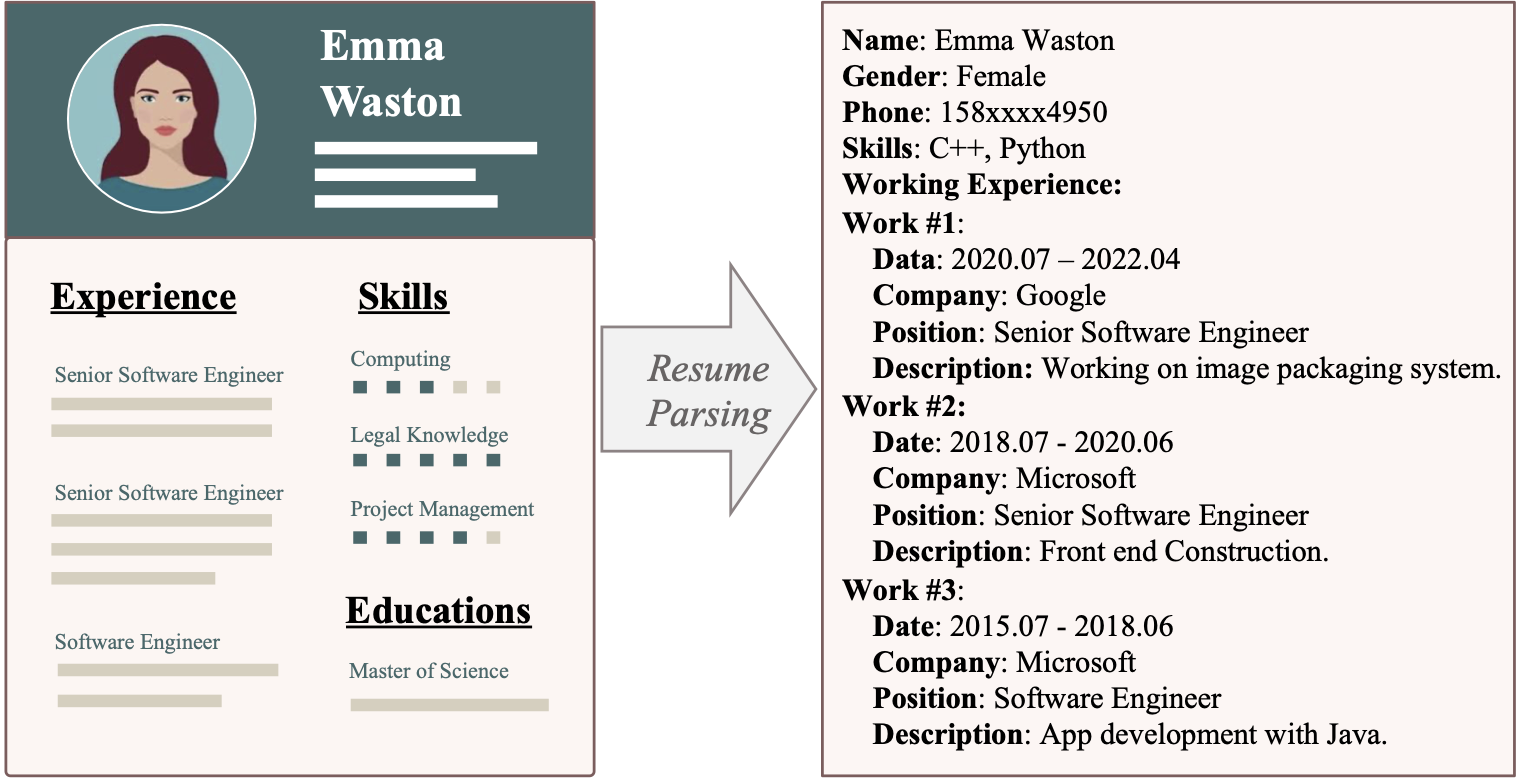}
		\caption{An example of parsing the resume.}
		\label{fig:resume}
	\end{minipage}
	\\ 
	\begin{minipage}[t]{0.95\linewidth}
		\centering
		\includegraphics[width=1.0\textwidth]{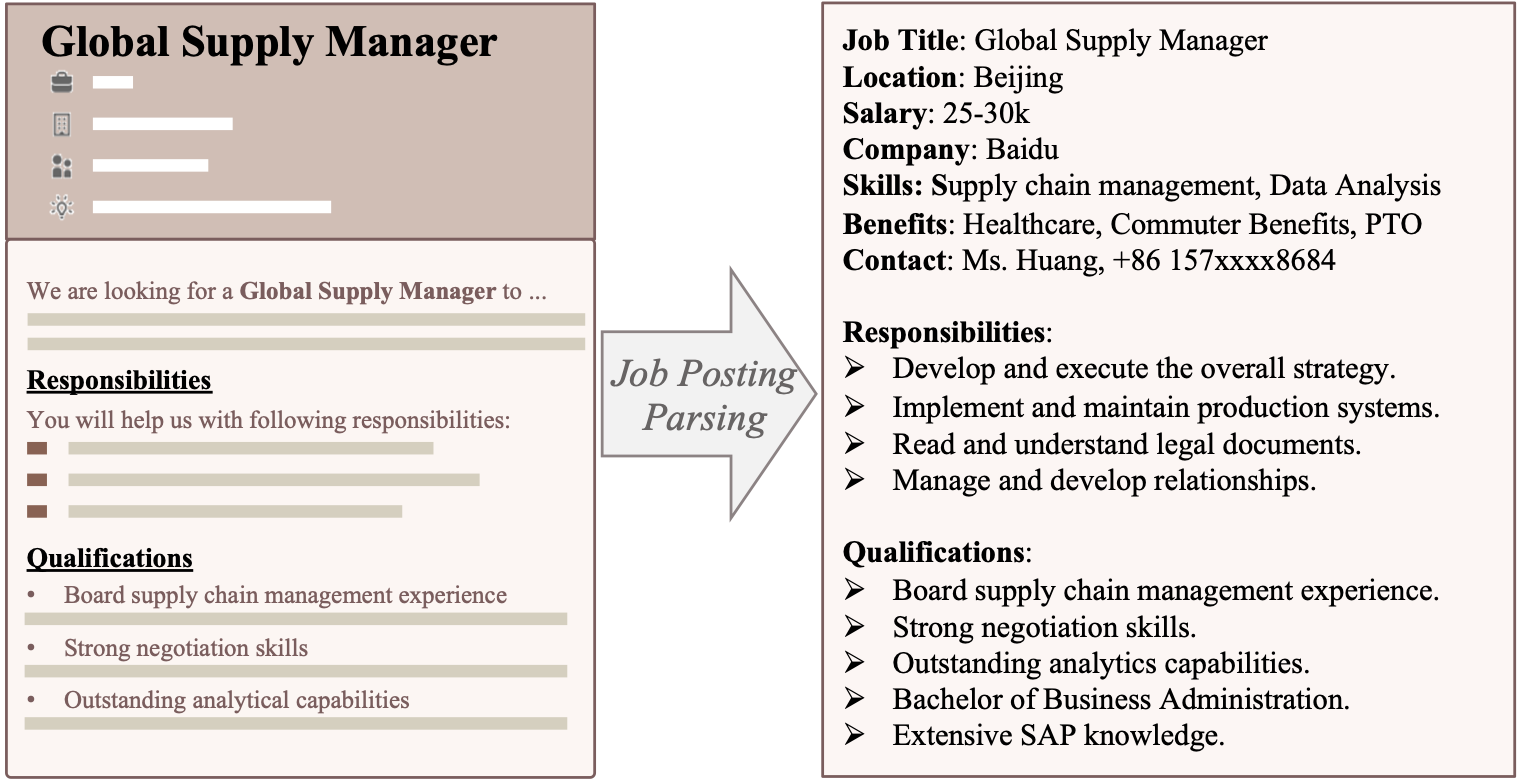}
		\caption{An example of parsing the job posting.}
        \vspace{-5mm}
		\label{fig:job_posting}
	\end{minipage}
\end{figure}

Recruitment data in pre-employment mainly includes the following types:

\noindent\textbf{Resume}:
A resume or Curriculum Vitae (CV) is a document that outlines a person's background, skills, and accomplishments, which plays a vital role in the recruitment process as it serves to facilitate talent screening and assessment~\cite{qin2018enhancing,shen2021joint}. {It is an important tool for job seekers to demonstrate their qualifications and fit for the position. Driven by the development of online recruitment, a large amount of resume data, typically in Word or PDF format, has recently been accumulated.}
As shown in Figure~\ref{fig:resume}, a resume typically comprises structured information such as gender, age, and education, as well as semi-structured information like educational experience, work experience, and project experience. 
Accordingly, several resume parsing techniques have been developed to extract the redundant information ~\cite{chen2016information,yao2023resume}.
{On this basis}, substantial efforts {have been made} in talent analytics with resume data from different perspectives.
For instance, \textsl{Yao et al.}~\cite{yao2021interactive} introduced a keyphrase extraction approach to explore job seekers' skills in resumes, and \textsl{Pena et al.}~\cite{pena2020bias} used image information in resume data to improve screening performance and explore {the fairness issue}. 
Moreover, several studies have proposed leveraging the text mining techniques to determine the matching degree between jobs and job seekers {based on} their resumes~\cite{qin2018enhancing,zhu2018person}. 
In addition, the resumes also encompass the career trajectories of the job seekers. As illustrated on the right side of Figure~\ref{fig:resume}, the candidate's profile showcases three job experiences, including a job change at Microsoft and work experience at Google. In this phase, \textsl{Zhang et al.} introduced the ResumeVis system to visualize the individual career trajectory and mobility within different organizations~\cite{zhang2018resumevis}. And the researchers further analyzed the sequential patterns of the career trajectory and proposed personalized career development recommendations~\cite{meng2019hierarchical}.

{The presence of substantial sensitive personal information in resume data poses significant challenges to the creation of publicly accessible datasets.} While some researchers have used publicly available information from sources such as academic homepages and online professional networks to build alternative datasets~\cite{dai2020joint,mittal2020methodology}, the severe lack of open datasets continues to hinder broader research in this area.



\noindent\textbf{Job Posting}:
A job posting is an advertisement for a vacant position that provides job seekers with information on the job description and requirements. The posting offers applicants a clear understanding of what the position is responsible for and what qualifications are necessary. Recently, the proliferation of online recruitment services {has made it increasingly common for job postings to be published as web pages.} Figure~\ref{fig:job_posting} illustrates a typical job posting that comprises structured information, such as the salary range and education requirements, as well as semi-structured content that includes job duty descriptions and ability requirements.
Nevertheless, it is still difficult to deal with such a large corpus of data {for} Human Resource (HR) experts manually. 
To this end, researchers have attempted to reduce the dependence on manual labor by using neural network-based techniques, particularly NLP, on voluminous job postings. 
As mentioned before, considerable effort {has been made} in Person-Job Fit ~\cite{qin2018enhancing, qin2020enhanced, bian2019domain}, which aims to match the job postings with suitable resumes. Moreover, \textsl{Shen et al.} leveraged the latent variable model to jointly model the job description, candidate resume, and interview assessment, which can further benefit several downstream applications such as person-job fit and interview question recommendation~\cite{ shen2021joint}. To reduce the expense of manual screening, researchers also extracted the job entities from postings and generated interview questions automatically~\cite{qin2019duerquiz, qin2023automatic, li2023ezinterviewer}. 
Apart from these in-firm applications, some studies are carried out to provide comprehensive insights into the global labor market. For instance, researchers have proposed several data-driven methods for salary analysis across different companies and positions~\cite{kenthapadi2017linkedin, meng2018intelligent}. \textsl{Zhang et al.} utilized large-scale job postings from one of the largest Chinese online recruitment websites and forecast fine-grained talent demand in the recruitment market~\cite{zhang2021talent}. {Moreover, several studies seek to measure the popularity of job skills and predict their evolving trends over time~\cite{xu2018measuring}.} Along this line, \textsl{Sun et al.} further focus on measuring the values of job skills based on massive job postings, contributing to the quantitative assessment of job skills~\cite{sun2021market}.

\noindent\textbf{Interview-related Data}:
{Data collected during interviews, whether conducted in person or via video, plays a critical role in evaluating applicants' overall qualifications for their intended positions. Textual and video-based assessments derived from these processes enable comprehensive candidate evaluations and promote the integration of AI into HR management practices.}
To address the subjectivity of traditional interviews, \textsl{Shen et al.} utilized the latent variable model to explore the relationship among job descriptions, candidate resumes, and textual interview assessments~\cite{shen2018joint, shen2021joint}. The results provide an interpretable understanding of job interview assessments. 
Indeed, textual interview assessments within a company are usually private and sensitive, whereas video assessments draw more attention.
For instance, several studies extract multimodal features from the videos for automatic analysis of job interviews~\cite{chen2016automated}. In addition, \textsl{Hemamou et al.} proposed a hierarchical attention model to predict the employability of the candidates using multimodal information, including text, audio, and video~\cite{hemamou2019hirenet}. Along this line, \textsl{Chen et al.} leveraged a hierarchical reasoning graph neural network to automatically score candidate competencies using textual features in asynchronous video interviews~\cite{chen2020hierarchical}. It is important to note that the use of interview data must account for potential ethical risks. Although researchers typically state that they have obtained permission to use data in their papers, {usage of} interview data {should be approached with caution in real-world scenarios.}


\subsubsection{Employee Data}
\begin{figure}[t!]
\centering
\includegraphics[scale=0.2]{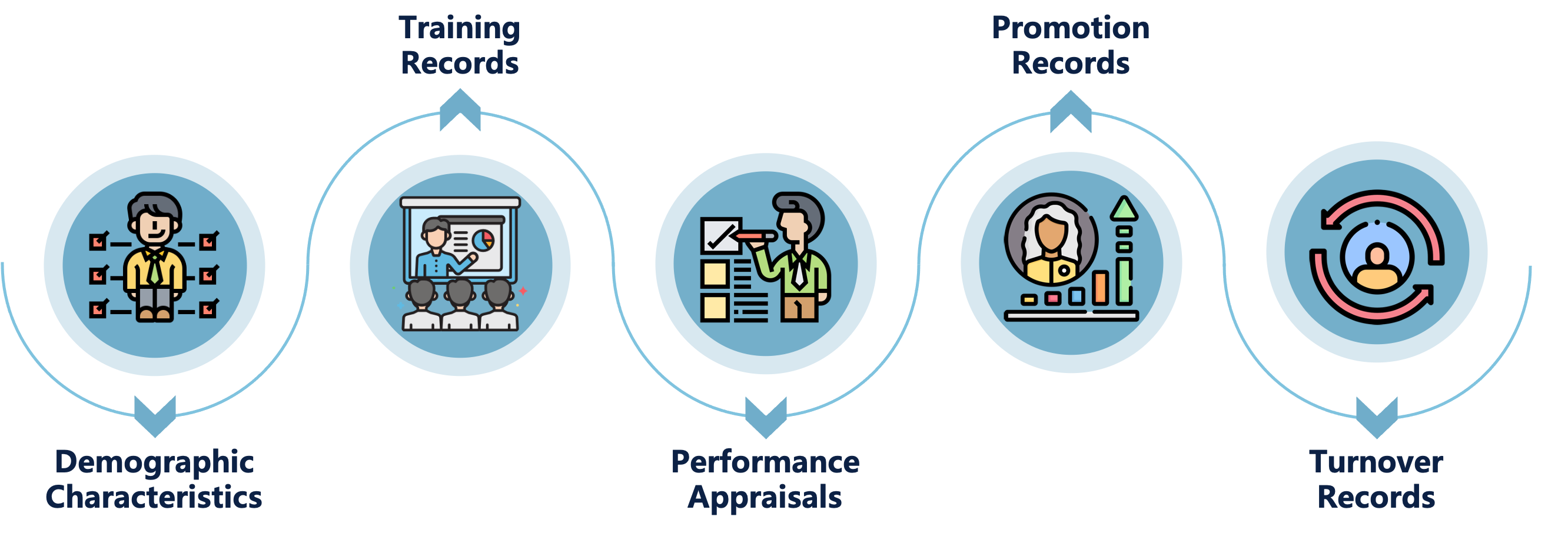}
\caption{An overview of employee-related data.}
\vspace{-6mm}
\label{fig:employee} 
\end{figure}

Regarding the development of employees within a company, a significant amount of employee data has been accumulated, {such as training records and individual work outcomes, as illustrated in Figure~\ref{fig:employee}.} 

\noindent\textbf{Employee Profiles}: Employee profiles typically describe an individual based on two main aspects: demographic characteristics and individual work outcomes. {Specifically}, the former branch includes characteristics such as age, gender, and education levels, which can be used to enhance employee representations and benefit various downstream analyses, such as career mobility prediction~\cite{qutub2021prediction} and performance forecasting~\cite{mahmoud2019performance}. In addition to these static variables, individual work outcomes depict the dynamic career development from different dimensions, such as performance appraisals, promotion, and turnover records. 
{In particular, a performance appraisal refers to the systematic evaluation of an employee's job performance and productivity, typically conducted by line managers. In addition, promotion and turnover records reflect employee movements within and across companies, respectively.}
All of this information contributes to further insights into employee dynamics.
For instance, researchers have leveraged performance appraisals to identify the high-potential talents within a company~\cite{ye2019identifying}. \textsl{Li et al.} utilized {employees'} static profiles, performance appraisals, and reporting lines to model career development within a company, focusing on turnover and career progression~\cite{li2017prospecting}. \textsl{Sun et al.} proposed to capture the dynamic nature of person-organization fit based on individual profiles, reporting lines, and communication records~\cite{sun2019impact}. To investigate the contagious effect of turnovers, researchers have utilized both employee profiles and turnover data~\cite{teng2019exploiting}. Furthermore, \textsl{Hang et al.} leveraged five kinds of standardized data, including employee turnover records, to predict the turnover probability and period~\cite{hang2022outside}.

\noindent\textbf{Training Records}: Employee training is a program designed to improve the performance of employees by equipping them with specific skills. Ongoing employee training has proven to be crucial in attracting and retaining top talent. Typically, the training record describes the learning path of an employee, which is a sequence of different skills. 
Based on these training records, considerable effort has been devoted to exploring the learning patterns of employees. For instance, \textsl{Wang et al.} utilized both learning records and skill profiles of employees from a high-tech company in China to develop a personalized online course recommendation system~\cite{wang2020personalized}. Along this line, \textsl{Srivastava et al.} collected employees' training and work history from a large multinational IT organization to provide personalized next training recommendations~\cite{srivastava2018s}. {Furthermore, several researchers have contributed insights into the study of employee competencies~\cite{li2020data,liu2009application}.} For instance, multi-dimensional features, including learning and training dimensions, were collected from a Chinese state-owned enterprise to provide competency assessment for employees~\cite{liu2009application}.

{Due to legal regulations, employee profiles and training records, which encompass both employee privacy and the company’s proprietary training strategies, should not be made available. This limits the ability of the field to gain broader research attention. Furthermore, many existing studies on employee datasets lack discussion on gender and racial diversity, which hinders the ability to perform fairness assessments within these datasets. This also makes it challenging to identify and address potential biases that may arise in career mobility predictions, performance evaluations, or training outcomes for different demographic groups. As a result, research aimed at understanding the disparate impact of HR decisions across diverse employee populations remains constrained, hindering efforts to ensure equitable career development opportunities and unbiased workforce management strategies.}

\subsubsection{Organizational Data}

\begin{figure}[t!]
\centering
\includegraphics[width=1.0\columnwidth]{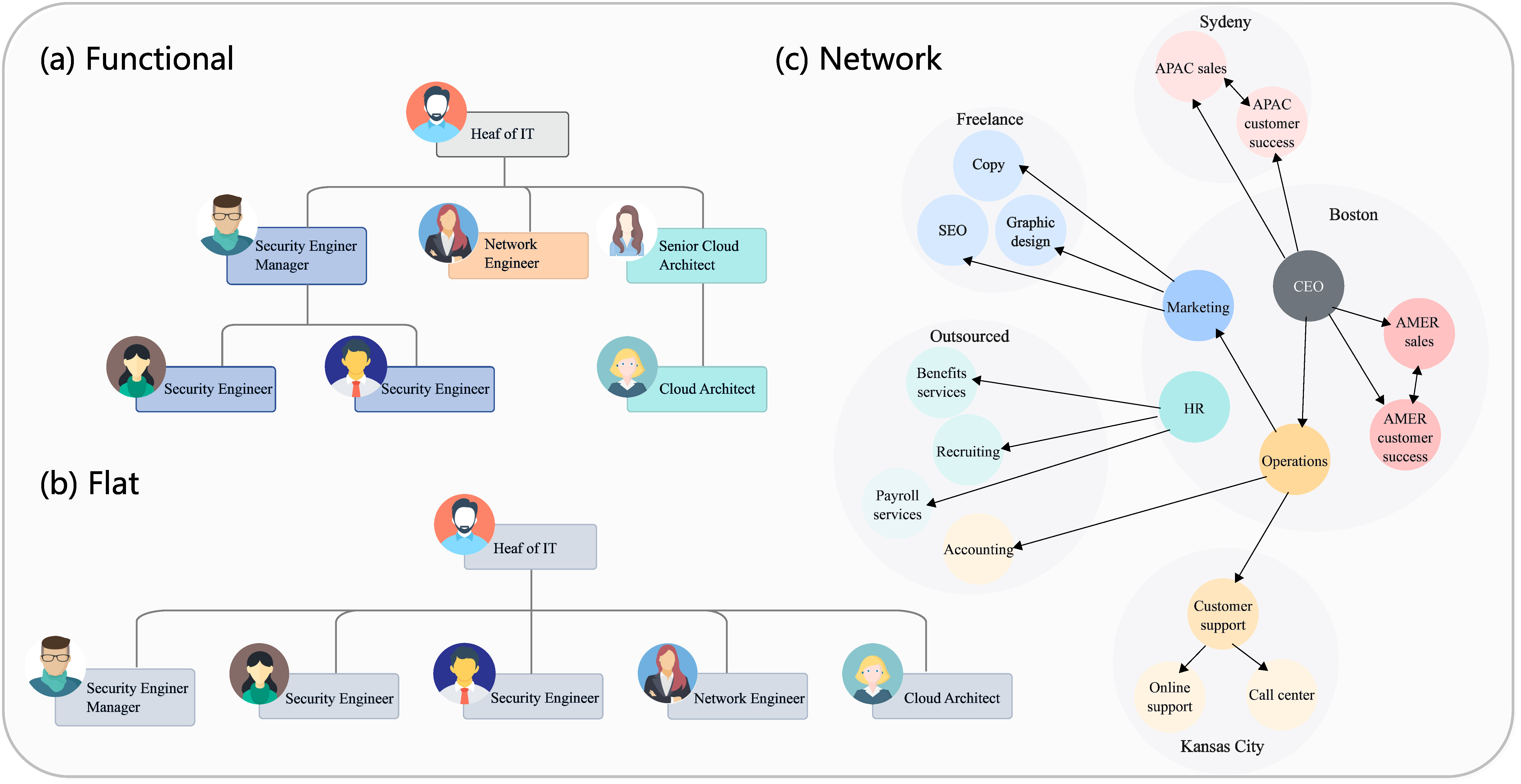}
\caption{Three common types of organization structure.}
\vspace{-4mm}
\label{fig:org_structure} 
\end{figure}

An organizational structure is a system that outlines how activities are directed toward the achievement of organizational aims~\cite{pugh2007organization}, which plays an important role in decision-making and knowledge management. Generally, an organization is commonly represented as a hierarchical tree structure, which can take on diverse forms, such as matrix, flat, and network structures. Figure~\ref{fig:org_structure} shows several common types of organizational structures. 
Typically, existing studies explore these complex structures from various dimensions, such as reporting lines and in-firm social networks.

{Reporting lines are generally the most representative aspect of an organizational structure, as they delineate how authority and responsibility are allocated within an organization.} 
Regarding this point, \textsl{Sun et al.} developed an organization {structure-aware} convolutional neural network to hierarchically extract compatibility features for measuring person-organization fit and its impact on talent management~\cite{sun2019impact, sun2021modeling}. 
Nevertheless, due to privacy concerns, mainstream studies utilize in-firm social networks for human resource management. In general, an in-firm social network can be formed from email or Instant Messaging (IM) records across employees.
For example, text-based communication has been used by several machine learning classifiers to identify group mood~\cite{klunder2020identifying}. Besides, \textsl{Cao et al.} leveraged the lasso regression model to explore team viability using text conversations of online teams~\cite{cao2021my}. In addition to social networks, researchers have also taken other information into account.
For example, \textsl{Ye et al.} utilized both email communication and a high-potential talent list to identify employees with high potential~\cite{ye2019identifying}. Along this line, \textsl{Teng et al.} further utilized datasets from three sources for organizational turnover prediction, including profile and turnover, social network, and job levels~\cite{teng2021exploiting}.

\vspace{-2mm}
\subsection{External Data}

Apart from the aforementioned internal data, external sources also contribute to a comprehensive understanding of the labor market, which can be broadly classified into two categories: social media platforms and job search websites.


\noindent\textbf{Social Media}:
{Widely used} social media platforms that contribute to a comprehensive understanding of the labor market include {Twitter, officially known as X since 2023}, Facebook, and news reports.
With the help of NLP techniques, numerous studies have been carried out to explore the semantic information in this corpus.
For example, more than 60,000 tweets related to nine energy companies were collected for sentiment analysis {from} Twitter~\cite{ikoro2018analyzing}. To gain further insight into the impact of public opinion, \textsl{Spears et al.}~\cite{spears2021impact} collected earnings reports and news articles spanning eight years from four companies. 
The results indicate that companies may experience a decline in valuation when they receive negative publicity.

\noindent\textbf{Job Search Websites}:
Recent years have witnessed the rapid growth of job search websites, such as \textit{Indeed}, \textit{LinkedIn}, and \textit{Glassdoor}. Specifically, Indeed and Glassdoor allow users to comment on a company, providing an overall understanding of the employer brand. For instance, \textsl{Lin et al.}~\cite{lin2017collaborative, lin2020enhancing} collaboratively modeled both textual (i.e., reviews) and numerical information (i.e., salaries and ratings) for learning latent structural patterns of employer brands. In addition, \textsl{Bajpai}~\cite{bajpai2019aspect} leveraged the data from Glassdoor to perform aspect-level sentiment analysis. Along this line, large-scale reviews of Fortune 500 companies are collected to identify topics that matter to employees~\cite{schmiedel2019topic}. Differently, LinkedIn provides a wide range of business services, including job listings, professional profile creation, and career development services, with personal profiles being the most analyzed, as they describe users' employment history. For example, \textsl{Park et al.}~\cite{park2019global} used LinkedIn’s employment history data from more than 500 million users over 25 years to construct a labor flow network of over 4 million firms worldwide, demonstrating a strong association between the influx of educated workers and financial performance in detected geo-industrial clusters. 
{Based on the public job advertisements collected in China between 2021 and 2023, \textsl{Job-SDF}~\cite{xi2024job} {is} a dataset designed to forecast job skill demand, offering new perspectives by evaluating forecasting models across different granularities, such as occupation and region. The benchmarking of various models on this dataset provides valuable insights for further research in skill demand forecasting.}


Furthermore, {several third-party business investigation platforms} offer detailed information about companies and their board members' relationships, such as \textit{Crunchbase} and \textit{Owler}. These details can be viewed as complementary information to the job search websites. Building upon this foundation, one can gain deeper insights into the aligned companies and conduct more relevant research, such as analyzing cooperative and competitive relationships and providing investment target recommendations.

\vspace{-2mm}
\section{Talent Management}\label{talent}

Talent management, which focuses on placing the right person in the right job at the right time, has emerged as a predominant human capital topic in the early twenty-first century~\cite{cascio2008research}.
Specifically, the management process includes the whole process of talent entering the organization to development. Accordingly, we first begin by describing various intelligent talent recruitment scenarios, such as job posting generation~\cite{qin2022towards} and talent searching~\cite{ha2017query, manad2018enhancing, gupta2018automation}. After entering the organization, to ensure the sustainable development of talent, timely and accurate feedback is critical. As a result, second, we discuss two primary issues in talent assessment: interview question recommendation~\cite{shen2021joint, qin2019duerquiz, shi2020learning} and assessment scoring~\cite{chen2020hierarchical, li2020data}. Finally, from the whole development process, career development is important for managing human capital resources and individual development. Accordingly, we outline several post-employment career development problems, including course recommendations~\cite{wang2020personalized} for employee training and employee dynamics analysis~\cite{zhao2018employee, long2018prediction}.
In the following sections, we will delve into these issues in detail.

\vspace{-2mm}
\subsection{Talent Recruitment}

Talent recruitment is a critical component of talent management, as it involves identifying the right candidates for positions within an organization. The quality of this function can significantly impact the organization's future development, which is why considerable human and material resources have been invested in ensuring the efficiency and effectiveness of related procedures. According to a Forbes article, US corporations spend nearly 72 billion annually on various recruiting services, and the global amount is likely three times larger. 
However, traditional talent recruitment methods rely heavily on the personal knowledge and experience of recruiters, which may introduce bias due to the subjective nature of the process. This potential bias can be exacerbated by varying levels of experience and personal qualities among different recruiters. Fortunately, the rapid development of online recruitment platforms, such as LinkedIn and Lagou, has ushered in a new era of data-driven talent recruitment, empowered by AI technologies.

\subsubsection{Job Posting Generation}

A job posting comprises both job duties, which describe the responsibilities and tasks of the role to candidates, and job requirements, which outline the professional experience, skills, and domain knowledge that an employer expects from the ideal candidate to perform the role. 
Job duties are tailored to each specific position. However, identifying the necessary capabilities and aligning them with job duties can be challenging, particularly when recruiters have limited experience and domain knowledge.
{A report by Allegis Group shows that only 39\% of candidates find job descriptions clear, highlighting the challenges organizations face in writing them~\cite{qin2022towards}.}

\noindent \textbf{Job Requirement Generation.} To improve the quality of job postings, many researchers are focusing on generating job requirements from job duties to more accurately attract suitable candidates~\cite{qin2022towards}. They approach this task from a data-driven perspective by collecting high-quality training data and framing it as a text generation problem. Specifically, the task of generating job postings from the perspective of job requirements can be formalized as follows:


\begin{figure}[t!]
\centering
\includegraphics[width=1\columnwidth]{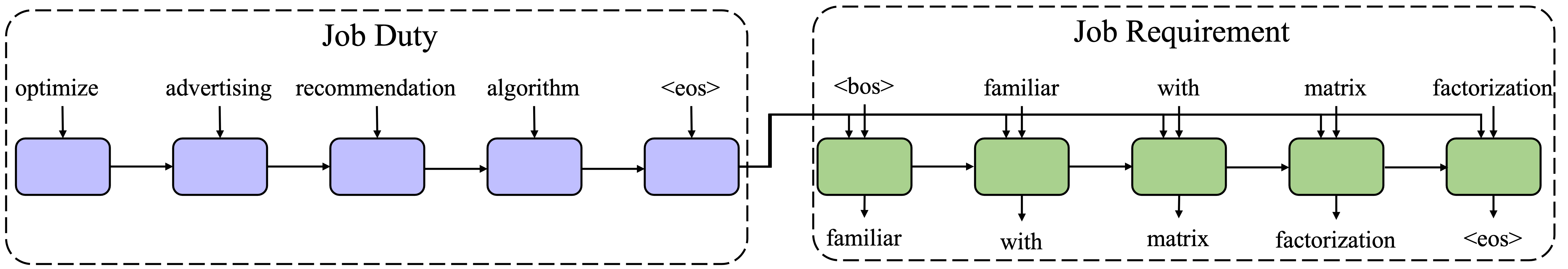}
\caption{The illustration of job requirement generation.}
\vspace{-5mm}
\label{fig:JobPostingGeneration} 
\end{figure}

\begin{definition}[Job Requirement Generation]
Given a set of job postings, denoted as $\mathcal{C}$, where each posting $C_i\in \mathcal{C}$ includes a job duty $X_i$ and a job requirement $Y_i$, the goal of job requirement generation is to train a model $M$. This model should be capable of generating a fluent and rational job requirement $Y_{new}$ when provided with a new job duty $X_{new}$.

\end{definition}

Technically, this task can be viewed as a text-to-text generation problem, where the job duties and job requirements are typically long sequences of text.
To address this task, sequence-to-sequence models using an encoder-decoder architecture are commonly employed, as illustrated in Figure~\ref{fig:JobPostingGeneration}.
For instance, \textsl{Liu et al.} applied two Long-Short Term Memory (LSTM) layers as the encoder and decoder, respectively, to extract the key information from job duties and generate the job requirement, respectively~\cite{liu2020hiring}.
Since it is important to precisely use and organize skill-related keywords in job requirements, they implemented the decoder in a two-pass manner. The first-pass decoder focuses on predicting skill-related keywords, while the second-pass decoder uses these predicted skills to guide the generation of fluent text. In particular, the attention mechanism is employed to integrate hidden states from the LSTM layer of the encoder with context information, which enhances skill prediction in the decoder.
Furthermore, \textsl{Qin et al.} trained a neural topic model in job duties to capture the global topic information~\cite{qin2022towards}. The topic distribution of each job duty is used as the context information to guide the generation of each word in the LSTM-based decoder.

\noindent \textbf{Writing Assistants.} Additionally, unlike the approaches that require models to learn the relationship between job duties and job requirements, some researchers focus on developing job posting writing assistants with an emphasis on content quality~\cite{lorincz2022transfer}. Most existing research adopts a data-to-text perspective to construct job posting generators. These models convert structured inputs, such as salary and working hours~\cite{lorincz2022transfer}, into natural language text that adheres to the conventional style of job postings, which can significantly alleviate the workload of human resources personnel. For instance, with inputs like ``MINSALARY = 12k, MAXSALARY = 15k, SALARYTYPE = per month'', the model can generate a sentence for a job posting such as ``You will receive a monthly salary between \$12,000 and \$15,000.'' Leveraging the advanced text understanding and generation capabilities of large language models (LLMs), \textsl{Lorincz et al.} fine-tuned the mT5 model to generate the benefits section of job postings from structured data~\cite{lorincz2022transfer}. They transformed structured data into a format compatible with mT5 by concatenating feature names with their corresponding values.

Moreover, \textsl{Borchers et al.} attempted to generate job postings directly from job names using LLMs~\cite{borchers2022looking}. Furthermore, the authors employed prompt engineering and gathered high-quality, gender-unbiased training data to enable the generator to produce unbiased job postings. {Additionally, some researchers focus on detecting bias within job descriptions to enable HR departments to identify and mitigate these issues before posting the position~\cite{bohm2020analysing,mao2023developing}. For instance, \textsl{Böhm et al.} proposed a tool called BetterAds, which not only identifies gender-stereotypical language and calculates a corresponding bias score but also offers suggestions for rephrasing the text in a more gender-neutral manner~\cite{bohm2020analysing}. \textsl{Mao et al.} explored the application of language models such as RoBERTa to detect gender and age biases in job descriptions~\cite{mao2023developing}. They also experimented with using ChatGPT to generate job descriptions in both gender-neutral language and feminized language to examine whether these approaches could specifically attract female applicants.}

\noindent \textbf{Evaluation.} The performance of job posting generation models can be assessed primarily from several perspectives: validity, fluency, realism, and bias.
\begin{itemize}[left=0pt]
    \item \textbf{Validity.} Does the generated job posting include appropriate recruitment information, such as skill requirements and benefits? \textsl{Qin et al.} evaluated the performance of various models through human assessments~\cite{qin2022towards}, while \textsl{Liu et al.} assessed the skills listed in generated job postings against a predefined skill vocabulary, measuring discrepancies with ground truth job postings~\cite{liu2020hiring}. They measured in terms of precision, recall, and F1-value.
    \item \textbf{Fluency.} Does the generated job posting read smoothly and fluently? This dimension is typically assessed through human evaluation~\cite{qin2022towards}.
    \item \textbf{Realism.} Can the generated job postings be distinguished from those written by humans? Realism, a concept first introduced by \textsl{Borchers et al.}, is employed to evaluate the quality of generated job postings~\cite{borchers2022looking}. The authors developed a Multinomial Naive-Bayes (MNB) classifier to serve as a discriminator, assessing whether it can distinguish between job postings written by humans and those generated by machines.
    \item \textbf{Bias.} Does the generated job posting contain potential biases, such as gender bias? \textsl{Borchers et al.} leveraged several existing lists of gender-laden words, including GenderCoded Word Prevalence, Superlative Prevalence, GenderLaden Scoring, Connotation Frames, and NRC VAD Lexicon, to determine whether generated job postings exhibit a bias toward one gender over another.
\end{itemize}
Furthermore, as job posting generation constitutes a text generation task, most existing studies evaluate the quality of generated postings by comparing them to ground truth job postings. These ground truth postings, typically crafted by human resources experts and refined through data filtering, serve as benchmarks for automated assessment. Specifically, the standard ROUGE metric, including ROUGE-1, ROUGE-2, and ROUGE-L, can be used. These measure the overlap of unigrams, bigrams, and the Longest Common Subsequence (LCS) between the ground truth and generated job postings. Additionally, the BLEU metric can be employed for evaluation, which assesses the co-occurrences of n-grams between the ground truth and the generation. 

Recent studies have demonstrated that ChatGPT and other {LLM-based models} can be utilized for text evaluation, achieving performance close to that of human evaluators~\cite{gilardi2023chatgpt}. This advancement opens new possibilities for the automated assessment of fluency, realism, and other dimensions in generated job postings.

\noindent \textbf{$\diamondsuit$ Takeaway.} 
\begin{itemize}[left=0pt]
    \item \textbf{Advantages of AI technology:} \textsl{(1) Existing AI-based methods for generating job postings not only effectively reduce the workload of human resources staff but also produce high-quality job postings that help attract suitable candidates. (2) With the rapid advancement of LLM technology, there has been a notable enhancement in both the ease of implementation and the quality of generated job postings. (3) AI-based generative models can potentially create unbiased job postings.}
    \item \textbf{Limitations and future directions:} \textsl{{(1) Research has demonstrated that job advertisements in the recruitment market frequently exhibit latent biases, including those related to gender, age, and race. However, most existing research on job posting generation relies on models trained using historical data, which can perpetuate or even introduce new biases in the generated job postings. While some studies have proposed methods for detecting bias in job descriptions~\cite{bohm2020analysing}, there is still a notable absence of end-to-end approaches capable of generating entirely unbiased job postings. (2) Furthermore, static historical recruitment data overlooks the dynamic nature of job requirements in the labor market. A potential research direction is to explore how models that reflect changing market demands~\cite{xi_pretrain_2024} can be incorporated to generate job descriptions that are responsive to current market trends.} 
    (3) Existing research has focused only on specific individual scenarios in job posting generation, lacking a unified model that can handle multiple generation scenarios simultaneously. Recent advancements in instruction learning have equipped LLMs with robust multitasking capabilities~\cite{chen2023minigpt}, making the development of more powerful job posting writing assistants feasible. (4) Existing research lacks an analysis of how job seekers' behaviors are influenced by reading AI-generated job postings.}
\end{itemize}

\subsubsection{Resume Understanding}
Resume understanding, also known as resume parsing, aims to extract semantically structured information from resume documents, which can facilitate a wide range of intelligent talent analysis applications, such as talent searching~\cite{ha2016search} and person-job fitting~\cite{qin2018enhancing}. As illustrated in Figure~\ref{fig:resume}, a resume typically includes the candidate’s personal information (e.g., name, phone number), educational background (e.g., major), work experience (e.g., company name), and various other relevant details. Specifically, this task can be formalized as follows:

\begin{definition}[Resume Understanding]
Given a set of resume documents $\mathcal{R}$, the target of resume understanding is to learn a model $M$, which can extract relevant segment $S_{i,j}$, corresponding to specific type of factual information $Y_j$, such as names and experience, from each resume $R_i \in \mathcal{R}$.
\end{definition}

Typically, resume understanding begins by converting various forms of resume files into textual formats. This conversion is accomplished using technologies or tools such as Optical Character Recognition (OCR) and PDF parsers. Consequently, resume understanding becomes a text-mining task, enabling the extraction of structured information from the textual data. The advancement of resume understanding technology is closely linked to innovations in information extraction techniques. Initially, \textsl{Kopparapu} utilized keyword searches and rule-based matching to extract six major fields of information from resumes: ``name'', ``software skills'', ``qualification'', ``experience'', and ``email''~\cite{kopparapu2010automatic}. Considering the hierarchical structure of resume information, for example, the ``personal information'' usually includes details such as ``name'' and ``gender''. In~\cite{yu2005resume}, \textsl{Yu et al.} first divided a resume into different blocks (e.g., ``personal information'' and ``education information'') using a Hidden Markov Model (HMM). They subsequently employed both Support Vector Machines (SVM) and Hidden Markov Models (HMM) to extract detailed information from these blocks. By leveraging traditional machine learning models such as HMM~\cite{yu2005resume}, SVM~\cite{yu2005resume}, and Conditional Random Fields (CRF)~\cite{chen2016information}, resume understanding has achieved an accuracy rate of over 80\% across most fields.

Although these traditional models achieve good performance, they incur the cost of extensive feature engineering. With the development of deep learning and its broad application in information extraction, many researchers have shifted to utilizing deep learning to construct more effective resume understanding models~\cite{ayishathahira2018combination,pham2018study,jin2019lstm,li2021method}. Most deep-learning approaches view resume understanding as a Named Entity Recognition (NER) task. \textsl{Ayishathahira et al.} utilized a Bidirectional LSTM-CNN architecture to perform resume information extraction~\cite{ayishathahira2018combination}. \textsl{Van et al.} leveraged Convolutional Neural Networks (CNNs) to learn the character-level representations of words in resumes, and utilized the classic neural NER architecture, BiLSTM-CRF, to enhance resume information extraction performance~\cite{pham2018study}. Meanwhile, \textsl{Jin et al.} integrated a highway network and self-attention mechanism into BiLSTM-CRF~\cite{jin2019lstm}. Subsequently, the introduction of pretrained language models such as BERT has significantly improved the representation of resume text, further enhancing the performance of deep learning-based resume understanding models~\cite{li2021method}. 

Prior research has primarily transformed resume documents into plain text inputs, thereby overlooking the multimodal information inherent in these files, such as layout information, which is crucial for resume understanding. \textsl{Wei et al} combined the pre-trained model RoBERTa with a Graph Convolutional Network (GCN) to extract structured information from resumes, taking into account layout and positional features~\cite{wei2020robust}. Recently, the integration of multi-modal features with pretrained models for document understanding has gained widespread adoption~\cite{jiang2024efficient,xu2021layoutxlm}. Inspired by the BERT architecture, LayoutLM was the first to enhance the masked language modeling task by incorporating the 2D coordinates of each token as layout embeddings~\cite{xu2020layoutlm}. This approach enables the model to capture interactions between text and layout information, leading to an improvement of approximately 15\% compared to text-only methods. Subsequently, advancements in methods such as LayoutXLM~\cite{xu2021layoutxlm} have progressively enhanced the performance of multimodal pretrained models in document understanding. These technologies have also been extensively adapted for resume understanding. Given that resume documents are more text-centric and voluminous compared to traditional document types like receipts, \textsl{Yao et al.} proposed a hierarchical multi-modal pre-training model tailored for long document understanding~\cite{yao2023resume}. Additionally, to tackle the scarcity of annotated training data for resumes, the authors developed a distantly supervised sequence labeling method. This method, trained within a self-distillation-based self-training learning framework, significantly improved the model's performance in resume understanding with limited training data. \textsl{Jiang et al.} introduced a novel layout-aware multi-modal fusion transformer that encodes resume segments by integrating textual, visual, and layout information~\cite{jiang2024efficient}. Additionally, they developed a multi-granularity sequence labeling task to address the hierarchical relationships among the fields to be parsed.

\noindent \textbf{Evaluation.} The performance of resume understanding models is typically evaluated at the instance level~\cite{yao2023resume,jiang2024efficient}. This evaluation assesses whether an extracted instance $S_{i,j}$ for the specific type of factual information $Y_j$ is considered correct, which is only when it is identical to a hand-annotated instance. 

\noindent \textbf{$\diamondsuit$ Takeaway.} 
\begin{itemize}[left=0pt]
    \item \textbf{Advantages of AI technology:} \textsl{(1) AI-based resume understanding underpins ATS and online recruitment, enabling a range of intelligent recruitment services. (2) Multimodal pre-trained models have emerged as the SOTA solution paradigm. }
    \item \textbf{Limitations and future directions:} \textsl{(1) Although \textsl{Yao et al.} considered performance optimization of resume understanding models under limited annotated data~\cite{yao2023resume}, there is still a lack of systematic research in low-resource scenarios. (2) Recent advancements show that generative LLMs possess strong capabilities for zero/few-shot learning and text annotation. The potential of LLM-based resume parsing models remains an area ripe for exploration.}
\end{itemize}

\subsubsection{Talent Searching}
Talent searching is designed to identify suitable candidates based on search queries provided by recruiters or hiring managers. Formally, considering the candidate set $\mathcal{R} = \{R_1, .., R_n\}$, where each candidate $R_i$ denotes his/her resume consisting of work experience, education experiences, {and so on.} 
Then, talent searching can be approached as an information retrieval task:

\begin{definition}[Talent Searching]
Given the candidate set $\mathcal{R}$ and a searching query $q$ consisting of search criteria, the goal of talent searching is to determine a subset of candidates $\mathcal{U} \subset \mathcal{R}$ satisfying the search criteria, and rank those candidates based on the fitness.    
\end{definition}

The key to talent search lies in measuring the fitness between the search query $q$ and each resume $R_i$. \textsl{Manad et al.} established connections between queries and candidates through skill matching. They extracted skill information from both queries and resumes, then ranked candidates based on scores that reflect the level of skill proficiency demonstrated in the resumes~\cite{manad2018enhancing}. 
\textsl{Wang et al.} focused on queries composed of competency keywords, such as skill and activity keywords. They utilized the BERT model to extract competency keywords from resumes and applied the weighted average method to calculate candidates' scores for different competency keywords. Meanwhile, the authors enhanced the effectiveness of talent search by leveraging a structured Competence Map (CMAP) to explore the relationships between various competency keywords~\cite{wang2021analysing}. In~\cite{apatean2017machine}, \textsl{Anca et al.} developed several independent classifiers based on traditional machine learning models, such as KNN and LDA, to categorize candidate attributes in areas such as Education (e.g., University), Programming Languages (e.g., Python), and Foreign Languages (e.g., English, French, German), thus enabling recruiters to search suitable candidates based on these aspects.

As talent search systems accumulate substantial historical data, researchers attempt to enhance retrieval performance through supervised training. For instance, \textsl{Ozcaglar et al.} proposed a two-level ranking system to integrate structured candidate features by combining Generalized Linear Mixed (GLMix) models and Gradient Boosted Decision Tree (GBDT) models, using recruiter actions as supervised information to learn candidate ranking~\cite{ozcaglar2019entity}. \textsl{Geyik et al.} involved recruiters' immediate feedback on recommendation results to cluster candidates according to recruiters' intents~\cite{geyik2018session}. Subsequently, they developed a multi-armed bandit-based approach to select the appropriate intent cluster for each recruiter, which was then used to rank the candidates within that cluster. Considering that queries in talent search scenarios can be quite complex, combining several structured fields (such as canonical skills and company names) and unstructured fields (such as free-text and keywords), \textsl{Ramanath et al.} have developed deep learning-based embedding models for both queries and candidates, and they explored learning to rank approaches with DNN models~\cite{ramanath2018towards}. \textsl{Yang et al.} further proposed a three-stage cascaded ranking model, integrating DNN and BERT, to enhance a deep learning-based talent search system, and accounted for personalized preferences of different users in the final stage~\cite{yang2021cascaded}.

Considering the challenge recruiters face in crafting effective search queries, \textsl{Ha et al.} developed a novel talent search system on the LinkedIn platform that allows users to select ideal candidate examples as queries~\cite{ha2016search,ha2017query}. They extracted keywords related to titles, skills, and companies from these candidates to reconstruct queries, built training data for a Query-By-Example retrieval system using historical data from a Query-By-Keyword system, and employed the Coordinate Ascent algorithm to optimize the model. 

\noindent \textbf{Evaluation.} Since talent searching is a classic retrieval task, its performance is typically evaluated using ranking metrics such as Precision@N and NDCG@N~\cite{ozcaglar2019entity,ha2016search,ha2017query}, where N denotes the top-N results produced by each model.

\noindent \textbf{$\diamondsuit$ Takeaway.} 
\begin{itemize}[left=0pt]
    \item \textbf{Advantages of AI technology:} \textsl{(1) AI-based talent search models can significantly enhance recruiters' efficiency in selecting candidates and are crucial components in intelligent ATS, online recruitment platforms, and employment-focused social media platforms. (2) Existing AI-based studies enable talent search tasks that incorporate queries for jobs, keywords, and various complex requirements.}
    \item \textbf{Limitations and future directions:} \textsl{(1) Existing supervised talent search models heavily depend on abundant training data and struggle in cold-start or data-sparse scenarios. Potential solutions, such as data augmentation and transfer learning techniques, merit further exploration. (2) Technologies related to novel talent search scenarios, such as conversational search, require further investigation. (3) With the rapid development of LLMs recently, there has been increasing research aimed at leveraging LLMs to enhance information retrieval systems, such as search engines. However, there is still limited research in the context of talent search scenarios. }
\end{itemize}

\subsubsection{Person-Job Fitting}
{Person-job fitting (PJF) is a fundamental task in the recruitment process, aimed at assessing the degree of alignment between the specifications of a job posting and the qualifications detailed in a candidate’s resume.}
Effective PJF assists employers in selecting the right talent for appropriate positions, thereby enhancing employee performance and job satisfaction. This ultimately contributes to the mutual success of both the organization and its employees. However, the traditional solution relies heavily on recruiters' domain expertise and subjective judgment, making it challenging to achieve both efficiency and objectivity~\cite{qin2018enhancing}. Recently, with the accumulation of extensive historical recruitment data through HRIS like ATS, many researchers have leveraged AI technology to develop supervised models for PJF. This advancement has significantly improved the efficiency and effectiveness of the recruitment process~\cite{qin2018enhancing,zhang2023fedpjf}. {Specifically, the task of PJF can be formally defined as follows: }
\begin{definition}[Person-Job Fitting]
Given a set of job applications $\mathcal{S}$, where each application $S_{i,j} \in \mathcal{S}$ contains a resume $R_i$ and a job posting $J_j$, along with a corresponding recruitment result label $y_{i,j}$, the target of PJF is to learn a predictive model $M$ for measuring the matching degree between $J$ and $R$, enabling the prediction of the result label. 
\end{definition}
\noindent Note that the recruitment result label $y_{i,j}$ varies across different recruitment contexts. When building intelligent recruitment systems for companies, the recruitment result label can be set based on the full progression of a job applicant's process, including stages such as interview, offer, and onboarding stages. For instance, in~\cite{qin2018enhancing}, $y_{i,j}$ is a binary label, where $y=1$ indicates that the candidate has been selected for further interviews. Conversely, within online recruitment platforms, the recruitment result label $y_{i,j}$ is typically configured to indicate whether a user $R_i$ has applied for a specific position $J_j$~\cite{saito2023multi}. Additionally, some online recruitment scenarios account for the reciprocal outcomes between employers and job seekers. This encompasses whether the employer of a job $J_j$, intends to advance the recruitment process after receiving an application from the job seeker $R_i$, and whether a job seeker $R_i$ accepts an invitation to interview for a specific position $J_j$ after being approached by the employer~\cite{fu2021beyond}. 

Early research efforts in the field of PJF using AI-based technologies can be traced back to~\cite{malinowski2006matching}. \textsl{Malinowski et al.} proposed that the compatibility between a job and a candidate often hinges on underlying factors not explicitly stated in the job posting or the candidate's resume. To address this issue, they developed a latent variable model to represent job requirements and candidate abilities, and formulated PJF as a bilateral matching problem from a demands-abilities perspective. Since then, several studies have explored the use of AI-based technologies to extract job and candidate profiles from textual data. For instance, \textsl{Zhu et al.} introduced a CNN-based neural network to extract representation vectors from job postings and resumes~\cite{zhu2018person}. They then evaluated the fit between a candidate's qualifications and the job requirements by measuring the similarity between these vectors, achieving an average AUC performance of around 75\% on real-world data from 2013 to 2016. In~\cite{zhao2021embedding}, the authors utilized CNNs combined with an attention layer to enhance the representation of textual information for both positions and candidates. \textsl{Qin et al.} used LSTM to model the text sequence and applied ability-aware attention strategies to measure the importance of each job requirement or candidate's ability on the final PJF decision, resulting in an improvement of approximately 10\% compared to previous methods~\cite{qin2018enhancing,qin2020enhanced}. Similarly, in~\cite{bian2019domain} the authors treated job postings and resumes as multi-sentence documents and utilized Bidirectional Gated Recurrent Units (BIGRU) and word-level attention to model sentences and documents. \textsl{Wang et al.} utilized co-attention mechanisms and Graph Neural Networks (GNN) to better leverage historical recruitment data, thereby enhancing the representations of jobs and resumes and subsequently improving PJF performance~\cite{wang2022person}. \textsl{Luo et al.} expanded upon these efforts by integrating LSTM, CNN, and attention models to process various types of structured textual information, such as skills and experiences~\cite{luo2019resumegan}. They proposed an adversarial learning-based framework to enhance the expressive representations of this data.

Recently, the rapid advancement of pretrained language models like BERT has significantly enhanced textual data representation, enabling researchers to develop more effective PJF models~\cite{lavi2021consultantbert,abdollahnejad2021deep,liu2022job,kaya2023exploration,fang2023KDD}. For instance, \textsl{Abdollahnejad et al.} serialized concatenated each pair of job $J_j$ and resume $R_i$ as inputs to BERT, using the representation of the $[\text{CLS}]$ token at the end the sequence to signify the joint representation of the application $S_{i,j}$~\cite{abdollahnejad2021deep}. Subsequently, they employed a fully-connected layer to predict the recruitment label $y_{i,j}$, ultimately fine-tuning the BERT-based PJF model with historical interaction data. A similar model structure has also been employed by \textsl{Kaya et al.}~\cite{kaya2023exploration}. \textsl{Lavi et al.} fine-tuned the BERT model using the Siamese networks and matched jobs and candidates with both classification and similarity-based objectives~\cite{lavi2021consultantbert}. Distinct from previous approaches that directly utilized BERT’s architecture, \textsl{Fang et al.} proposed a skill-aware, prompt-based pretraining framework that significantly enhanced the representation of recruitment domain corpora~\cite{fang2023KDD}. {Compared to BERT and other models, this framework improved performance on several downstream tasks, including PJF.}

Based on a wealth of high-quality historical data, the aforementioned supervised PJF models are powerful. However, in many recruitment scenarios, job-resume interaction data is sparse and noisy, which significantly undermines the effectiveness of PJF models~\cite{bian2019domain,bian2020learning,yu2024confit}. To address this, researchers have approached the problem from various perspectives. For instance, in~\cite{bian2020learning}, the authors constructed a Job-Resume Relation Graph to enhance the representation of jobs and resumes, where the edges in the graph were created based on historical interactions, category labels, and keywords. Additionally, they designed a Multi-View Co-Teaching Network to learn text-based and relation-based matching modules simultaneously. \textsl{Bian et al.} addressed this issue from the perspective of transfer learning, utilizing potential semantic relations among different job categories and domain adaptation technologies to enhance the performance of PJF models in target domains with sparse interactions~\cite{bian2019domain}. In~\cite{yu2024confit}, the authors employed data augmentation techniques based on EDA and ChatGPT to construct additional synthesized job-resume pairs, effectively addressing the issue of data sparsity.

In addition to capturing textual features from application data, various studies have incorporated additional related information to enhance the performance on the PJF task. 
{For instance, numerical and categorical attributes of jobs and candidates, such as career level, education, company name, tags, and region, have been identified as crucial factors that contribute significantly to model performance~\cite{jiang2020learning,he2021finn,he2021self}.}
Consequently, several neural network-based approaches have been designed to capture the comprehensive interaction of different types of data, such as Factorization Machine~\cite{jiang2020learning}, CNN~\cite{he2021finn}, and Self-Attention~\cite{he2021self}. Furthermore, many researchers have enhanced the performance of PJF models by incorporating knowledge graphs from the recruitment domain as side information~\cite{yao2022knowledge}. Moreover, user interaction records on online recruitment services serve as important complementary features to represent both jobs and candidates. Specifically, \textsl{Yan et al.} and \textsl{Jiang et al.} integrated historical interviewed applicants for a job posting and historically applied jobs for a particular talent to complement and enhance the representation learning for jobs and candidates~\cite{jiang2020learning}. The hidden idea is that historical interview choices and job applicants reveal the preferences of jobs and candidates for each other. Meanwhile, in~\cite{hou2022leveraging}, the authors leveraged job seekers' click behaviors and search histories on recruitment websites to better model their job-seeking intentions, thereby enhancing the performance of the PJF task. Inspired by existing multi-behavior recommendation methods, \textsl{Saito et al.} utilized job seekers' interactions—including viewing, favoriting, and applying—to implement a multi-behavior job recommendation system~\cite{saito2023multi}. In~\cite{fu2022market}, the authors integrated the supply-demand situation of various companies at specific times as supplementary labor market information into the PJF model. They utilized hierarchical reinforcement learning to address the dynamic components of the PJF task, particularly by incorporating the temporal factors of the application process.

\noindent\textbf{Applications for Talent Recommendation.}
One intuitive application of person-job fit is talent recommendation, which involves finding suitable candidates for a specific job. 
Given a specific job posting $J$ and a set of candidates $\{R_1, R_2,..., R_n\}$, a well-trained PJF architecture can be used to measure the fitness of each pair $(J_j, R_i)$ and rank the candidates. 
While the PJF models mentioned previously can all be applied in this context, some studies have attempted to enhance candidate ranking with a more comprehensive evaluation.
For instance, personality traits have been identified as critical success factors for job performance and organizational effectiveness~\cite{shen2023exploiting}. Researchers have mined these traits through linguistic analysis of social media text data~\cite{faliagka2014line}. Moreover, by integrating various measurements and data sources, researchers and companies have developed e-recruitment systems for more effective and efficient recruitment, particularly in the talent pre-screening stage~\cite{faliagka2014line}. These systems have been demonstrated to impact recruitment processes positively~\cite{van2019marketing}.

\noindent\textbf{Applications for Job Recommendation.}
As a dual problem of talent recommendation, job recommendation aims to recommend jobs for a specific candidate. 
The PJF architecture, which can be built on any PFJ models mentioned previously, can output the fitness of each job in the job set $\{J_1, J_2, ..., J_n\}$ for the given candidate $R_i$. 
This can be valuable in recruitment scenarios where job application redistribution is necessary, such as position assignments for candidates in campus recruitment or for employees in internal position adjustments. Job recommendations can also be used in online recruitment services to assist job seekers in finding suitable job opportunities~\cite{yu2024disco}. 

Furthermore, some online recruitment systems offer recommendation services to both employers and job seekers. In response, researchers have developed joint models to address these dual recommendation scenarios~\cite{yang2022modeling,fu2021beyond}. For instance, \textsl{Yang et al.} proposed a unified dual-perspective interaction graph that incorporates two distinct nodes for each candidate (or job) to characterize both successful and failed matching~\cite{yang2022modeling}. Subsequently, the authors employed BERT and GCNs to construct the PJF model. It was trained using a quadruple-based loss and a dual-perspective contrastive learning loss to enable bidirectional recommendation~\cite{zhao2021coea}. In~\cite{fu2021beyond}, the authors investigated the dynamic preferences of users, including browsing, clicking, and online chat behaviors, within dual recommendation scenarios. They introduced a Bilateral Cascade Multi-Task Learning framework to implement dynamic PJF effectively.

\noindent \textbf{Evaluation.} The performance of PJF is evaluated using two categories of metrics: classification (Accuracy, F1-score, AUC) and ranking (NDCG@N, Precision@N, Recall@N).

\noindent \textbf{$\diamondsuit$ Takeaway.} 
\begin{itemize}[left=0pt]
    \item \textbf{Advantages of AI technology:} \textsl{(1) The use of AI technology to solve PJF tasks has garnered extensive attention from both academia and industry, leading to its widespread application in corporate recruitment systems and online recruitment platforms, effectively achieving precise matching between job seekers and positions. (2) Current mainstream AI-based PJF models can consider not only the match between job seekers and positions based on rich textual information but also historical interaction data and personalized preferences. Consequently, advancements in PJF models are influenced by developments in representation techniques and recommendation system technologies. (3) AI-based PJF models play a vital role in addressing some social challenges, such as integrating migrants and refugees into society~\cite{ntioudis2022ontology}.}
    \item \textbf{Limitations and future directions:} \textsl{(1) Most current PJF models prioritize enhancing model accuracy, often overlooking interpretability. While some researchers employ strategies like attention mechanisms to highlight key features~\cite{zhang2021explainable}, the actual benefits to recruitment system users remain uncertain. Consequently, there is a significant gap in systematic research focused on improving the interpretability of PJF models to boost the operational efficiency of recruitment systems. (2) \textsl{Cardoso et al.} identified a Matching Scarcity Problem (MSP) in talent/job recommendation systems, characterized by candidates or jobs experiencing a lack of matches within the system~\cite{cardoso2021matching}. \textsl{Mashayekhi et al.} associated this issue with congestion problems prevalent in recommendation systems~\cite{mashayekhi2023recon}. Indeed, while intelligent PJF services facilitate new avenues for information exchange between labor supply and demand, they also pose the risk of creating imbalances in information distribution. There is a notable gap in the existing research regarding the systematic integration of intelligent PJF with resource allocation strategies in the labor market. (3) With the gradual implementation of AI regulatory frameworks in regions like Europe, the inherent risks of AI-based PJF models—including concerns related to fairness~\cite{kumar2023fairness} and user privacy~\cite{zhang2023fedpjf}—are increasingly under scrutiny by researchers. As a result, the development of responsible AI practices in PJF technology has become a critical focus of research. (4) With the rapid advancement of generative LLM technologies, recent studies have begun to leverage their robust text understanding and generation capabilities to develop more powerful PJF models~\cite{zheng2023generative,du2024enhancing}. However, this field is still in its infancy and requires further research. Additionally, issues such as hallucinations and low training and inference efficiency in LLM technology also impact the effectiveness of current LLM-based PJF models.}
\end{itemize}

\vspace{-3mm}
\subsection{Talent Assessment}
Talent assessment is a crucial process for companies to identify the competency of candidates. 
In this paper, we discuss two primary branches of talent assessment, i.e., interview question recommendation and assessment scoring.

\subsubsection{Interview Question Recommendation}
Job interviews aim to assess the fitness of candidates and the job positions by evaluating their skills and experiences. A critical aspect of this process is designing appropriate questions to comprehensively assess the competencies of potential employees. In this phase, personalized question recommendation has emerged as a feasible approach that selects the right questions from the question set based on the candidate and the job position.    
Along this line, \textsl{Qin et al.} proposed to recommend personalized question sets for various applications, taking into consideration both job requirements and candidates' experiences~\cite{qin2019duerquiz}. In particular, to enhance the performance of the recommendation system, a knowledge graph of job skills was built using query logs from the biggest search engine, Baidu. \textsl{Datta et al.} further took into account the difficulty of interview questions and designed an interview assistant system using knowledge graphs and Integer Linear Programming techniques~\cite{datta2021generating}. To cope with the substantial risk of bias arising from the subjective nature of traditional in-person interviews, \textsl{Shen et al.} proposed to learn the representative perspectives of in-person interviews from the successful job interview records in history~\cite{shen2018joint}. With the help of topic models, they represented job postings, resumes, and interview assessment reports in an interpretable way. The potential relationships among them are also mined to recommend questions or skills that should be estimated during interviews~\cite{shen2018joint,shen2021joint}. Substantially, \textsl{Shi et al.} proposed an automated system for recommending personalized screening questionnaires based on job postings~\cite{shi2020learning}. They encoded the job posting with the BERT model, selected the question templates with a Multi-Layer Perception (MLP) Classifier, and extracted necessary parameters from the templates using feature-based regression models. In~\cite{qin2023automatic}, the authors introduced a scalable, skill-oriented interview question generation model that leverages external knowledge from an online Knowledge-Sharing Community (KSC). Additionally, they developed a GNN-based interview question recommendation system that tailors interview questions to user queries. 

\noindent \textbf{Evaluation.} Generally, the effectiveness of interview question recommendation systems is validated through online experiments, which examine whether recommended questions improve talent screening~\cite{qin2019duerquiz} or are adopted by interviewers and recruiters~\cite{shi2020learning}. Moreover, offline experiments can evaluate the performance of recommendation algorithms by constructing a standardized test set using historical assessment reports from interviewers on various candidates~\cite{shen2018joint,shen2021joint}.


\noindent \textbf{$\diamondsuit$ Takeaway.} 
\begin{itemize}[left=0pt]
    \item \textbf{Advantages of AI technology:} \textsl{AI algorithms can assist interviewers in efficiently preparing personalized interview questions for candidates, facilitating tailored talent.}
    \item \textbf{Limitations and future directions:} \textsl{(1) Recently, some studies have furthered the development of fully automated interview robots by employing LLMs to devise algorithms that generate follow-up questions in conversational interview settings~\cite{pal2022weakly,jiang2024enhancing}. Nonetheless, research that integrates a comprehensive database of interview questions with conversational interviewing techniques is still insufficient. (2) In the field of intelligence education, Computerized Adaptive Testing (CAT) is widely studied for assessing students' knowledge mastery using a minimal number of test items~\cite{ma2023novel}. Despite its applicability to talent assessment scenarios, where it shares several methodological parallels, there is a notable deficiency in studies exploring CAT methodologies specifically tailored for interviewing contexts. (3) Given the scarcity of high-quality interview-related data, there is a lack of systematic research into effectively addressing the recommendation of interview questions in low-resource scenarios using data augmentation and transfer learning techniques. (4) Currently, due to significant variations in data types across specific scenarios, there is no unified definition for interview question recommendation. In the future, the development of standardized datasets will significantly advance research in this field.}
\end{itemize}

\subsubsection{Assessment Scoring}
Assessment scoring is another critical problem in talent management, evaluating the competency of candidates or employees based on their performance.
Typically, this problem can be formulated as a binary classification problem as follows:

\begin{definition}[Assessment Scoring]
Based on the observed attributes $x_i$ of the employee $R_i$, the target is to predict whether he/she is competent for the current job, i.e., $P(Y_i=1 \mid x_i)$.
\end{definition}

Substantial effort has been invested during the interviewing phase~\cite{naim2015automated, chen2016automated, chen2017automated, hemamou2019hirenet, hemamou2019slices, chen2020hierarchical}. Within this context, a major area of research involves using recorded face-to-face job interviews or simulated asynchronous video interview (AVI) data to assess candidates' competencies in various aspects. For instance, \textsl{Nguyen et al.} were the first to focus on the automated prediction of employment interview outcomes based on both audio and visual nonverbal cues from the interviewee and interviewer~\cite{nguyen2014hire}. \textsl{Naim et al.} extracted 82 features from 138 recorded interview videos, covering three dimensions: prosodic, lexical, and facial information~\cite{naim2015automated}. They employed regression models such as Support Vector Regression (SVR) and Lasso to predict overall interview scores. \textsl{Chen et al.} focused on monologue-style responses to Structured Interview (SI) questions and constructed an AVI dataset~\cite{chen2016automated,chen2017automated}. Subsequently, they utilized the Linguistic Inquiry Word Count (LIWC) to extract lexical features from text-based interview content, used automated speech analysis and transcription to generate features assessing various dimensions of speaking, such as fluency, rhythm, intonation \& stress, and pronunciation, and extracted visual features related to facial expressions using Visage SDK’s FaceTrack and FACET SDK. Finally, several shallow classification models, including Support Vector Machine (SVM) and Random Forest (RF), were employed to predict interviewee traits such as agreeableness, conscientiousness, emotional stability, extraversion, and openness. \textsl{Hemamou et al.} collected a corpus of over 7,000 candidates who participated in asynchronous video job interviews for real positions, recording themselves answering a set of questions~\cite{hemamou2019hirenet}. Utilizing this extensive dataset, they designed a hierarchical attention model to predict candidates' hireability. {Additionally, the authors developed attention mechanisms to extract fine-grained temporal information and pinpoint relevant answer segments~\cite{hemamou2019slices}.}

With the European Union's approval of the world's first AI technology regulation, researchers are now reconsidering the ethical risks associated with AI-based assessment scoring methods used in interview processes. For instance, unlike previous work that utilized multimodal information from interview recordings, \textsl{Chen et al.} solely employed automatic speech recognition transcriptions from AVI data~\cite{chen2020hierarchical}. They introduced GNNs to construct dependency relations between questions, enabling the learning of a model that scores multiple question-answer pairs. Besides, \textsl{Singhania et al.} were the first to investigate fairness concerning gender and race in video interviews~\cite{singhania2020grading}. Moreover, in~\cite{hemamou2021don}, the authors proposed an approach using adversarial methods to remove sensitive information from the latent representations of neural networks, which was applied to interview assessments to promote fairness in job selection.

In addition to the interviewing phase, there is still a focus on predicting competency based on employee profile information. For instance, \textsl{Liu et al.} and \textsl{Hong et al.} applied the SVM model to predict the competence level of civil servants and highway construction foremen, respectively~\cite{liu2009application}. Furthermore, \textsl{Li et al.} evaluated the performance of traditional classifiers, including SVM, RF, and Adaboost, in competency evaluation and found that prediction results based solely on structured personal static data were suboptimal~\cite{li2020data}. For improved performance, incorporating more unstructured and dynamic data, such as textual data or social networks, may represent a significant direction for future research. 

\noindent \textbf{Evaluation.} Generally, the performance of assessment scoring models is evaluated by calculating the difference between predicted results and expert scores. Therefore, classification metrics (e.g., F1-score) and regression metrics (e.g., RMSE) are commonly used. 

\noindent \textbf{$\diamondsuit$ Takeaway.}

\begin{itemize}[left=0pt]
    \item \textbf{Advantages of AI technology:} \textsl{(1) Existing AI-based assessment scoring models offer an automated and objective method for evaluating candidates' abilities in professional skills, social skills, and other overall competencies. (2) AI-based assessment scoring models can analyze candidates' performances in simulated interview scenarios, aiding them in honing their interview skills and increasing their chances of success in actual job interviews. }
    \item \textbf{Limitations and future directions:} \textsl{(1) Although some researchers are increasingly aware of the potential ethical risks associated with the use of data and models in AI-based assessment scoring, there is still a lack of comprehensive research in this area. (2) Most existing research focuses on training supervised models trained with original scores from interviewers or managers for automated scoring predictions. However, in-person interviews often introduce significant bias due to their subjective nature. Existing research frequently overlooks the inherent biases embedded in the data. Therefore, further research is needed on how to train unbiased assessment scoring models from biased data. {Naturally, unsupervised methods should also be considered~\cite{zhang2024can}. Furthermore, enhancing the usability of assessment scoring models by allowing them to provide the reliability of their predictions or the reasoning behind their decisions could significantly improve fairness in the interview process and help mitigate the risk of biased outcomes.} (3) Cognitive diagnosis has been widely validated in computational education for learning the knowledge profiles of students and predicting their future exercise performance~\cite{zhang2023relicd,yu2024rdgt}. However, research concerning the application of these techniques to competency diagnostics in talent assessment contexts, such as interviews, remains limited.}
\end{itemize}






\vspace{-4mm}
\subsection{Career Development}
Career development is the process of acquiring and experiencing planned and unplanned activities that support the attainment of life and work goals. In this paper, 
from the development process view, talent training is the first step of career development, and behavior management is the next step. Therefore, we investigate the training process, e.g., course recommendation, and behavior management, including promotion, turnover, and career mobility.

\subsubsection{Course Recommendation}
The organization offers diverse training programs, including technical, project management, quality, leadership, specialized, and soft skills training. Hence, it is crucial to assist employees in selecting the training courses that align closely with their background and objectives~\cite{srivastava2018s}. Course Recommendation aims to provide personalized courses based on the different preferences and needs of users in various aspects. Much research in this direction is concerned with student education~\cite{lin2022hierarchical}. Recently, some effort has also been made in the talent management field~\cite{srivastava2018s,wang2020personalized,patki2021personalised,zhengzhi2023recsys,yangyang2023TOIS}. There is a growing focus on the advancement of enterprise Learning Management Systems (LMS), designed to enhance both individual and organizational performance by customized online training programs aimed at augmenting employees' skills and knowledge~\cite{zhengzhi2023recsys}. Typically, we formalize this problem as follows:

\begin{definition}[Course Recommendation]
Given the sequential history learning record $\mathcal{H}_i$ of the employee $R_i$ and her/his profile $\mathcal{S}_i$, the target is to predict the rating $Y_{i,j}$ of employee $R_i$ on course $C_j$, i.e., $E[Y_{i,j} \mid \mathcal{H}_i, \mathcal{S}_i, C_j]$.
\end{definition}

As a recommendation task, most of the methods in the recommendation system can be utilized to conduct course recommendations. However, unlike traditional item recommendation, where the decision is determined by the users' rating, more attention should be paid to mining employees’ competencies and their needs for further development from side information such as the employees' profiles. 
Therefore, \textsl{Wang et al.} used a topic model to extract the latent interpretable representations of the employee's current competencies from their skill profiles, as well as a recognition mechanism to explore the personal demands from learning records~\cite{wang2020personalized}. Then, they integrated the collaborative filtering algorithm with the Variational AutoEncoder (VAE) to develop an explainable course recommendation system.
In addition to learning records, \textsl{Srivastava et al.} also introduced work history into the learning content recommendation and defined a Markov Decision Process (MDP) to extract the past training patterns~\cite{srivastava2018s}. Moreover, \textsl{Sharvesh et al.} proposed an adaptive recommendation model that incorporates employees' styles and goals to suggest the most suitable courses, considering their interactions and performance~\cite{patki2021personalised}. \textsl{Zheng et al.} proposed a novel generative recommendation framework called Generative Learning for Adaptive Recommendations (GLAD)~\cite{zhengzhi2023recsys}, which integrates reinforcement learning. It concurrently considers enhancing employee performance and ensuring the rationality of generated recommendations. \textsl{Yang et al.} introduced a contextualized knowledge graph embedding to recommend training courses to the talent in an explainable manner~\cite{yangyang2023TOIS}.

\noindent \textbf{Evaluation.} Typically, the performance evaluation metrics for course recommendation models align with those utilized in recommender systems, including AUC, Recall@N, Precision@N, F1@N, Hit@N, NDCG@N, and MAP@N. The N denotes the top-N results produced by each model.

\noindent \textbf{$\diamondsuit$ Takeaway.}

\begin{itemize}[left=0pt]
    \item \textbf{Advantages of AI technology:} \textsl{(1) The AI-based course recommendation system integrates employee and course profiles to provide tailored course recommendations, thereby enhancing learning outcomes for employees. (2) Several AI-based course recommendation systems not only offer personalized course suggestions to employees but also offer suitable justifications for these recommendations, thus facilitating employees' decision-making in selecting training courses. (3) The AI-based course recommendation system recommends courses to employees and also assesses their competency level, thereby facilitating their subsequent career development.}
    \item \textbf{Limitations and future directions:} \textsl{(1) While many AI-based course recommendation models prioritize courses based on employee preferences, these preferences may not always align with the employee's career development needs. As a result, the recommended courses may not effectively improve work performance~\cite{zhengzhi2023recsys}. (2) While existing AI-based course recommendation systems offer personalized recommendations to employees, they often fail to consider the complex motivations driving employees' course selections~\cite{yangyang2023TOIS}. Further research is needed to understand the diverse motivations behind employees' course choices. (3) Most current course recommendation systems prioritize student course recommendations, leaving limited research on recommending courses for employee training. As employee training advances, there is a pressing need for further research on employee course recommendation systems. {(4) Current course recommendation systems fail to model changes in employees' skill levels. Integrating these systems with corporate cognitive diagnostics and knowledge-tracing techniques~\cite{zhang2023relicd,yu2024rdgt,yu2024rigl} {should} better capture shifts in employee capabilities, thereby enhancing the effectiveness of course recommendation systems.}}
\end{itemize}
\subsubsection{Promotion Prediction}

Promotions serve two essential roles in the organization: assigning individuals to the jobs for which they are best suited and providing incentives for lower-level employees. 
Traditionally, promotion is decided by only managers. {Promotion prediction is primarily used to better assess employees’ development potential, thereby enabling more effective work allocation and enhancing overall organizational productivity and performance.} This prediction or judgment is generally decided through questionnaires, which are labor-intensive.
To this end, AI-based research is concerned with identifying the features that are correlated with promotion and applying machine learning methods to predict promotion. Typically, the problem can be formulated as a classification task:

\begin{definition}[Promotion Prediction]
Given the history record $\mathcal{H}_i^{t}$ in a certain period $t$ of the employee $R_i$, the target is to predict the probability of a promotion $P(Y_i = 1 \mid \mathcal{H}_i^{t})$.
\end{definition}

In this phase, several fundamental machine learning algorithms have been employed to address this problem, including KNN, SVM, random forest, {and} Adaboost, among others~\cite{ilwani2023machine}. Generally, the majority of machine learning methods achieve an accuracy of over 90\% in promotion prediction. {At the same time, to address the issue of data imbalance in promotion datasets, various sampling techniques have been employed, including random oversampling, under-sampling, and hybrid sampling methods that combine multiple strategies~\cite{csahinbacs2022employee,shafie2023prediction}.} With increasingly complex features introduced,
\textsl{Yuan et al.} suggested that work-related interactions and online social connections are strongly predictive and correlate with promotion and resignation~\cite{yuan2016promotion}. Along this line, \textsl{Long et al.} used Random Forest to predict promotions based on demographic and job features~\cite{long2018prediction}. However, these methods may not fully capture the complexity and dynamism of career development. To address this, \textsl{Li et al.} proposed a novel survival analysis approach, where predicting promotion events is transformed into estimating the expected duration of time until promotion occurs~\cite{li2017prospecting}.
{\textsl{Liu et al.}~\cite{liu2021data} defined development potential across multiple dimensions, such as the number of areas of expertise. By leveraging dimension distribution analysis and clustering algorithms, their model classified development status into four categories, based on which promotion predictions were generated.} \textsl{Guarino et al.} designed a method for optimizing the growth path with Deep Q-Learning~\cite{guarino2022adaptive}. In this case, experience results, which are the reward value, are evaluated by the competencies development. Through this system, employees' development potential is judged so that we can confirm whether this employee should be promoted.

\noindent \textbf{Evaluation.} Generally, {promotion prediction is a classification problem. On the one hand, in the case of a predicted promotion, it} is a binary classification problem, so accuracy, precision, recall, and f1-score are often used~\cite{long2018prediction,ilwani2023machine}. On the other hand, in the case of a predicted development potential, promotion prediction is transformed into a development potential judgment with various categories in different works. Hence, accuracy, macro-recall, macro-precision, and macro-F1 are used~\cite{liu2021data}. Moreover, in this case, offline experiments on the development potential or development path optimization system could be used to evaluate the managers' or employees' satisfaction~\cite{guarino2022adaptive}.

\noindent \textbf{$\diamondsuit$ Takeaway.} 
\begin{itemize}[left=0pt]
    \item \textbf{Advantages of AI technology:} \textsl{(1) AI-based algorithms can assist managers in assigning employees to the right jobs, facilitating the improvement of organizational efficiency. (2) AI-based methods could accurately judge the development potential of employees, which is helpful for many downstream tasks in the field of HR, such as high-potential talent assessment and position system design.}
    \item \textbf{Limitations and future directions:} \textsl{(1) Generally, there are many kinds of promotion, such as horizontal or vertical~\cite{ilwani2023machine}, and how to judge the development potential of different types of promotion processes should be considered. (2) Given the inherent bias in promotion data, addressing data imbalance using data mining methods has been a continuously explored research direction~\cite{ma2022employee, konar2023employee}. {Moreover, the quality and comprehensiveness of the data significantly impact the reliability of promotion predictions. However, existing research lacks an in-depth analysis of this issue and exploration of potential solutions. (3) Given the complexity and dynamism of career development, while some studies~\cite{li2017prospecting,liu2021data,guarino2022adaptive} have shown improved modeling performance compared to traditional machine learning methods such as KNN, SVM, and random forests, the application of more powerful temporal modeling techniques, such as Transformers and their variants, remains an area for further exploration. (4) Existing research on promotion prediction often relies solely on individual career development data. In reality, companies make promotion decisions based on comprehensive comparisons across employees. Furthermore, a company's promotion strategies, including the determination of promotion quotas, can significantly affect promotion decisions. Thus, it is essential to develop more advanced approaches to model such information, which would yield more accurate prediction results.} (5) {Promotion is a continuous and dynamic process. An important issue is how to account for changes in development potential over time after promotion, and how to sustain this potential continuously.} (6) LLM-based methods could be more extensively considered for promotion assessments, such as evaluating experience completion.}
\end{itemize}

\subsubsection{Turnover Prediction}

In the stable development of any organization, employee turnover stands as a critical influencing factor. Whenever a fully integrated employee departs from the organization for any reason, it leads to varying degrees of structural incompleteness, thereby resulting in a decline in productivity. 
However, early identification of employee turnover tendencies, combined with strategic preparation, can substantially alleviate the productivity losses linked to employee turnover. In recent years, an expanding body of literature on talent management and organizational behavior has emerged, with a plethora of machine learning methods being employed to predict employee turnover, yielding remarkable results~\cite{gollapalli2023data}. Mathematically, the turnover prediction problem can be formulated as follows:

\begin{definition}[Turnover Prediction]
Given the history record $\mathcal{H}_i^{t}$ in a certain period $t$ of the employee $R_i$, the target is to predict the probability of a turnover $P(Y_i = 1 \mid \mathcal{H}_i^{t})$.
\end{definition}

Many data mining techniques have been introduced to investigate employee turnover prediction~\cite{alao2013analyzing, zhao2018employee}. For instance, \textsl{Nagadevara et al.} used five data mining techniques to predict turnover, including MLP, Logistic Regression, Decision Tree, etc~\cite{nagadevara2008establishing}. The results reveal that absenteeism and lateness, job content, demographics, and experience in the current team are strong predictors of turnover. 
Considering the potential noise introduced by a multitude of employee characteristics, some studies employ techniques such as feature filtering algorithms~\cite{juvitayapun2021employee} and feature weighting~\cite{chaudhary2022comparative} to eliminate or reduce unnecessary information. Meanwhile, oversampling techniques such as synthetic minority over-sampling technique, adaptive synthetic, and borderline synthetic minority oversampling technique have been validated to effectively address the issue of data imbalance in turnover prediction~\cite{ma2022employee, konar2023employee}.
Additionally, some scholars classify employees before conducting turnover prediction to effectively enhance predictive accuracy by accounting for inter-individual variability~\cite{wild2021prediction}. 
Tailoring feature modeling to the specific characteristics of particular industries has also yielded favorable outcomes in turnover prediction.


Besides predicting turnover based on static features, considerable attention is invested in capturing the dynamic factors, especially the evolving neural network-based methods, which have demonstrated significant expressive ability~\cite{li2017prospecting}. In this phase, \textsl{Teng et al.} investigated the contagious effect of employee turnover on an individual and organizational level, respectively~\cite{teng2019exploiting, teng2021exploiting}. Specifically, they developed two LSTM cells to process peers' turnover sequence and environmental change, as well as a global attention mechanism to evaluate the heterogeneous impact on potential turnover behavior. The experiments conducted on the dataset provided by a high-tech company in China demonstrate the effectiveness of their proposed framework, including profile information and turnover records. Similarly, the authors in ~\cite{zhang2021attentive} utilized vast career trajectory data to build a heterogeneous company job network, integrating macro-level company position information into individual career transition prediction using the Dual-GRU model. On this basis, \textsl{Hang et al.} further modeled employee turnover from both internal and external views~\cite{hang2022outside}. For the internal component, they captured the influence of close collaborators and colleagues with similar skills using a graph convolutional network. From the external-market view, they connected employees and external job markets through shared job skills. Finally, internal and external information is fed into BiLSTM and survival analysis for turnover predictions. 
Furthermore, \textsl{Subha et al.} used principal component analysis for feature extraction, followed by CRF-BiLSTM-CNN, yielding significantly improved performance~\cite{subha2023predictive}.

\noindent\textbf{Job Satisfaction}
In this phase, another problem also receives considerable attention, namely, how to measure the employee's job satisfaction.
Employee job satisfaction is a significant determinant of employee turnover. Effectively measuring employee job satisfaction helps reduce the likelihood of turnover and maintains organizational stability.
Traditional approaches in this direction are based on self-reported questionnaire surveys, which are time-consuming and cannot be applied to large organizations. Recently, AI-based technologies have been introduced to automatically analyze job satisfaction from various aspects. Overall, the prediction of job satisfaction can be formulated as a binary classification problem, defined as follows:

\begin{definition}[Job Satisfaction Prediction]
Given a set of independent variables $X_i$ of the employee $R_i$, the target is to predict the probability of employee job satisfaction $P(Y_i = 1 \mid {X_i})$.
\end{definition}

For instance, \textsl{Arambepola et al.} explored the influence of job-specific factors on job satisfaction level by combining both the employee's background data and company-related factors, where several classifiers, including Random Forest, Logistic Regression, and SVM, were used for the prediction of the job satisfaction level of software developers~\cite{arambepola2021makes}. \textsl{Saha et al.} proposed to assess job satisfaction by leveraging large-scale social media, i.e., employees' Twitter post dataset, where word frequency statistics, lexical analysis, and sentiment analysis have been conducted to extract features from textual data~\cite{saha2021social}. Accordingly, multiple classifiers, such as SVM and MLP, were used to predict employees' job satisfaction.
Furthermore, \textsl{Saleh et al.} utilized an online questionnaire for data collection and employed the artificial neural network and decision tree for predicting employee job satisfaction~\cite{saleh2021predicting}.
In addition to the aforementioned classification models, \textsl{Mcjames et al.} developed a causal inference machine learning approach to identify practical interventions for improving job satisfaction~\cite{mcjames2022factors}. This approach was based on the TALIS 2018 dataset, which provides a representative sample of school teachers. 
\textsl{Devi et al.} achieved improved prediction of employee job satisfaction by employing a weighted averaging ensemble method with logistic regression~\cite{devi2022prediction}.

\noindent \textbf{Evaluation.} Generally, both turnover prediction and job satisfaction prediction are binary classification tasks. Common evaluation metrics include accuracy, precision, recall, F1-score, and ROC-AUC.

\noindent \textbf{$\diamondsuit$ Takeaway.} 
\begin{itemize}[left=0pt]
   \item \textbf{Advantages of AI technology:} \textsl{(1) AI-based algorithms can help organizations prepare for employee turnover in advance, thus reducing the productivity loss resulting. (2) They can assist organizations in perceiving employee satisfaction, enabling them to adapt talent strategies to retain valuable employees, thereby maintaining the stability of organizational development.}
   \item \textbf{Limitations and future directions:} \textsl{(1) Current turnover prediction research relies on limited real-world data, creating a gap between research and industrial application, significantly impacting organizational practicality. (2) {Current research on this topic has limitations in the factors it considers. This would affect the practical application of prediction results, such as how to intervene in the attrition of talent. Future investigations could explore the impact of globalization, digitalization, online business factors, as well as personal reasons, on individual turnover rates and job satisfaction for a more comprehensive understanding~\cite{maertz2012integrating}. (3) Existing research often divides historical records into discrete time intervals, which complicates achieving finer-grained turnover predictions over continuous time. Utilizing continuous time series forecasting models may offer a promising approach in this regard. Furthermore, it could also aid in modeling the impact of sporadic events on individual turnover probabilities. (4) Most existing research models turnover as an isolated career development behavior, neglecting the opportunity for joint modeling with other related tasks, such as promotion, as discussed earlier.}}
\end{itemize}

\subsubsection{Career Mobility Prediction}
In the rapidly evolving labor market, the deployment of AI for career mobility prediction significantly enhances personalized career planning. By analyzing extensive datasets on career trajectories in the labor market, AI-based methods can provide tailored recommendations that align with future employment opportunities. Following~\cite{zhang2021attentive}, we define the career mobility prediction problem as follows:

\begin{definition}[Career Mobility Prediction]
Given the career path $\{\mathcal{H}_i, \mathcal{I}_i\}$ of employee $R_i$, where $\mathcal{H}^t_i = \{c^t_i, p^t_i, d^t_i\}$ records the work experience of $R_i$ at company $c^t_i,$, position $p^t_i$ with duration $d^t_i$, and $\mathcal{I}_i$ stands for the personal information. The goal of career mobility prediction is to learn a model $M$, which can predict the employee $R_i$'s next $K$-step career move, including future company $c^{t+K}_i$, position $p^{t+K}_i$, duration $d^{t+K}_i$, and other relevant information.
\end{definition}

\begin{figure}[t!]
\centering
\includegraphics[width=0.82\columnwidth]{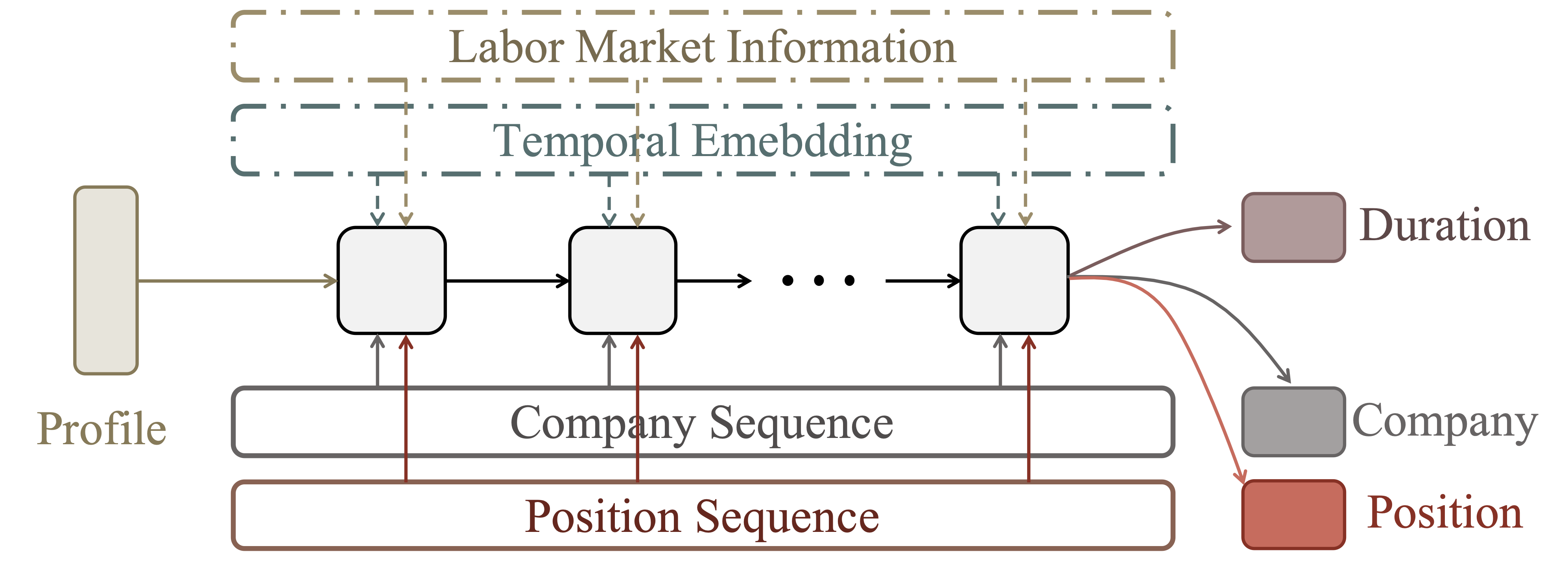}
\caption{The illustration of Career Mobility Prediction.}
\vspace{-6mm}
\label{fig:CareerMobility} 
\end{figure}

Generally, as illustrated in Figure~\ref{fig:CareerMobility}, the task of career mobility prediction can be addressed through time series analysis, where the career path (also known as career trajectory) is treated as a sequence of events~\cite{li2017nemo,meng2019hierarchical,yamashita2022looking}.
Most existing work focuses on predicting next-step career mobility, i.e., $K=1$. Specifically, \textsl{Li et al.} were the first to design a contextual LSTM model that integrates the profile context and career path dynamics simultaneously to predict the next company/position of talents~\cite{li2017nemo}. To provide a fine-grained prediction, researchers have developed several methods for modeling career trajectories that predict both the next employer and the corresponding job duration. \textsl{Meng et al.} employed a hierarchical neural network structure with an embedded attention mechanism to characterize internal and external job mobility~\cite{meng2019hierarchical}. Moreover, \textsl{Wang et al.} introduced a temporal encoding mechanism that handles dynamic temporal information~\cite{wang2021variable}. Macro-level job transition behavior may also impact individual career choices. To address this, \textsl{Zhang et al.} constructed a heterogeneous company-position network based on massive career trajectory data and integrated macro information from the company-position into personal career move prediction~\cite{zhang2021attentive}. Meanwhile, \textsl{He et al.} expanded the prediction tasks to include forecasting salary levels and company sizes, utilizing LSTM and CNN to construct the model~\cite{he2021your}. Recently, several studies have focused on employing pre-training techniques to model career trajectory representations, aiming to enhance performance in career mobility prediction~\cite{decorte2023career,zha2024towards}. For instance, \textsl{Decorte et al.} developed CareerBERT, a model built upon the BERT framework, to analyze sequences of work experiences—including job titles and descriptions—alongside corresponding ESCO occupation sequences within career trajectory data~\cite{decorte2023career}. They utilized contrastive learning to fine-tune this pre-trained model. Following this, the authors introduced a mapping network designed to predict future ESCO occupations based on sequences of work experiences.

Unlike previous work that focused solely on predicting the ``immediate'' next career move, \textsl{Yamashita et al.} utilized the transformer architecture to predict one's future career pathway as a ``sequence''—specifically, the next $K$ steps of career movement~\cite{yamashita2022looking}. Furthermore, some researchers have incorporated various trajectory rewards, such as company ratings, staying probabilities, and salary ranges, into the prediction of future career paths. {For instance, \textsl{Guo et al.} propose an intelligent sequential career planning system via stochastic subsampling reinforcement learning to find globally optimal career paths~\cite{guo2022intelligent}.}

\noindent \textbf{$\diamondsuit$ Takeaway.}
\begin{itemize}[left=0pt]
    \item \textbf{Advantages of AI technology:} \textsl{(1) AI-based methods can leverage the career trajectory data inherent in the labor market to provide personalized career path predictions and recommendations for individuals. (2) Existing AI-based career mobility prediction models can forecast career movements over short and long-term time spans. }
    \item \textbf{Limitations and future directions:} \textsl{(1) {As emerging events like COVID-19 significantly disrupt the labor market~\cite{sun2024large}}, {the challenge of developing cross-industry or cross-domain career planning has emerged as a pressing issue}. Given that career trajectory data are often sparse in cross-industry mobility contexts, existing methods may struggle to handle such scenarios. Consequently, further exploration of career mobility prediction techniques, particularly those based on transfer learning and few-shot learning, is necessary. (2) Due to unique events or reasons, some career trajectories often lack generalizability. Moreover, career trajectory data frequently faces challenges related to missing information or a lack of up-to-date records. Addressing such noise in data represents a critical research direction.} (3) Existing research often overlooks the interpretability of recommended results. Although some studies~\cite{schellingerhout2022explainable} have discussed this issue, a significant gap exists in systematic research on this topic. (4) After analyzing career mobility, existing methods lack relevant upskilling pathways that could assist institutions and employers in developing proactive training programs.
\end{itemize}

\vspace{-3mm}
\subsection{Summary}
In summary, AI-based approaches have been applied in talent management for recruitment, assessment, and career development. Specifically, the majority of research efforts have focused on talent recruitment, which may be attributed to the rapidly growing demand brought about by the development of HRIS like ATS and online recruitment platforms. Moreover, the application of AI-based approaches to comprehensively assess talent and plan career development has also attracted increasing attention in this era of information. Compared to traditional methods, AI-based models enhance the efficiency, objectivity, and accuracy of various decision-making scenarios in talent management by leveraging not only extensive historical data but also powerful technologies such as deep learning and pretrained language models. Although AI-based models are rapidly progressing in a range of talent management applications, there remains a substantial gap in research concerning the fairness and interpretability of these models, as well as issues related to data privacy protection and the challenges associated with data sparsity and bias in some specific contexts.
\vspace{-4mm}
\section{Organization Management}
{Organization management is often regarded as the art of fostering collaboration among talents and steering the organization toward a shared, predefined objective. Organization management broadly encompasses the management of processes, structures, technologies, identities, forms, and people, with human behavior being primarily reflected through the study of organizational processes.}
At the same time, since {the} organizational network is an important formulated element of {these studies}, we first summarized the AI-related analysis in organizational networks. Generally, the complex relationships among employees and organizations will naturally form a network structure, and the AI-related techniques for organizational network analysis aim to help understand the importance of critical connections and flows in an organization by modeling the special network structure~\cite{ye2022mane}, thus {serving as} the downstream management applications, such as organizational turnover prediction~\cite{teng2021exploiting} and high-potential talents identification~\cite{ye2019identifying}. Next, AI-related analysis in organizational structures is what we focus on. Because the stability and development of organizational structure are the most important parts, we mainly focus on organizational stability analysis. To study the stability of the organization, some studies propose to analyze the composition of the organization from the formation and optimization perspectives~\cite{duncan1979right}. Besides, several studies explore the compatibility between employees and organizations~\cite{sun2019impact,sun2021modeling}. Finally, we have summarized an important part of organizational forms, which is the incentive. To motivate the talents in the organization to perform well, some AI-related studies {conducted} organizational incentive analysis, which mainly focuses on two important tasks in human resources, namely job title benchmarking~\cite{zhang2019job2vec} and job salary benchmarking~\cite{meng2018intelligent} respectively.

\vspace{-2mm}
\subsection{Organizational Network Analysis}

In modern organizations, it is common for employees to build informal ``go-to'' teams to facilitate business collaboration beyond the organizational structure. Organizational social networks often emerge spontaneously, forming communicative and socio-technical connections. In this context, Organizational Network Analysis (ONA) serves a crucial role. It aids in understanding the significance of these critical connections and information flows within an organization, making leaders aware of the importance of vibrant communities and employees to be more targeted and effective in business operations. In this subsection, we will introduce AI-related techniques for organizational network modeling and introduce a classic application for ONA, namely, high-potential talent identification.

\subsubsection{Organizational Network Modeling}
\begin{figure}[t!]
\centering
\includegraphics[width=1\columnwidth]{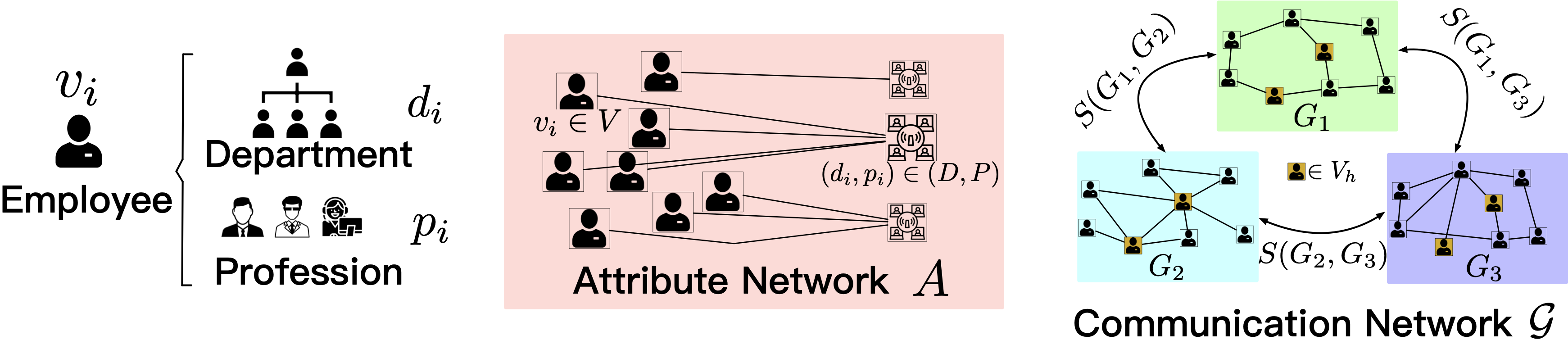}
\vspace{-6mm}
\caption{The illustration of the organizational network.}
\vspace{-6mm}
\label{fig:orgainzational_network} 
\end{figure}

In {real-world scenarios}, abundant talent data can be utilized to construct an organizational network, unveiling the intricate relationships among employees as they form project teams or forge alliances across different groups. An illustration of an organizational network is shown in Figure~\ref{fig:orgainzational_network}. However, extracting useful information to construct the network and obtaining meaningful knowledge to support managerial decisions are still difficult. Traditionally, the organizational network is usually constructed by questionnaires like formal or informal relationships, which are labor-intensive and subjective. To improve the accuracy and efficiency of organizational network construction, {many researchers are focusing on using network embedding techniques to model the organizational network.}
Without loss of generality, the generalized organization network can be defined as follows:

\begin{definition}[Organizational Network Modeling]
Organizational network is defined as $G = (V, E)$, where $V$ denotes the node set representing employees and departments (or organizations), and $E$ denotes the edge set representing the relationship between nodes, such as the belonging relationship between employees and department, the frequency of communication (e.g., email or instant message) and the reporting lines between employees. Besides, the employee node has a specific profession, which indicates the type of their work (e.g., engineer and product). Meanwhile, each employee and department node has some work-related attributes (e.g., length of service, job level). 
Based on the organizational network, organizational network modeling aims to capture knowledge from this network to support talent management-related tasks.
\end{definition}

The network embedding describes the network as low-dimensional vectors and further serves the downstream applications. For instance, \textsl{Ye et al.} proposed a multiplex attentive network embedding approach for modeling organizational networks in a holistic way~\cite{ye2022mane}. In their work, the organizational network is composed of multiple communication interactions among employees. They generated embeddings for employees based on the random walk strategy~\cite{perozzi2014deepwalk} with the k-core and approximated shortest path algorithm. Furthermore, they proposed a relational transition-based approach to represent each department. In this way, the learned representation can be leveraged for several talent management tasks, including employee turnover, performance prediction, and department performance prediction. Besides, \textsl{Teng et al.} exploited the network fusion technique for organizational turnover prediction~\cite{teng2021exploiting}. They concentrated on modeling the relationships among organizations. Specifically, they demonstrated the correlation between the topology of organizational networks and organizational turnover. To this end, they constructed a turnover similarity network based on the multiple organizational social networks and took advantage of the GNN to learn comprehensive knowledge from these topological structures, which were further used for organizational turnover prediction. 

\noindent \textbf{Evaluation.} {Generally, the performance of Organizational Network Modeling is evaluated through various downstream tasks in the field of human resource management, such as turnover prediction and department performance assessment~\cite{ye2022mane}. These tasks are typically formulated as binary classification problems, for which common evaluation metrics include Accuracy, F1-score, and AUC. Furthermore, turnover prediction is not only a benchmarking task but also a practical application scenario of organizational network modeling, where metrics such as MAE and MSE are used to evaluate the predicted proportion of employees who leave the organization~\cite{teng2021exploiting}.}

\noindent \textbf{$\diamondsuit$ Takeaway.} 
\begin{itemize}[left=0pt]
    \item \textbf{Advantages of AI technology:} \textsl{(1) Existing AI-based methods for organizational network modeling efficiently represent employees or departments and reduce subjective bias, aiding HR in decision-making. (2) AI-based methods could generate complete social information for employees or departments rather than traditional direct communication targets. (3) AI-based modeling methods could avoid subjective biases due to individual preference.}
    \item \textbf{Limitations and future directions:} \textsl{(1) Existing methods mostly represent the edge of an organizational network as communication, but there are various relationships in the organization, such as formal and informal communications, project cooperation, and so on. {This limitation can result in an incomplete understanding of organizational structures and dynamics. Therefore, integrating heterogeneous graph representation learning and graph denoising methods to enable more reliable organizational network modeling represents a crucial direction for future research.} (2) Existing modeling evaluation is mainly connected with classic downstream tasks such as turnover prediction, but an important application of modeling is the visual presentation after modeling, which should involve more expert evaluation metrics. (3) In the Human-AI symbiosis era, the interaction and corresponding evaluation should be considered more. (4) Although the dynamic network is considered in some research, the environment inside and outside the organization changes rapidly, more fine-grained time slicing or continuous-time organizational network modeling should be paid more attention to in the future.}
\end{itemize}

\subsubsection{High-potential Talents Identification}

High-potential talents (HIPOs) possess leadership abilities, business acumen, and a strong drive for success, making them more likely to emerge as future leaders within organizations when compared to their peers. {Identifying HIPOs has long been a critical challenge in human resource management, as it plays a vital role in executing organizational strategy and optimizing organizational structure~\cite{silzer2010identifying}.} At the same time, they are strategic assets for companies to achieve sustainable competitive advantages~\cite{zenger2017companies}. Due to their strategic importance, the identification and retention of high-potential talent are regarded as critical components of business strategy.

The traditional methods for HIPOs identification usually rely on the subjective selection of HR experts. They primarily focus on evaluating certain talent factors, such as communication skills, teamwork, and self-learning~\cite{gelens2014talent}. However, these manually selected factors may lead to unintentional bias and inconsistencies. Recently, with the development of ONA, objective data-driven HIPOs identification has become possible. The rationale behind this is that HIPOs usually perform more actively and have higher competencies than their peers to accumulate their social capital during their daily work~\cite{ye2019identifying}. We can detect HIPOs implicitly through social information. Formally, the HIPOs identification problem based on the organizational network can be formulated as:

\begin{definition}[HIPOs Identification]
Given a new employee $v$, who joined the company in the $t$-th time slice, and a set of organizational {networks} $G = \{G_{t}, G_{t+1}, ... G_{t+k}\}$, where $G_{t}$ represents organizational network of time slice $t$. The objective is to develop a model $f(v, G)=y$ to predict whether $v$ is a HIPO (i.e., $y=1$) or not (i.e., $y=0$).
\end{definition}

To solve this problem, \textsl{Ye et al.} proposed a neural network-based dynamic social profiling approach for quantitative identification of HIPOs, which focuses on modeling the dynamics of employees' behaviors within the organizational network~\cite{ye2019identifying}. In particular, they applied GCN and social centrality analysis to extract both local and global information in the organizational network as social profiles for each employee. Then they adopted LSTM with a global attention mechanism to capture the profile dynamics of employees during their early careers. Finally, they evaluated their model on real-world talent data, which clearly validates the effectiveness and interpretability of the proposed model.
This method combines longitudinal social network data to embed employees' work performance during different periods, reflecting the employees' development and growth over time, like social capital increasing.
Furthermore, there are other network construction methods using the cooperation experience. For example, as for high-potential scholars identification, \textsl{Yin et al.} constructed the innovation potential with the disruption of technology and science of each paper, then used the co-authorship network in each year to model the development of scholars. Combining social capital theory and LSTM, they measured social capital in different years to predict future innovative potential~\cite{yin2022understanding}.

\noindent \textbf{Evaluation.} As introduced in Definition 4.2, HIPOs identification is a binary classification problem, therefore, the evaluation metrics are usually Accuracy, Precision, Recall, and F1 score~\cite{yin2022understanding,ye2019identifying}.

\noindent \textbf{$\diamondsuit$ Takeaway.}
\begin{itemize}[left=0pt]
    \item \textbf{Advantages of AI technology: }\textsl{(1) Existing AI-based methods could significantly reduce the labor cost in 360-degree feedback of HIPOs and improve the decision speed. (2) HIPOs identification usually includes leaders' biases, AI-based modeling methods could avoid subjective biases with fixed features or networks.}
    \item \textbf{Limitations and future directions:} \textsl{(1) {Existing methods primarily focus on social information extraction. However, there are many other indicators for identifying HIPOs, such as abilities, competencies, traits, and leadership skills. These soft skills should also be incorporated into the model.} (2) {Similar to organizational network} modeling, although a dynamic network is constructed in months, years, and other time slices, more fine-grained time slicing or continuous-time social network information extraction should be paid more attention to in the future. (3) As for evaluation, HIPOs could be assessed by various aspects like promotion, turnover, or performance after selection~\cite{cheng2021effectiveness}, therefore, selection results are usually not the final real result of if these employees are HIPOs, and the real work performance of these employees or expert evaluation should be used as the real evaluation result to optimize the prediction model.}
\end{itemize}

\subsection{Organizational Stability Analysis}
Generally, organizational stability means the stability of the structure of the organization itself and the compatibility between employees and the organization. An organizational structure defines how activities such as task allocation, coordination, and supervision are directed toward the achievement of organizational aims~\cite{pugh1971organization}. Considering the availability of data, existing AI-related research has largely explored the formation and optimization of organizations (e.g., teams) under certain goals. In this part, we will introduce AI-related techniques for organizational stability from three perspectives, namely team formation, team optimization, and person-organization fit.


\begin{figure}[t!]
\centering
\includegraphics[width=0.82\columnwidth]{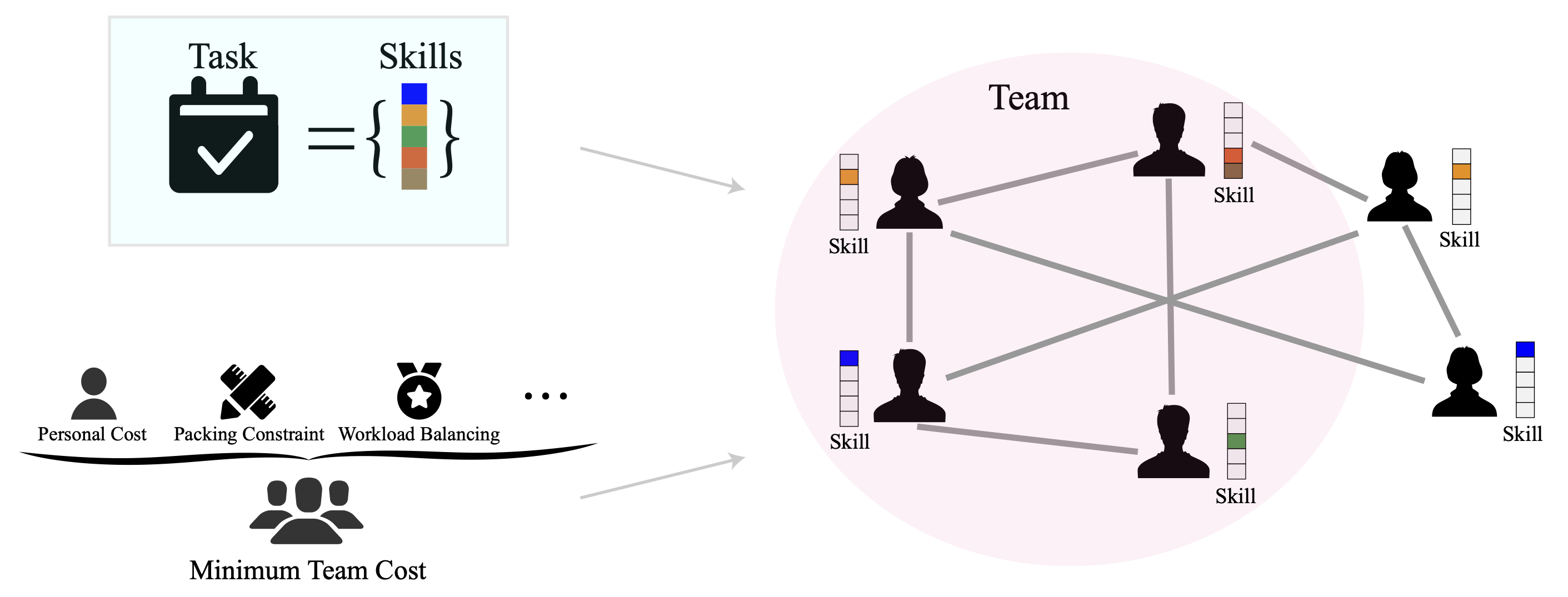}
\caption{The illustration of team formation.}
\vspace{-4mm}
\label{fig:team formation} 
\end{figure}

\subsubsection{Team Formation.}
{Team formation plays a pivotal role in organizational management. It not only enables organizations to respond swiftly to urgent tasks, but also contributes significantly to enhancing overall organizational efficiency.}
In brief, team formation could be defined as discovering a team of experts that collectively cover all the required skills for a given project, as shown in Figure~\ref{fig:team formation}.
Whereas it is proven to be NP-hard~\cite{lappas2009finding}, this requirement still needs to be solved in many real-world scenarios, such as team discovery in a social network, which contains professionals who provide specialized skills or services. 

\textbf{\textit{Definition 4.2 (Team Formation).}}
Given a team collaboration in the $t$-th time slice as $\mathcal{C}_{t}=\left\{(s, e)_{t} \mid s \subseteq \mathcal{S}, e \subseteq \mathcal{E}\right\}$, where $\mathcal{S}$  and  $\mathcal{E}$ represent the sets of skills and experts. The target of Team Formation is to minimize costs in teamwork (e.g., communication costs) or to succeed in their goals.

For instance, \textsl{Kargar et al.} proposed a method to find the object team with minimal communication cost as well as personnel cost of the project~\cite{kargar2012efficient}. Specifically, they used a graph to model a social network where nodes represent experts and formulate the task as a constrained bi-criteria optimization problem. Since it is proven that the problem of minimizing the combined cost function is still NP-hard, the authors efficiently {solved} the problem with an approximation algorithm and three heuristic algorithms in polynomial time. 
Later, \textsl{Zihayat et al.} took both communication cost and experts' authority into account and proposed greedy algorithms to solve the optimization problem~\cite{zihayat2016authority}. 
Since these team formation algorithms are based on very different criteria and performance metrics, \textsl{Wang et al.} implemented these algorithms using a common platform and evaluated their performance with several real datasets~\cite{wang2016ustf}.
However, these studies have limitations in terms of scalability and fail to effectively manage the dynamic nature of expert networks. To this end, instead of searching over the graph representation of the expert network, \textsl{Hamidi et al.} searched for variational distributions of experts and skills in the context of a team~\cite{hamidi2020learning}. 

In many scenarios, the purpose of team formation is to anticipate future expert teams by combining sequences of expert skills and the evolution of collaborative relationships over time. \textsl{Hamidi et al.} defined an optimal team as a team observed in the past, and proposed a variational Bayesian neural network to estimate the mapping function from the skill power set to the expert power set, finally forming an optimal expert team e for a required subset s of skills~\cite{hamidi2023variational}. \textsl{Fani et al.} presented a streaming-based neural model training strategy to estimate the mapping function from the stream of the collaborative set and the skill subset to the expert subset to find the best possible team~\cite{fani2024streaming}. {When recommending future teams based on the expert and skill distributions of past successful teams, a common phenomenon is that a small number of experts contribute to most successful collaborations, whereas the majority of experts are rarely involved. Such a long-tailed distribution in the training data often results in model overfitting; to mitigate this issue, \textsl{Dashti et al.} proposed three negative sampling heuristics~\cite{dashti2022effective}.}


Another scene that draws wide attention is the fast-growing Online Labor Marketplaces, which provide a sharp decrease in communication costs. For example, \textsl{Liu et al.} first implemented team formation in crowdsourcing markets with consideration of the impact of teamwork~\cite{liu2015efficient}. The study designs a mechanism that combines the greedy selection rule and a special payment scheme, obtaining various desirable properties, such as efficiency, profitability, and truthfulness. In addition, \textsl{Barnabo et al.} considered the fairness of algorithms related to these online marketplaces~\cite{barnabo2019algorithms}. {They formalized the \textit{Fair Team Formation} problem as identifying the minimum-cost team capable of completing the task while ensuring an equal representation of two non-overlapping classes.} 
Consequently, four algorithms are designed to solve the problem and experiment on real-world data to confirm their effectiveness.
{Nevertheless, the majority of existing research focuses on the offline variant of the team formation problem, in which all tasks are known a priori.} 
To this end, \textsl{Anagnostopoulos et al.} implemented the problem of online cost minimization, where the goal is to minimize the overall cost (paid on hiring, outsourcing, and salary costs) of maintaining a team that can complete the arriving tasks~\cite{anagnostopoulos2018algorithms}. {Moreover, the study considers a more complex outsourcing case where hiring, firing, and outsourcing decisions are made by an online algorithm, leading to cost savings over alternatives.}


\noindent \textbf{Evaluation.} Broadly speaking, the performance of team formation is assessed in two main ways: communication cost and collaboration prediction.
\begin{itemize}[left=0pt]
    \item \textbf{Communication Cost}: {\textsl{Nemec et al.} proposed a team formation approach by minimizing communication costs, achieved through minimizing a distance function over a graph network of experts, and evaluated the effectiveness of the formed teams~\cite{nemec2021rw}.}
    \item \textbf{Collaboration Prediction}: {\textsl{Fani et al.} predicted the need for experts to achieve successful collaboration by learning from past teams that collaborated successfully and modeling a mapping function from collaborations and skills to experts~\cite{fani2024streaming}.}
\end{itemize}

\noindent \textbf{$\diamondsuit$ Takeaway.}
\begin{itemize}[left=0pt]
    \item \textbf{Advantages of AI technology:} \textsl{(1) Existing methods can help teams accurately match candidates' skills and experience with their needs, leading to intelligent hiring and talent recommendations. By analyzing large amounts of data, AI-based methods can identify the best candidates and provide personalized recruiting solutions to improve the efficiency and quality of recruiting. (2) Based on historical data and trend analysis, AI-based approaches can predict team member turnover and team expansion needs, helping teams make timely staffing adjustments and expansion plans. This can help teams better cope with changes and challenges and maintain a competitive edge.}
    \item \textbf{Limitations and future directions:} \textsl{(1) While AI technology can provide efficient tools and systems, it may lack the human touch and human interaction. Human emotions and communication are crucial in teamwork, and AI technology cannot completely replace this aspect. (2) Current research may have difficulty explaining its results and decision-making process, which may reduce the team's trust and acceptance of the algorithm. The lack of interpretability may make it difficult for team members to understand why particular team members were chosen or particular actions taken. Future technologies will need to focus more on human emotions and emotional intelligence. This may include developing algorithms that understand and respond to human emotions, as well as designing user interfaces and interactions that are more human and approachable. {It is also important to develop algorithms that offer explainability and transparency.} This means that algorithms need to be able to explain their results and decision-making processes so that team members can understand and trust the algorithm's decisions.}
\end{itemize}

\subsubsection{Team Optimization.} 
Team optimization means the optimization of existing teams, which is beneficial for responding to changes in team tasks. Indeed, there are two ways to optimize teams, including team member replacement and team expansion.

\textbf{\textit{Definition 4.3 (Team Optimization).}}
For a team $T$, the purpose of team optimization is to recommend the group of candidate members $i$ based on their fit with the existing team $T$ when the team is expanding or to recommend the group of new members (sub-teams) $i$ to replace departing members (sub-teams) $i^{\prime}$ in the team when members are replacing them using the GNN approach.

\begin{figure}[t!]
\centering
\includegraphics[width=0.82\columnwidth]{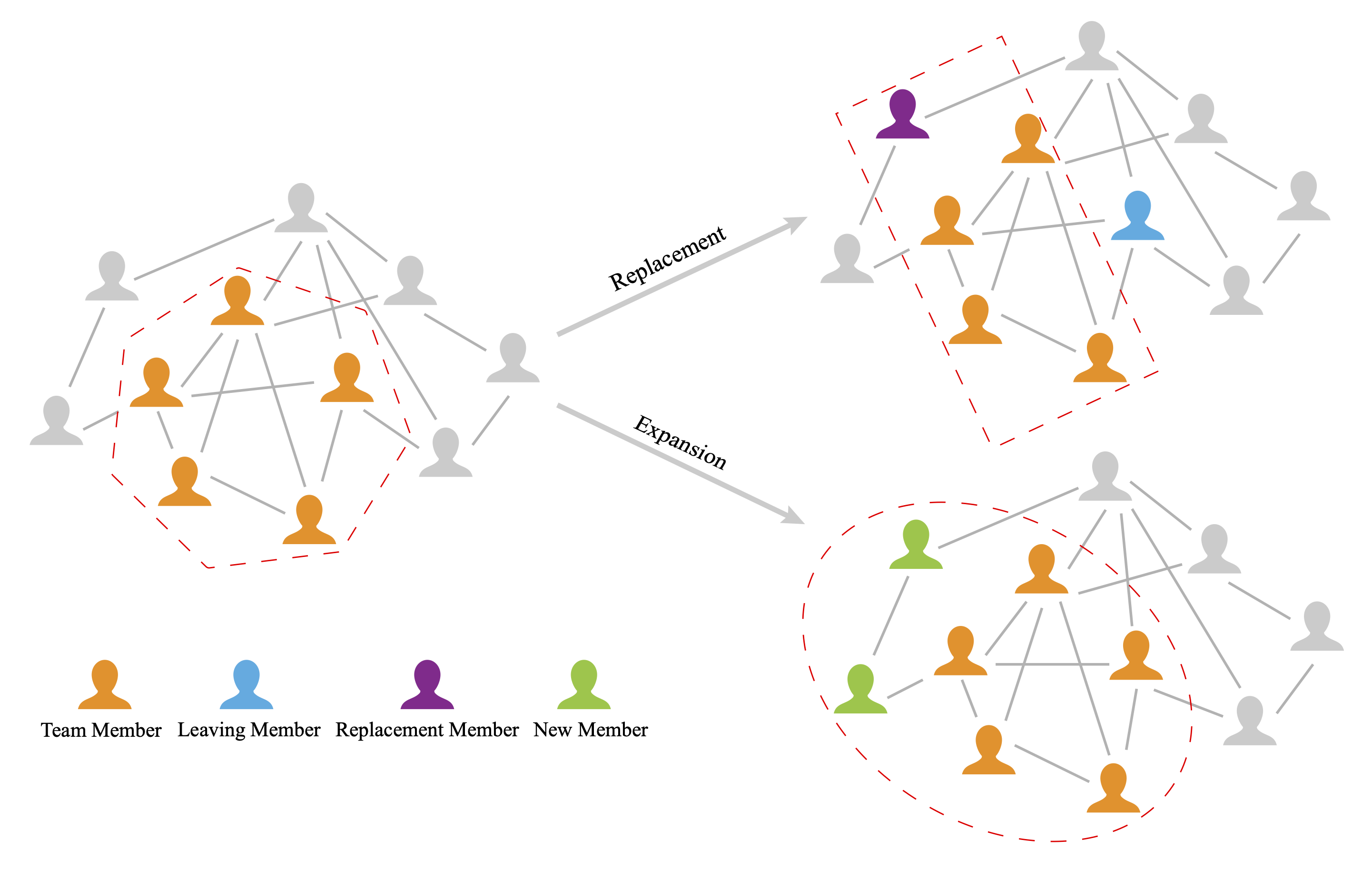}
\vspace{-1mm}
\caption{The illustration of team optimization.}
\vspace{-4mm}
\label{fig:team optimization} 
\end{figure}

Two key problems within the scope of team optimization are \textit{team member replacement} and \textit{team expansion}, as illustrated in Figure \ref{fig:team optimization}.
Specifically, \textit{team member replacement} is put forward first by \cite{li2015replacing}, which aims to find a good candidate to best replace a team member who becomes unavailable to perform the task. 
{To address this challenge, they introduced graph kernels that consider both skill and structural matching requirements, and proposed a series of effective and scalable algorithms for efficient team formation.}
In \cite{li2016enhancing}, the authors further took the synergy between skill similarity and structural similarity into consideration, instead of considering the two aspects independently.
Later, \cite{li2021reform} proposed a new graph kernel that evaluates the superiority of candidate subteams in a holistic way, which can be freely adapted to the needs of the situation. An effective pruning strategy is used in the algorithm to reduce the kernel computation by exploiting the similarity between the structures of the candidate teams, which can output more human-agreeable recommendations compared to previous studies. What's more, \cite{hu2022genius} used a clustering-based GNN framework to capture team network knowledge for flexible subteam replacement; and incorporated a self-supervised positive team comparison training scheme into the model for improved team-level representation learning and unsupervised node clustering to reduce candidate objects for fast computation.

In addition, {some efforts have been made toward} \textit{team expansion}. \textsl{Zhao et al.} formally defined the problem in collaborative environments and proposed a neural network-based approach, considering three important factors (team task, existing team members, and candidate team member) as well as their interactions simultaneously~\cite{zhao2018team}.
However, most works on team optimization treat teams as a static system and recommend a single action to optimize a \textit{short-term} objective. To this end, \textsl{Zhou et al.} proposed a deep reinforcement learning-based framework to continuously learn and update its team optimization strategy by incorporating both skill similarity and structural consistency~\cite{zhou2019towards}. 

\noindent \textbf{Evaluation.} An important metric in team optimization is the fitness score. In general, selecting members to join a team is similar to a recommendation problem while the team is expanding. In recommending candidate members to the team, \textsl{Zhao et al.} determined the members to join the team by calculating the fitness scores between the candidate members and the team~\cite{zhao2018team}.

\noindent \textbf{$\diamondsuit$ Takeaway.}
\begin{itemize}[left=0pt]
    \item \textbf{Advantages of AI technology:} \textsl{(1) Existing methods can intelligently match the right team members based on individual skills, experience, and preferences. By analyzing large amounts of data, the most suitable candidates can be found more accurately, leading to improved team performance and efficiency. (2) Compared with traditional methods, the extant method can help ensure diversity and inclusion in teams. It can identify potential biases and tendencies and provide objective recommendations to ensure diversity in team members, thus promoting innovation and better decision-making.}
    \item \textbf{Limitations and future directions:} \textsl{(1)Although AI technology can analyze large amounts of data and patterns, it still lacks human intuition and judgment when dealing with complex situations and interpersonal relationships. For example, in team member turnover, AI may not be able to fully take into account individual emotions, motivations, and interpersonal relationships, resulting in less comprehensive or humanized decision outcomes. (2)Training data for existing methods may be biased and unbalanced, {which may affect the predictive accuracy and fairness of the model~\cite{zhao2023ensemble}}. For example, in team member turnover, if an AI-based model overly relies on historical data, it may perpetuate past biases and inequalities, resulting in unequal opportunities for new team members. Future AI technologies should focus more on interpretability and transparency so that team members can understand and trust the decision-making process of AI-based models. By developing explainable AI-based models and algorithms, team acceptance of AI technology can be increased, promoting human-machine collaboration and co-development.}
\end{itemize}

\subsubsection{Person-Organization Fit}
\begin{figure}[t]
\centering 
\subfigure[A classic P-O fit modeling process.]{
\includegraphics[width=0.44\linewidth, height=2.7cm]{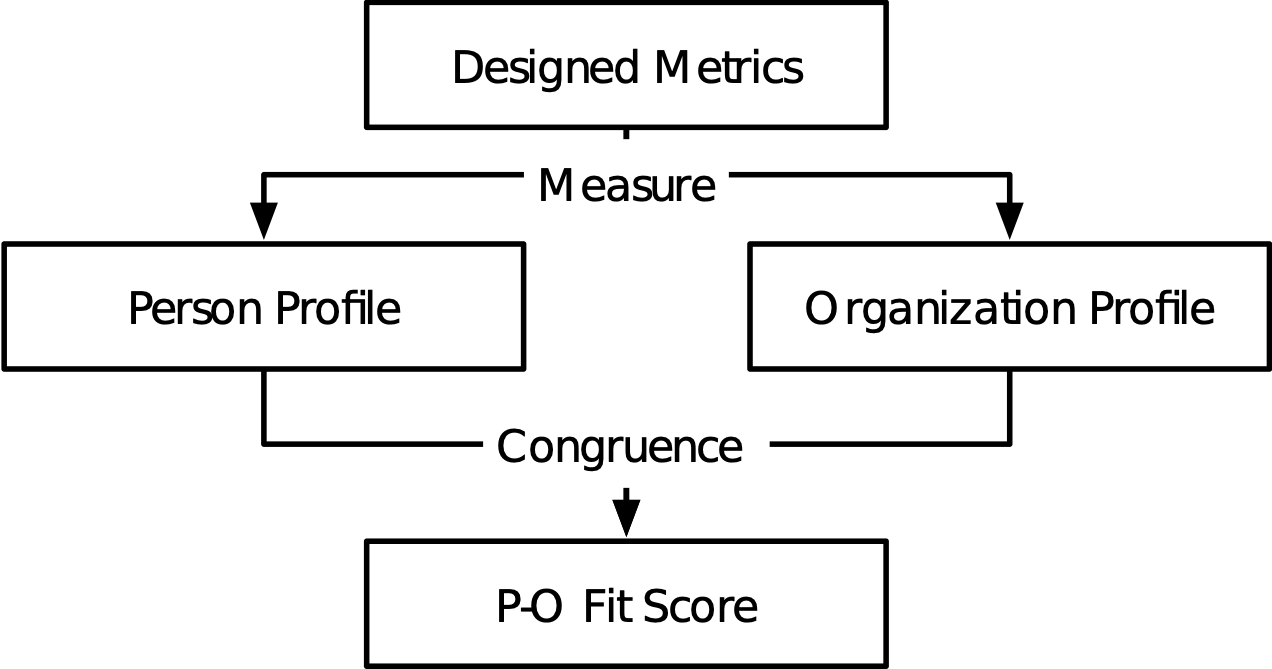}\label{fig:profit1}}
\subfigure[AI-driven P-O fit modeling process.]{
\includegraphics[width=0.44\linewidth,height=2.7cm]{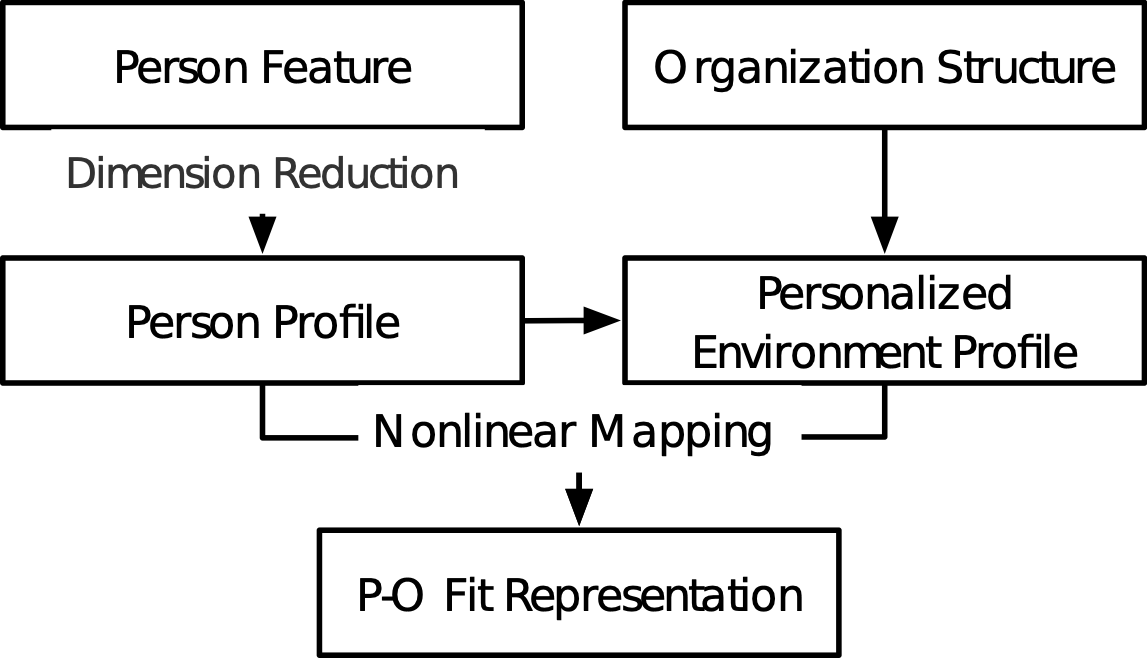}\label{fig:profit2}}
\caption{The overview of P-O fit modeling.}
\vspace{-1mm}
\label{p-o fit}
\end{figure}

Person-organization fit (P-O fit) refers to the compatibility between employees and their organizations. In fact, P-O fit has been widely recognized as an effective indicator of proactive talent management, and it has a significant impact on outcomes such as work attitudes, turnover intentions, and job performance~\cite{lauver2001distinguishing}. 
In the domain of organizational behavior, most studies measure P-O fit based on the similarity between the organizational profile and the employees' profile. Figure~\ref{fig:profit1} presents the classical P-O fit modeling process. At first, experts collect information with questionnaires to extract employee and organization profiles and manually design metrics. Then, the statistical methods are applied to measure the congruence between an employee and an organization as the P-O fit score. However, this process is labor-intensive and subjective, which makes it difficult to apply to real-world applications. {To this end, AI-driven techniques are proposed to extract profiles and model P-O fit in a dynamic, quantitative, and objective way. Specifically, the P-O fit problem is defined as follows:}

\begin{definition}[P-O Fit Problem]
Given a sequence of time periods, each associated with an organization network $G_{t}=(V, E_{t})$, where the nodes $V$ are employees and links $E^t$ indicate their relationships (e.g., reporting relationship) in the time period $t$. Each node $v_i$ has a feature vector $x_{t,i}$, representing its traits and behaviors in the $t$-th time period. The target of Person-Organization Fit is to learn to model the compatibility of each node on the tree with their local environment, seize their dynamic nature and patterns, and accordingly predict relevant talent outcomes $y$.
\end{definition}

To solve this problem, \textsl{Sun et al.} proposed a new P-O fit modeling process based on AI technology~\cite{sun2019impact}, as shown in Figure~\ref{fig:profit2}. Specifically, they first extracted features automatically from collected employees’ in-firm data and generated person profiles by dimension reduction on these features. Then, they exploited the organization profile by combining the organization’s structure with the profiles of the employees and extracted a unique environment profile for each employee based on their corresponding positions. Finally, they applied a deep neural network to achieve a more complicated mapping from person and environment profiles to a P-O fit representation. To capture the dynamic nature of P-O fit and its consequent impact, they exploited an adapted Recurrent Neural Network with an attention mechanism to model the temporal information of P-O fit. Later, in~\cite{sun2021modeling}, \textsl{Sun et al.} further proposed the attentional features extraction layers that can distinguish individualized relation-level and individual-level influence differences for different nodes on the organizational tree. This largely enhanced the performance of person-organization compatibility modeling and improved the interpretability.
Combining the P-O fit theory, \textsl{Artar et al.} used the K-Nearest Neighbors algorithm to cluster the employees through P-O fit representations like the number of former employers, and the number of years in the organization~\cite{artar2024improving}.

\noindent \textbf{Evaluation.} Generally, P-O fit is to learn the compatibility of each node on the tree with its local environment and predict relevant talent outcomes like turnover prediction and performance prediction. These tasks are all classification problems, therefore, Cross-Entropy and AUC are usually used as the evaluation metrics~\cite{sun2021modeling}.

\noindent \textbf{$\diamondsuit$ Takeaway.}
\begin{itemize}[left=0pt]
    \item \textbf{Advantages of AI technology:} \textsl{(1) Traditionally, employee movement among departments has been task-oriented, and existing AI-based approaches take better account of adaptability so that employee performance can be improved over the long term. (2) Compared with traditional linear modeling, the AI-based method is more accurate {in modeling the nonlinear relationship} in P-O matching.}
    \item \textbf{Limitations and future directions:} \textsl{(1) Existing methods mostly concentrate on {organizational network information extraction including communication and report chain, however, there are many other indicators in profiles for P-O fit like age, working years and so on, these indicators should be considered by the model in the future.} (2) {Existing methods overlook the dynamic changes of employees and organizations. For example, while long-term matching may exist, short-term tasks can disrupt {the} alignment between employee and organizational performance.}}
\end{itemize}

\subsection{Organizational Incentive Analysis} 
{Compensation and benefits (C\&B) represent one of the most important branches of human resource management, playing an indispensable role in attracting, motivating, and retaining talent.} It includes the process of determining how much an employee should be paid and deciding what benefits should be offered. In the past few decades, considerable efforts have been made in this research direction from the management perspective. Recently, the accumulation of massive job-related data has enabled a new paradigm for organizational incentive analysis in a data-driven view. In this part, we will introduce two classic data-driven tasks in C\&B, namely job title benchmarking and salary benchmarking, respectively.

\subsubsection{Job Title Benchmarking}
Job title benchmarking (JTB), as an important function in C\&B, aims at matching job titles with similar expertise levels across various organizations (i.e., companies), which provides precise and substantial facilitation of job and salary calibration/forecasting for both talent recruitment and job seekers. Traditional JTB mainly relies on manual market surveys, which are expensive and labor-intensive. Recently, the 
popularity of online professional networks has helped to accumulate {a massive set of career records}, which provides the opportunity for a data-driven solution. Formally, JTB can be defined as follows:

\begin{definition}[Job Title Benchmarking]
JTB is a process that matches job titles with similar expertise levels across various companies. Formally, given two job title-company pairs, i.e., ($title_i, company_i$) and ($title_j, company_j$), the objective is to determine whether the given pairs are on the same level. 
\end{definition}

To handle this problem, \textsl{Zhang et al.} proposed to construct a Job-Graph by extracting information from large-scale career trajectory data, where nodes represent job titles affiliated with the specific companies and edges represent the number of transitions between job titles~\cite{zhang2019job2vec}. They redefined JTB as a link prediction task on the Job-Graph by assuming that the benchmarked job title pairs should have a strong correlation with the link. Along this line, they proposed a collective multi-view representation learning model to represent job titles from multiple views, including graph topology view, semantic view, job transition balance view, and job transition duration view. Subsequently, they devised a fusion strategy to generate a unified representation from a multi-view representation. Finally, they leveraged the similarity between these representations as an indicator for job title benchmarking. Following this line, subsequent studies have refined the job transition network. Specifically, Zhu and Hudelot~\cite{zhu2022towards} proposed {an} enhanced job transition network with the bi-directional edges when job nodes have the same tag, for example, ``automotive shop manager" and ``purchasing manager" have the same tag named ``manager". At the same time, bi-directional edges between a job title and a tag, representing the ``has/in" relationship, have been added into the job transition network. Furthermore, it uses an open dataset from a Kaggle competition containing a collection of working experiences.

{Another way to solve the job title benchmarking problem is through embedding. In these studies, job titles are typically embedded with auxiliary information such as skills, employees, and other related attributes.} \textsl{Zbib et al.} embedded the auxiliary skills under job titles, through the feedback based on the similarity calculation with the text encoding of the job title itself, the reasonable skill embedding under the job title can be learned, and different job titles can be integrated~\cite{zbib2022learning}. Finally, the effect is verified by skill-based retrieval and text-based retrieval. \textsl{Malandri et al.} used the job contents under job title and generated embedding of job titles to cluster similar jobs~\cite{malandri2021meet}. Besides, \textsl{Zha et al.} aggregated job titles by decomposing the semantics of different modules of job titles~\cite{zha2023career}.

The information provided by the job transition graph and embedding can also be combined to perform job title benchmarking. For example, JAMES proposed by \textsl{Yamashita et al.} combined the hyperbolic graph embedding generated from the job transition records, BERT embedding of original job titles, and syntactic embedding generated by the average similarity of original job titles and occupations in job taxonomy like European Skills/Competences, qualifications, and Occupations (ESCO) taxonomy~\cite{yamashita2023james}. Liu and Ge jointly trained the job embedding and employee embedding, which used the job contents. 
{Specifically, the job context learning module generates employee embeddings conditioned on job embeddings by leveraging employees’ job transition histories, where observed positions are considered positive samples and positions selected via negative sampling are considered negative samples~\cite{liu2022job}.}

\noindent \textbf{Evaluation.} Generally, job title benchmarking could be evaluated by calculating the job text similarity or some downstream tasks as follows.
\begin{itemize}[left=0pt]
    \item \textbf{Similarity of Job.} Job title embedding should improve the job similarity, in the ~\cite{liu2022job}, job similarity is obtained from Amazon Mechanical Turk (MTurk), the human-labeling task is \textsl{The similarity between jobs A and B is higher than the similarity between jobs A and C.}, \textsl{The similarity between jobs A and B is lower than the similarity between jobs A and C.} and \textsl{The similarity between jobs A and B is almost the same as the similarity between jobs A and C.} Finally, the average accuracy of the real comparison of the similarity between jobs A and B and the similarity between jobs A and C is adopted as a metric to evaluate the model effectiveness.
    \item \textbf{Clustering Effectiveness.} {There are two specific types of evaluation: intrinsic evaluation, such as the Mann–Whitney U test, and extrinsic evaluation, which involves classification metrics including precision, recall, F1-score, and accuracy~\cite{malandri2021meet}.}
    \item \textbf{Job Title Classification.} {Due to the existence of job taxonomies such as ESCO, the classification of job titles into standardized taxonomies is also adopted as a means to evaluate the model’s effectiveness.} In this case, Precision@N and NDCG@N are used, with N being the top-N results produced by each model~\cite{yamashita2023james}. {Similarly, some scholars adopt alternative classification schemes derived from the dataset, thereby using additional evaluation metrics such as Macro-F1 and Micro-F1~\cite{zhu2022towards}.}
    \item \textbf{Link Prediction.} {Due to the use of job transition networks in job title embedding, link prediction has become one of the most common tasks for such graphs. In this context, AUC is commonly used as the primary evaluation metric~}\cite{yamashita2023james,zhang2019job2vec}.
    \item \textbf{Job Mobility Prediction.} Different from the link prediction, job mobility prediction is a job-domain downstream task, generally, MAP@10, mean average precision at 10 jobs, is adopted~\cite{yamashita2023james}. More strictly, the accuracy of the next job prediction is adopted in some other research~\cite {zhu2022towards}. Besides, some commonly used metrics for job duration prediction, such as RMSE and MAE, are employed~\cite {zha2023career}.
    \item \textbf{Text Ranking.} Job recommendation is an essential application of job title embedding. Therefore, precision at 5 or 10 is commonly used in the evaluation of tasks such as retrieval~\cite{zbib2022learning}.
\end{itemize}

\noindent \textbf{$\diamondsuit$ Takeaway.}
\begin{itemize}[left=0pt]
    \item \textbf{Advantages of AI technology:} \textsl{(1) AI-based job title benchmarking can effectively help organizations sort out the positions in the market and within the organization. However, the traditional method is manual and cannot comprehensively analyze all the positions in the organization, which avoids subjective biases. (2) Existing AI-based approaches can help improve the efficiency of the external market analysis for organizations, while providing the foundation for some meaningful downstream tasks of talent management, such as job pricing, job grading, and labor market competition analysis.}
    \item \textbf{Limitations and future directions:} \textsl{(1) Job taxonomy currently {is} quite diverse, such as ESCO, Occupational Information Network (O*NET), and modeling considering the adaptation of different taxonomies and research context is necessary for future research. (2) The existing methods are mostly based on the previous job transition network or {the embeddings of published job postings}, and do not take into account the dynamic evolution of jobs over time. In the future, more consideration should be given to the lifelong models. (3) In the process of external environment changes, many new jobs will appear. How to model these new jobs is also the future development direction.}
\end{itemize}

\subsubsection{Job Salary Benchmarking}
Job salary benchmarking (JSB) refers to the process by which organizations obtain and analyze labor market data to determine appropriate compensation for their existing and potential employees. Traditional approaches for JSB mainly rely on the experience {of} domain experts and market surveys provided by third-party consulting companies or governmental organizations~\cite{porter2004dynamics}. However, fast-developing technology and industrial structure lead to changes in positions and job requirements, making it difficult to conduct salary benchmarking in a timely manner in dynamic scenarios. 

In recent years, the prevalence of emerging online recruiting services, such as Indeed and Lagou, has provided the opportunity to accumulate vast amounts of job-related data from a wide range of companies, enabling a new paradigm of compensation benchmarking in a data-driven manner. Formally, the job salary benchmarking problem can be formulated as follows:

\begin{definition}[Job Salary Benchmarking]
Suppose there are job positions $i=1,2,3...,I$ and location-specific company $j=1,2,3,...,J$. Each position $i$ has some features, e.g., bag-of-words, and each location-specific company $j$ can be described as a list of features, e.g., location and industry. Given a combination of position and company ($i,j$), the objective is to predict its salary $\hat{s}_{ij}$ so that the similarity between $\hat{s}_{ij}$ and real observation $s_{ij}$ is maximized.
\end{definition}

To address this problem, some scholars used statistical machine learning methods to combine the company characteristics to predict the average salary of industries or economic activities~\cite{matbouli2022statistical}. Then, from the job-company salary matrix view,
where each entry indicates the corresponding salary of a given job-company pair, the JSB problem can be regarded as a matrix completion task. Generally, matrix factorization (MF) is a widely used method for handling this task. It aims to factorize an incomplete job-company salary matrix into two lower-rank latent matrices, and use their dot product for estimating the possible salary of the missing entries. However, the intuitive method is too general to meet the various special needs of C\&B professionals. To this end, \textsl{Meng et al.} proposed an expanded salary matrix by expanding the original job-company salary matrix with locations and time information for the fine-grained salary benchmarking~\cite{meng2018intelligent}. Then they designed a matrix factorization-based model for predicting the missing salary information in the expanded salary matrix by integrating multiple confounding factors, including company similarity, job similarity, and spatial-temporal similarity. Further, \textsl{Meng et al.} designed a nonparametric Dirichlet-process-based latent factor model for JSB, which learns representations for companies and positions to alleviate the data deficiency problem. By conducting experiments on two large-scale real-world datasets, the effectiveness and interpretability of the proposed model have been proved~\cite{meng2022fine}.
Similarly, the matrix equation method is used in~\cite{hung2021aggregating} to minimize the unbiased salary, company competitiveness in the salary of the same job, and inflation of the same job. Following another way of thought, some scholars proposed to construct the auxiliary network of skill requirements under job postings to model the relationship between the job and salary. {Specifically, the auxiliary network can learn job representations through graph learning, enabling the grouping of similar jobs with comparable salary-related learning patterns, and thereby improving the effectiveness of salary prediction across different jobs~\cite{sun2021market}}.

\noindent \textbf{Evaluation.} {Generally, there are two main approaches to evaluating job salary benchmarking: job similarity assessment and salary prediction. As for the job similarity task, the aim is to minimize the salaries among similar jobs, so MAE could be used to test the biases. The Kendall Coefficient} is adopted to evaluate the similarity of generated vectors of competitiveness (or inflation) and the ground-truth competitiveness (or inflation)~\cite{hung2021aggregating}. As for salary prediction, due to this task being generally a regression problem, MAE, RMSE, and Pearson relationship (PR) have been used~\cite{sun2021market,meng2022fine}.

\noindent \textbf{$\diamondsuit$ Takeaway.}
\begin{itemize}[left=0pt]
    \item \textbf{Advantages of AI technology:} \textsl{(1) Existing methods can improve the efficiency and accuracy of job salary setting because they fully integrate market information. (2) Compared with the traditional method, it is difficult to achieve the salary alignment of all positions due to the huge labor cost and individual preference. The existing method provides a solution for the transparent management of C\&B in the organization, eliminating as much as possible the inequities in the different positions of the organization.}
    \item \textbf{Limitations and future directions:} \textsl{(1) The validation sets used by the existing methods come from various countries, and more comparison of differences across countries is needed. (2) {Existing methods often overlook the dynamic changes in job salaries across different time periods. For example, in the long term, talents with large language model (LLM) expertise tend to receive higher salaries. However, such skills were either nonexistent or not highly valued in the past labor market, resulting in lower historical salaries for similar roles.}}
\end{itemize}

\subsection{Summary}

In conclusion, AI-related techniques for organization management contain three aspects, including the organizational network analysis, organizational stability analysis, and organizational incentive analysis. Specifically, organizational network analysis aims to help understand the importance of critical connections and flows in an organization, which can serve downstream talent management applications. Then, organizational stability analysis focuses on analyzing the composition of the organization and exploring the compatibility between employees and organizations. Finally, organizational incentive analysis concentrates on leveraging data-mining techniques to solve the job title/salary benchmarking problem in human resources. 
Indeed, compared with the traditional manually selected factors, these methods could decrease the labor cost and subjective biases to a great degree. At the same time, AI-based methods could generate complete information about employees and organizations to develop more accurate results. While existing methods support time slicing, the accelerated evolution of organizations and talents calls for finer-grained time slicing and continuous-time information extraction in future research. {Additionally, as performance evaluations of organization management are typically conducted by top managers and involve multifaceted considerations, future approaches should combine data-driven techniques with expert evaluation methods. Finally, with the rapid advancement of LLMs, agent-driven organizational behaviors present an important direction for further study.}


\section{Labor Market Analysis}\label{cluture}

\begin{figure}[t!]
	\centering
    
	\includegraphics[width=1.0\columnwidth]{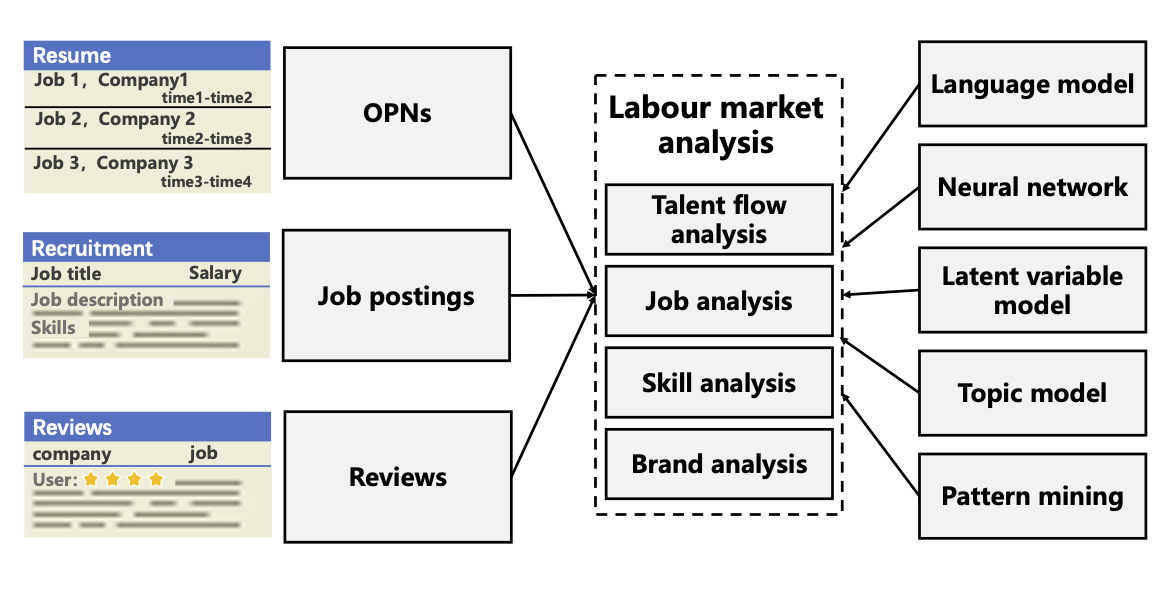}
	\vspace{-6mm}
	\caption{The overview of labor market analysis.}
	\vspace{-6mm}
	\label{fig:labour market} 
\end{figure}

Labor market analysis is crucial to the formulation of the strategy, which is an important part of intelligence talent management. 
Traditionally, most existing studies on the labor market generally also rely on expert knowledge, subjective surveys, and qualitative analysis from psychological, economic, and cultural perspectives. These methods make it challenging to uncover the complex associations among multi-source data and the hidden patterns in massive data. The efficiency is also limited by manual analysis. Moreover, some studies that rely on online collected data typically employ causal inference or statistical analysis methods. For example, \textsl{Jackson et al.} deployed psychometric measures with internet surveys to infer the reasons behind talent flow~\cite{jackson2005exploring}. \textsl{Hershbein et al.} analyzed concentration in labor markets from vacancy and employment data~\cite{hershbein2018concentration}. \textsl{Hershbein et al.} deployed various statistical approaches to analyze the different skill requirements of job postings in different economic situations~\cite{hershbein2018recessions}. 

Recently, the prevalence of Online Professional Networks (OPNs) and online recruitment websites has facilitated the accumulation of a large number of job reviews, company reviews, digital resumes, and job postings. These sources contain a wealth of intricate and diverse information about the labor market, including talent flow, talent demand, market trends, job skills, company branding, and more. These extensive datasets provide novel perspectives and opportunities for conducting a more fine-grained analysis of the labor market at a large scale. 
{However, traditional methods face significant challenges in uncovering complex market patterns from data and accurately forecasting market trends.}
AI and machine learning algorithms possess powerful pattern recognition, data generalization, and fitting capabilities, making them well-suited for exploring labor market data~\cite{lin2017collaborative,moniz2014sentiment,schmiedel2019topic,zhang2020large,sun2021market,zha2023career}. 
Many researchers have analyzed the labor market with AI methods mainly from four aspects: talent flow analysis, job analysis, skill analysis, and brand analysis.

We initially introduce AI-driven analysis of talent flow as a common behavior observed in the labor market. 
AI methodologies play a pivotal role in discerning talent flow inclinations and organizational competition, facilitating two downstream tasks as flow prediction and flow pattern analysis~\cite{zhang2019large, zhang2020large, liu_talent_2023, cheng2013jobminer, cheng2015mining, oentaryo2017analyzing,oentaryo2018talent, xu2016talent}.
Subsequently, we direct our attention to AI-driven job analysis works. Fundamentally, the labor market refers to the supply of jobs and demand for talent. Talent demand forecasting is an important part of job analysis. Several studies have proposed methodologies to scrutinize demand time series, discern fluctuations, and anticipate future demand trends~\cite{lovaglio2020comparing, karakatsanis2017data, zhang2021talent, guo_talent_2022}. Besides, topic trend analysis constitutes a vital aspect of job analysis, delving into recruitment evolution under different recruitment topics~\cite{zhu2016recruitment}.
Furthermore, we {introduce works} related to skill analysis, which is also a pivotal aspect of labor market analysis. As skills are inherent within job descriptions but not directly accessible, extracting and predicting potential skills contained therein constitute significant endeavors~\cite{colombo2018applying, akhriza2019constructing, patacsil2021analyzing, wowczko2015skills, wu2019trend, xu2018measuring, liu_2021_learning, walek_2021_data, leon_2024_hierarchical}. 
To further adapt to the rapid evolution of the labor market, AI delves deeply into changes of skill trends, captures dynamic characteristics of skill demand, and forecasts future skill demand~\cite{mahdavimoghaddam_2021_congruence, mahdavimoghaddam_2022_exploring, de_macedo_practical_2022, wolf_2023_generating, chao_cross-view_2024, xi_pretrain_2024}. Besides, the skill valuation represents a vital aspect of skill analysis, which can furnish job seekers and employers with clearer insights into skills ~\cite{rahman_worker_2015, sun2021market, stephany_what_2024}.
Lastly, we introduce brand analysis, crafting a comprehensive company profile through the mining of employee data and reviews about the company~\cite{lin2017collaborative,lin2020enhancing, bajpai2019aspect, spears2021impact, yuan_self-supervised_2021, bose_sentiment_2022, savin_topic-based_2023}. This approach not only elucidates users' perceptions but also reveals potential expectations and social responsibility placed on the company~\cite{chae2018corporate, mazza_text_2022, pilgrim_csr_2023, thakur_impact_2023}. Moreover, leveraging established sentiment classification methods within NLP enables the effective identification of employees' sentiments towards the company~\cite{moniz2014sentiment, ikoro2018analyzing,shi_listening_2021,ganga_employees_2022, gaye_sentiment_2021, rehan_employees_2022, mouli_unveiling_2023}, thereby preventing employee turnover.
Figure~\ref{fig:labour market} presents an overview of works related to labor market analysis.

\subsection{Talent Flow Analysis.} 

{Talent flow analysis includes prediction and pattern analysis tasks, primarily leveraging OPNs to reflect talent movement between companies and support strategic planning.}


Talent flow analysis encompasses two key components: talent flow prediction tasks and diverse flow pattern analyses. These analytical tasks are predominantly reliant on OPN data, which provides a detailed depiction of talent movement across different companies and enables a thorough examination of the talent flow dynamics within the market. By leveraging this data, enterprises can gain invaluable insights into the patterns and trends of talent migration, enabling them to formulate targeted strategies to address issues related to brain drain~\cite{zhang2019large}. Furthermore, governments can utilize talent flow analysis to cultivate a robust ecosystem for talent circulation~\cite{liu_talent_2023}, {fostering sustained economic growth and innovation.}

Traditional methods, reliant on subjective surveys for qualitative inference, lack objectivity and precision. They are time-consuming, struggle to adapt to market changes, and provide limited predictive insights. {The advent of OPNs enables large-scale, timely talent flow analysis through data-driven AI approaches.}

Following ~\cite{zhang2019large}, we denote talent flow as transition tensor $R^t\in \mathbb{R}^{N\times N \times M} $ for each time slice $t$, where $N$ denotes the number of companies, $M$ denotes the number of job positions, and each element $R^{t}_{ijk}$ is defined as the normalized number of corresponding job transitions: 

\begin{equation}
R^{t}_{ijk}=\frac{Num^t_{i,j,k}}{\sum_{j=1}^{N}Num^t_{i,j,k}},
\end{equation}
where $Num^t_{i,j,k}$ denotes the transition number from the job position $k$ of company $i$ to company $j$ at time slice $t$. By definition, talent flow analysis tasks mainly contain talent flow prediction and various flow pattern analyses.

\subsubsection{Flow Prediction}

The task of talent flow prediction mainly revolves around anticipating changes in the labor market, thereby offering guidance for talent strategies. {AI-driven techniques enhance the accuracy and flexibility of such predictions. Formally, the talent flow prediction problem is defined as follows:}


\begin{definition}[Talent Flow Prediction]
Given a set of talent flow tensors {\{$R^1$, ..., $R^T$\}, and some attributes of companies and market context, the goal of talent flow prediction is to predict the value of $R^{T+1}$.}
\end{definition}

To solve this problem,
\textsl{Zhang et al.} designed a dynamic latent factor-based Evolving Tensor Factorization (ETF) model for predicting future talent flows~\cite{zhang2019large}. In detail, they used $U^t_i$, $V^t_j$, $W^t_k$ to represent the latent vectors of origin company $i$, destination company $j$, and job position $k$ at time slice $t$, and evolved them to time slice $t+1$ for predicting talent flows at $t+1$. This model also integrates several representative attributes of companies as side information for regulating the model inference.
The authors also proposed a Talent Flow Embedding (TFE) model to learn the bi-directional talent attractions of each company~\cite{zhang2020large}. Subsequently, they explored the competition between different companies by analyzing talent flows using data from OPNs. In detail, the objective of this latent variable model is to learn two attraction vectors $S_u$ and $T_u$ from the talent flow network $G$, where $S_u$ is the source attraction vector of company $u$ and $T_u$ is the target attraction vector of company~$u$. Paired dots representing $S_u$ and $T_v$ indicate the talent flow from company $u$ to company $v$. The experimental results show the pairwise competitive relationships between different companies.
\textsl{Xu et al.} enriched the sparse talent flow data by exploiting the correlations between the stock price movement and the talent flows of public companies~\cite{xu2018dynamic}. They developed a fine-grained data-driven RNN model to capture the dynamics and evolving nature of talent flows, utilizing the rich information available in job transition networks. 
In addition to leveraging OPN data, \textsl{Liu et al.} transformed questionnaire data into graph data and achieved accurate talent flow prediction using the Graph Attention Network~\cite{liu_talent_2023}. This model incorporates an attention mechanism to mitigate information overload and decrease the network's reliance on temporal and spatial factors.
{Several studies have used online search data to capture human behaviors and mobility patterns~\cite{zhu2020rapid}. Along this line, \textsl{Sun et al.} analyzed four billion job search queries and 400 million job postings to study labor migration during COVID-19~\cite{sun2024large}. They built a city-level labor flow graph, identifying clusters of cities with significant labor flow, including cities with high labor inflow and high outflow. Using the HITS algorithm, they measured city centrality for labor attraction and exportation, and classified job postings by labor types to track changes in regional labor demand. Their findings show a decreased central role of major cities and reduced regional labor mismatches during COVID-19.}

\noindent \textbf{Evaluations.}
Generally, the evaluation of talent flow prediction encompasses two primary perspectives: value prediction and link prediction.
\begin{itemize}[left=0pt]
\item \textbf{Value Prediction}: \textsl{Xu et al.}~\cite{xu2018dynamic} and \textsl{Zhang et al.}~\cite{zhang2019large} utilized widely accepted regression metrics like RMSE, MAPE, and MAE to gauge the accuracy of predicted flow values. These metrics provide insights into the disparity between predicted and actual flow values, offering a comprehensive assessment of the model's predictive prowess.
\item \textbf{Link Prediction}: \textsl{Liu} extracted features from questionnaire data to predict the occurrence of talent flow among academics, introducing overall classification accuracy for evaluation~\cite{liu_talent_2023}. Additionally, in the context of talent movement, companies with greater competitive appeal tend to attract talent. \textsl{Zhang et al.} employed AUC to assess company attraction and NDCG to evaluate ranking effectiveness~\cite{zhang2020large}. AUC measures the model's ability to discriminate between attractive and less attractive companies, while NDCG evaluates the ranking list's efficacy in capturing the most desirable companies.
\end{itemize}

\noindent \textbf{$\diamondsuit$ Takeaway.} 
\begin{itemize}[left=0pt]
\item \textbf{Advantages of AI technology}: 
\textsl{
(1) Leveraging extensive job transition data, AI significantly enhances the prediction of talent flow across larger populations.
(2) AI-based methods offer objectivity compared to subjective survey-based approaches.
(3) Through analyzing the evolving patterns of talent flow over time, AI facilitates timely forecasts of dynamic talent movement.}
\item \textbf{Limitations and future directions}: 
\textsl{
(1) Job titles are often clustered without a standardized benchmark for evaluation, leading to a lack of assessment regarding the quality of the clustering process. Establishing improved benchmarks for job titles can enhance the standardization and normalization of this clustering procedure.
(2) Existing research predominantly focuses on talent flow within homogeneous job categories, neglecting the complexity of predicting talent movement across diverse job types. Expanding the analysis to encompass multiple dimensions of talent flow and achieving more nuanced predictions could offer valuable insights into job mobility across various occupations.
(3) Current research fails to delve into the underlying motivations driving talent flow, a challenge compounded by the opaque nature of neural networks. However, recent advancements in LLM have shown promising capabilities in explanation, thereby rendering the analysis of the reasons behind talent flow more attainable.}
\end{itemize}

\subsubsection{Flow Pattern Analysis}
Many researchers also explored other various flow pattern analysis tasks, such as competitiveness analysis, hopping behavior analysis, and talent circle detection. 
In~\cite{cheng2013jobminer}, the author initially gathered job-related information from various social media data. Subsequently, they developed a model called JobMiner, which focuses mainly on employing graph mining techniques to mine influential companies and uncover talent flow patterns. This method provides a better understanding of company competition and talent flow in professional social networks.
Yu Cheng further developed machine learning and analytical techniques for the purpose of mining OPNs data~\cite{cheng2015mining}. From OPNs, they can mine influential companies with the related company groups and evaluate the company's influence and competitiveness.
\textsl{Oentaryo et al.} developed a series of data mining methodologies to analyze job-hopping behavior between different jobs and companies using publicly available OPNs data~\cite{oentaryo2017analyzing}. In detail, they used a weighted version of PageRank to measure the competitiveness of jobs or companies, and constructed some metrics to measure the relationship between many properties of jobs and the propensity for hopping.
\textsl{Li et al.} undertook the quantification of the dynamic attractiveness of companies to various types of talent~\cite{li_measuring_2024}. Specifically, they utilized job title clustering to identify prevalent talent categories and then integrated the PageRank algorithm with job titles and talent flow graphs to determine companies' attractiveness to specific talent categories over time.
Then, \textsl{Oentaryo et al.} enhanced the data mining framework for analyzing talent flow patterns~\cite{oentaryo2018talent}. The results show that the factors influencing employee turnover can mainly be divided into four categories: employee personal factors, organizational factors, external environmental factors, and structural factors.
\textsl{Xu et al.} developed a talent circle detection model and designed the corresponding learning method maximizing the NDCG to detect a suitable circle structure~\cite{xu2016talent}. Each talent circle includes organizations with similar talent exchange patterns. Formally, a talent circle is a subset of neighbors of an ego node. In one circle, nodes are closely connected and similar to each other. The circles can be denoted as $\{C_m\in \mathbb{C}\}$, where $m = 1, 2, ..., M$ and $C_m \subseteq V$. $V$ represents the set of all organizations. The circles can be overlapped, and the appropriate talent circles mean similar flow patterns and are closely connected.
The detected talent circle can be used to predict talent exchange in the future and improve the recommendations in talent recruitment.

\noindent \textbf{Evaluations.}
In analyzing talent flow patterns, various metrics are employed to assess performance across different tasks. \textsl{Xu et al.} utilized precision and recall to evaluate talent exchange prediction~\cite{xu2016talent}. \textsl{Li et al.} compared models using clustering-validation metrics, including Silhouette score (higher is better), Intra-Dispersion (lower is better), and Inter-Dispersion (higher is better)~\cite{li_measuring_2024}. These metrics gauge separation and cohesion within clusters. F1-score is reported by \textsl{Li et al.} for Employee attrition and Career outcome prediction, assessing classification performance. \textsl{Oentaryo} delved into talent flow patterns using metrics like work experience and job age, along with higher-level metrics such as external hop fraction and job level aggregated at job or organization levels~\cite{oentaryo2017analyzing, oentaryo2018talent}. They also introduced network centrality metrics like in-degree centrality, out-degree centrality, and PageRank centrality to gauge node importance in job and organization graphs.

\noindent \textbf{$\diamondsuit$ Takeaway.} 
\begin{itemize}[left=0pt]
\item \textbf{Advantages of AI technology}: 
\textsl{(1) AI technology enables the analysis of vast amounts of data, such as detailed job activity data from online professional networks, at a scale that would be impossible with traditional methods like surveys. This scalability allows for a more comprehensive understanding of job-related insights.
(2) Real-time Insights: Unlike traditional surveys, which may have long turnaround times and cover only a small percentage of businesses or organizations, AI technology can provide real-time insights by continuously analyzing data as it becomes available. This real-time aspect is crucial for staying abreast of fast-changing trends.
(3) Network Analysis: AI technology facilitates the exploration of interconnected networks within the job market, capturing talent flows from one job to another and from one organization to another. This network analysis provides a more holistic understanding of talent flow patterns, competition among organizations for talent, and the impact of these dynamics on job creation and talent attraction. By leveraging AI for network analysis, researchers and practitioners can uncover valuable insights that were previously obscured by a lack of visibility into these complex interactions.}
\item \textbf{Limitations and future directions}: 
\textsl{
(1) Data Privacy and Ethics: As AI technology relies heavily on data, especially from online professional networks, ensuring data privacy and ethical use becomes paramount. Future directions should focus on developing robust frameworks for data anonymization, consent management, and ethical guidelines for AI-driven analyses to protect user privacy while still extracting valuable insights.
(2) Bias and Fairness: AI algorithms can inadvertently perpetuate biases present in the data they are trained on, leading to unfair outcomes. Future directions should prioritize the development of bias detection and mitigation techniques within AI systems, along with ensuring fairness and transparency in decision-making processes to mitigate societal biases.
}
\end{itemize}

\subsection{Job Analysis.}

Job analysis focuses mainly on the analysis of trends, such as demand trends, topic trends, and other relevant factors. These tasks mainly use job posting data, which contains information about the recruitment demand of jobs from companies, to analyze the situation of the recruitment market. 
Job analysis is essential to model labor market variations, enabling companies to adjust recruitment strategies effectively and empowering job seekers to plan their career paths proactively.
Traditional job analysis in recruitment usually involves domain experts employing basic statistical methods. However, with the advancement of online recruitment services and data mining technologies, researchers now predict labor market trends using data-driven methods.
Following~\cite{zhang2021talent}, the job recruitment data can be denoted as trend tensor $D^t\in \mathbb{R}^{N \times M} $ for each time slice $t$, where $N$ denotes the number of companies, $M$ denotes the number of job positions, and each element $D^{t}_{ij}$ is the number of job postings published at time slice $t$, from company $i$ and position $j$. 
Meanwhile, more context information on companies and job positions can be denoted as $\{C_1,...,C_N\}$ and $\{P_1,...,P_M\}$. 
Based on these data, a variety of trend analysis tasks can be explored.

\subsubsection{Talent Demand Forecasting}
Forecasting talent demand is pivotal for both job seekers and employers as it offers a crucial insight into economic trends. Leveraging AI-driven techniques can significantly boost the precision and adaptability of these forecasts. 
Formally, the talent trend forecasting problem can be defined as follows.
\begin{definition}[Talent Demand Forecasting]
Given a set of talent demand tensors $\{D^t_{ij} |t\in [1, T], i \in [1, N], j \in [1, M]\}$, and some side information about companies and job positions. The goal of talent demand forecasting is to predict the value of $D^{T+1}_{ij}$.
\end{definition}

The demand trend mainly focuses on the number of job recruitment, 
some researchers used classifier method SVM and decomposition methods STL (Seasonal and Trend decomposition using Loess) to analyze the demand time series characteristics from the web data and the official data~\cite{lovaglio2020comparing}. These studies show that web data can reflect the trend of the labor market.
\textsl{Karakatsanis et al.} suggested a data mining-based approach for identifying the most in-demand occupations in the modern job market~\cite{karakatsanis2017data}. In detail, a Latent Semantic Indexing (LSI) model was developed for online job posts with job description data. The analysis results can highlight most in-demand job trends and identify occupational clusters.
\textsl{Zhang et al.} provided Talent Demand Attention Network (TDAN), which can forecast fine-grained talent demand in the labor market~\cite{zhang2021talent}. Specifically, they constructed multiple-grained levels of information (e.g., market level, company level, job level et al.) and the intrinsic attributes of both companies and job positions from recruitment job post data. Then, they designed a transformer-based attentive neural network to automatically utilize this information to forecast the demand trend of each job in each company. 
\textsl{Guo et al.} introduced a Dynamic Heterogeneous Graph Enhanced Meta-learning (DH-GEM) framework to predict fine-grained talent demand-supply jointly~\cite{guo2022intelligent}. They employed a Demand-Supply Joint Encoder-Decoder (DSJED) and a Dynamic Company-Position Heterogeneous Graph Convolutional Network (DyCP-HGCN) to capture the correlation between demand-supply sequences and company-position pairs. Additionally, they proposed a Loss-Driven Sampling-based Meta-learner (LDSM) to optimize long-tail forecasting tasks with limited training data.


\noindent \textbf{Evaluations.}
Given the challenges associated with forecasting the value of individual points within the univariate time series $D^T_{ij}$, \textsl{Zhang et al.}\cite{zhang2021talent} and \textsl{Guo et al.}\cite{guo2022intelligent} opt to categorize the values into distinct trend categories, treating the prediction task as a time-series classification problem. Consequently, assessments of talent demand forecasting primarily rely on metrics such as ACC, F1, and AUC.

\noindent \textbf{$\diamondsuit$ Takeaway.} 
\begin{itemize}[left=0pt]
\item \textbf{Advantages of AI technology}: 
\textsl{
(1) AI revolutionizes talent demand analysis by leveraging large-scale data from online recruitment platforms. Unlike traditional survey-based methods, AI offers more comprehensive insights, enabling fine-grained forecasting at the level of specific positions within companies. This granular approach enhances prediction accuracy, facilitating real-time analysis of talent demand dynamics. 
(2) Moreover, AI techniques quantitatively model the dynamic nature of the recruitment market, crucial for understanding evolving trends. 
(3) Additionally, AI uncovers latent data dependencies not apparent through conventional models, identifying nuanced relationships between factors of talent demand.
}
\item \textbf{Limitations and future directions}: 
\textsl{Existing demand trend forecasting can only predict the categories of trend change, and it struggles to predict specific demand values. In the future, more consideration will be given to the correlation between companies and positions and the fusion and interaction of multivariate time series, which can further achieve accurate multivariate time series predictions.}
\end{itemize}

\subsubsection{Topic Trend Analysis}
The topic trend primarily focuses on text mining and language modeling techniques applied to job postings. For instance, 
\textsl{Marrara et al.} designed a language modeling approach for discovering novel occupations in the labor market, which can help the company catch the new trend of recruitment~\cite{marrara2017language}.
\textsl{Zhu et al.} developed MTLVM, which is a sequential latent variable model~\cite{zhu2016recruitment}. This model can capture sequential patterns of recruitment states. Moreover, it can automatically learn the latent recruitment topics by the Bayesian generative framework. In detail, it uses $c_{e,t}$ to represent the latent recruitment state of {company $e$} at time step $t$. Then, the transition probability between different states is learned to analyze the evolving rules of the recruitment trend, and the topic model is deployed to reveal the trend of different recruitment topics.
\textsl{Azzahra et al.} categorized job vacancies into groups like Administration, Finance, IT, and Marketing using a multilabel-classification method with word2vec and an RNN model~\cite{azzahra_text_2021}.
\textsl{Mahdavimoghaddam} used NER tools to explore online social topics related to jobs~\cite{mahdavimoghaddam_utility_2022}. They assessed social content's effectiveness in predicting future job requirements, analyzed the relationship between work-related emotions online and social demographics, and identified potential impacts of community support on users' well-being in the job market.

\noindent \textbf{Evaluations.}
In Topic Trend Analysis, \textsl{Zhu et al.} evaluated using Validity Metric (VM) and Coverage Metric (CM) to assess topic relevance and word coverage~\cite{zhu2016recruitment}. They also analyzed recruitment state prediction and trend forecasting accuracy using log-likelihood.
\textsl{Azzahra et al.} converted job vacancy tagging as a multi-label task, data evaluation framework~\cite{azzahra_text_2021}. It includes partition and ranking evaluation, along with label hierarchy utilization. Metrics used are accuracy, precision, F1-score, and Hamming loss, measuring different aspects of prediction performance.

\noindent \textbf{$\diamondsuit$ Takeaway.} 
\begin{itemize}[left=0pt]
\item \textbf{Advantages of AI technology}: 
\textsl{
(1) AI enables a fine-grained understanding of recruitment market trends, capturing nuances and fluctuations that traditional methods might overlook.
(2) Leveraging large-scale analysis on massive recruitment data, AI provides data-driven insights beyond relying solely on domain expert knowledge or classic statistical models.
(3) AI facilitates recruitment forecasting for companies over time, offering more precise predictions for future trends and developments.
}
\item \textbf{Limitations and future directions}: 
\textsl{
(1) The current landscape of labor market trend prediction primarily centers on forecasting trends for specific job vacancies, uncovering potential job opportunities, and predicting topic trends within particular companies. However, a unified framework for anticipating topic trends across the labor market is absent.
(2) Moreover, in the realm of company-specific topic trend prediction, the model's efficacy is restricted to companies present in the training dataset, thereby lacking the capability for few-shot learning. Addressing this limitation, future endeavors should explore methodologies to extrapolate topic trends from established companies to emerging ones, enabling accurate predictions even in low-resource scenarios. This transition toward effective topic trend prediction in few-shot settings represents a promising avenue for further research and development.
}
\end{itemize}


\begin{figure}[t!]
	\centering
	\includegraphics[width=0.8\columnwidth]{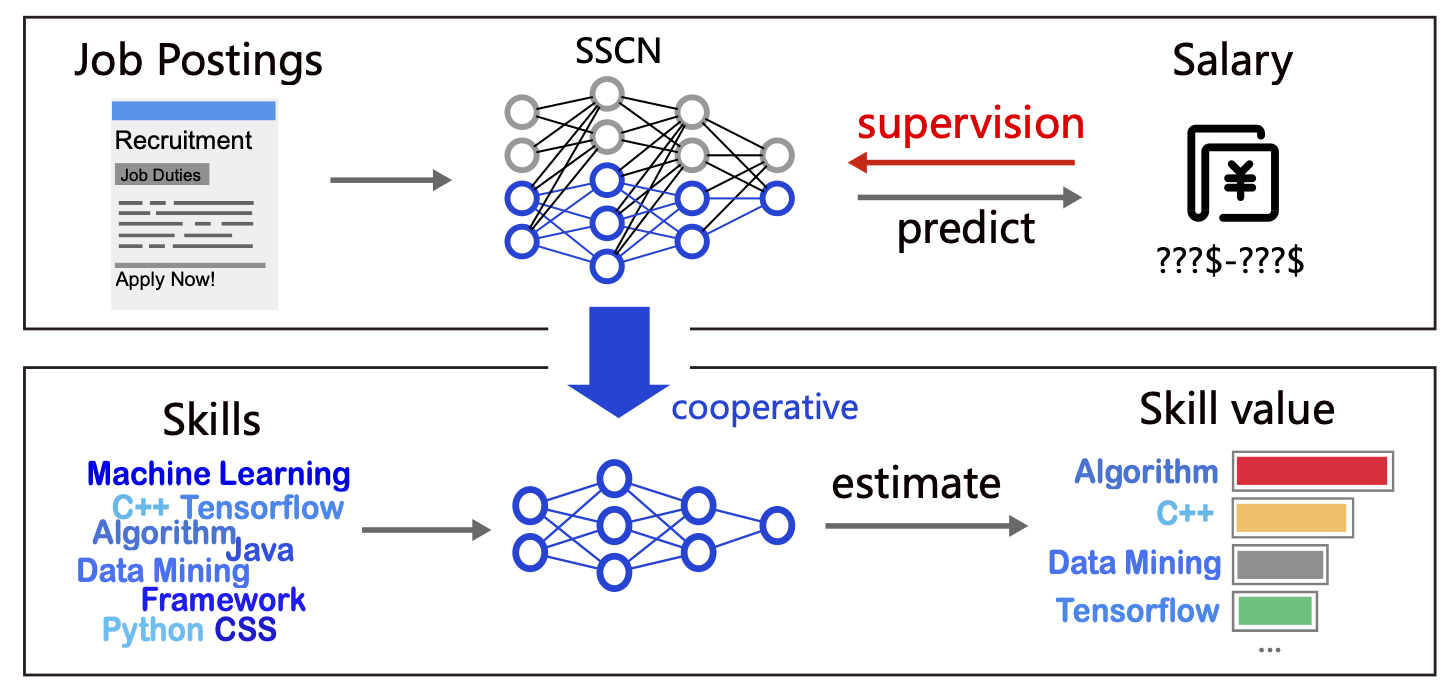}
	\vspace{-4mm}
	\caption{The overview of job skill valuation.}
	\vspace{-6mm}
	\label{SSCN} 
\end{figure}

\subsection{Skill Analysis.}
Tasks related to skill analysis mainly concentrate on exploring the relationship between jobs and skills, such as analyzing the skills required for different jobs and estimating the value of newly emerged skills. These tasks rely on job posting data, which contains information on skills, jobs, and salaries, to analyze the situation of job skills in the recruitment market. Following~\cite{xu2018measuring,sun2021market}, the job postings can be denoted as
$\mathcal{P}=\{(D_i,J_i,S_i,Y_i,T_i)\rvert i=1,2,...\}$, 
where $D_i$ denotes a set of job description, $J_i$ denotes the job title, $S_i$ denotes required skill set, $Y_i$ denotes the job salary, $T_i$ denotes the publish time. 

\subsubsection{Potential Skill Prediction}
The skill requirements can be inferred from job postings, and analyzing these requirements can provide valuable assistance in talent selection, job description formulation, and other related tasks.
The task of skills requirement prediction can generally be formulated as:

\begin{definition}[Potential Skill Prediction] Given a set of job postings $\mathcal{P}\{(J_i,S_i,(D_i,Y_i,T_i)^*)\rvert i=1,2,...\}$, where * indicates optional information. The goal of skills requirement prediction is to measure the required level or popularity for potential skills $S_i$ of job $J_i$ {in the labor market}.
\end{definition}

To solve this problem,
\textsl{Wowczko et al.} used k-NN clustering methods to identify key skill requirements in online job postings~\cite{wowczko2015skills}.        
\textsl{Colombo et al.} deployed language models and machine learning classification approaches, e.g., SVM, to calculate the skills requirements of jobs~\cite{colombo2018applying}. Furthermore, they also classified job skills into a standard classification system and measured the relevance of soft and hard skills, which is important for talent selection and culture cultivation.
\textsl{Xu et al.} proposed a Skill Popularity based Topic Model (SPTM) for modeling the generation of the skill network~\cite{xu2018measuring}. They used the neighbors of a skill on the skill network to generate the document for this skill. Then, the documents can be used to further analyze the popularity of the skill using topic models. This kind of topic model can integrate different criteria of jobs (e.g., salary levels, company size) and the latent connections between different skills. Then they effectively ranked the job skills based on the multi-faceted popularity.
\textsl{Wu et al.} designed a Trend-Aware Tensor Factorization (TATF) framework to analyze the skill demand of jobs~\cite{wu2019trend}. In detail, TATF constructs the relationship between skills and jobs as a special tensor with 4 dimensions, each element $e_{t,c,p,s}$ in this tensor reflects the demand trend of skill $s$ in job $p$, company $c$ at time $t$. Then, they enhanced tensor factorization with aggregation-based constraints, i.e., competition (among companies) and co-occurrence (among skills) based aggregations. Furthermore, they designed the temporal constraint based on previous models to output jobs and skills representations that can quantify the potential skill trends of jobs.
\textsl{Akhriza et al.} applied {the Apriori algorithm used with association rules} and used recommendation techniques based on the output of the skill association to determine the most sought-after IT skills in the industry~\cite{akhriza2019constructing}. 
\textsl{Patacsil et al.} applied the Frequent Pattern-growth (FP-growth) algorithm of the association rule to analyze the relationship of jobs and skills requirements, which provides a new dimension in labor market research~\cite{patacsil2021analyzing}. {These results can reveal the skill requirements of jobs, which are important for enhancing training strategies.}
\textsl{Liu et al.} devised multiple graphs, including J-Net, S-Net, and JS-Net, to jointly learn the semantics of job positions and skills using a three-layer graph neural network (GNN)~\cite{liu_2021_learning}. Their approach successfully mapped job position descriptions to their requisite skills.
\textsl{Walek et al.} introduced an automated detection method leveraging SDCA logistic regression to identify and accurately label all job requirements within job advertisements~\cite{walek_2021_data}.
\textsl{Leon et al.} proposed a classification methodology aimed at uncovering correlations between job ad requirements and cross-cutting skill sets~\cite{leon_2024_hierarchical}. They focused on predicting the necessary skills for individual job descriptions using Bert embeddings.

\noindent 
\textbf{Evaluation.}
\begin{itemize}[left=0pt]
\item \textbf{Ranking Metrics:} \textsl{Xu et al.} developed a method to measure the popularity of job skills using various criteria, facilitating the recommendation of suitable job skills based on given job descriptions~\cite{xu2018measuring}. They evaluated performance using log-likelihood as a measure.
\item \textbf{Classification Metrics:} \textsl{Liu et al.}, in their endeavor to identify correlations between job ad requirements and skill sets, utilized classification-based metrics including Micro-F1, Area Under the ROC Curve (AUC), and Hamming Loss~\cite{liu_2021_learning}. Additionally, \textsl{Leon et al.} evaluated the classification accuracy (ACC) of their proposed methodology through a comprehensive analysis~\cite{leon_2024_hierarchical}.
\item \textbf{Regression Metrics:} \textsl{Wu et al.} focused on predicting the demand for skills, employing commonly accepted regression metrics such as Mean Absolute Error (MAE) and Mean Absolute Percentage Error (MAPE)~\cite{wu2019trend}. MAE calculates the mean value of squared errors for all samples, while MAPE emphasizes the absolute ratio between the prediction error and the ground truth value.
\end{itemize}

\noindent \textbf{$\diamondsuit$ Takeaway.} 
\begin{itemize}[left=0pt]
\item \textbf{Advantages of AI technology}: 
\textsl{
(1) Traditionally, analyzing the skill demand from job postings relied on time-consuming expert recognition. With the significant advancements in AI technology, particularly in the NLP and {recommendation systems}, accurate extraction of skill information and the comprehensive construction of skill demand understanding have been achieved.
(2) Traditional methods struggle to capture the skill demand trend based on the observed statistics of skill demand. AI technology has significantly contributed to trend capturing, achieving notable performance improvements.}
\item \textbf{Limitations and future directions}:
\textsl{
Current efforts have primarily focused on the multi-label classification to map job descriptions to a set of skills or skill recommendations. However, they often lacked the capability to predict the demand for innovative skills brought by emerging technologies. Looking ahead, it is feasible to leverage the common skill development patterns along with transfer learning technology to predict the demand for emerging skills. This approach holds promise for enhancing the adaptability of prediction models to accommodate the rapidly evolving landscape of technological advancements.}
\end{itemize}

\subsubsection{Skill Demand Forecasting.}
Skill demand forecasting aims to anticipate the fluctuating demand for skills, thereby providing valuable insights for both employees and employers to stay ahead in the continuously evolving labor market landscape. Conventional approaches to this analysis predominantly rely on labor-intensive interview-based methods, which are susceptible to human biases. However, the emergence of online recruitment platforms has paved the way for a data-driven approach, leveraging AI technology to address previous challenges and demonstrate superior performance. Following~\cite{xi_pretrain_2024}, the task of skill demand forecasting can be broadly conceptualized as follows:
\begin{definition}[Skill Demand Forecasting]
Given a set of job postings $\mathcal{P}\{(J_i,S_i,(D_i,Y_i,T_i)^*)\rvert i=1,2,...\}$, where * indicates optional information, the skill demand can be formulated as $\mathcal{R}^t_{i} = \sum_{p\in \mathcal{P}^t} \mathbf{1}(s\in p)/{|\mathcal{P}^t|}$, where $|\mathcal{P}^t|$ is the number of job postings at timestamp $t$.
Given the historic skill demand sequences $\{\mathcal{R}^t| t \in [1, T]\}$, the goal of skill demand forecasting is to predict the future skill demand $\mathcal{R}^{T+1}$.
\end{definition}
To address the issue of skill demand prediction, 
\textsl{Mahdavimoghaddam} undertook the prediction of future in-demand skills derived from social content~\cite{mahdavimoghaddam_2021_congruence, mahdavimoghaddam_2022_exploring}. Initially, they employed Granger causality analysis to discern the relationship between online social content and in-demand skills\cite{mahdavimoghaddam_2021_congruence}. Furthermore, they utilized various classification algorithms, including Logistic Regression, Random Forest, and Linear Support Vector Classifier~\cite{mahdavimoghaddam_2022_exploring}, trained on Bert embeddings to forecast topic-based time series.
\textsl{Macedo et al.} employed various RNN methodologies, including LSTM, CNN combined with LSTM, and GRU on skill-share datasets~\cite{de_macedo_practical_2022}. They forecasted skill demand for the subsequent 6, 12, 24, and 36 months. \textsl{Wolf et al.} generated a synthetic skill demand time-series dataset using a Time-Generative-Adversarial-Network (TimeGAN) and trained an LSTM model on it, surpassing the performance of models trained only on real data~\cite{wolf_2023_generating}. \textsl{Chao et al.} concentrated on the interplay between skill demand and supply, introducing a Cross-view Hierarchical Graph Learning Hypernetwork framework for joint skill demand-supply prediction~\cite{chao_cross-view_2024}. This framework comprises a cross-view graph encoder to capture asymmetrical relationships, a hierarchical graph encoder to model high-level skill co-evolution trends, and a hyper-decoder for skill trend output based on historical demand-supply data. \textsl{Chen et al.} delved into fine-grained skill demand forecasting at the occupation level~\cite{xi_pretrain_2024}. {They pre-trained a Graph Autoencoder with initial job descriptions, using contrastive loss for sparse data and combined Tweedie and ranking losses for imbalanced demand. An efficient two-step optimization was also proposed to fine-tune the model for accurate skill demand prediction.}

\noindent 
\textbf{Evaluation.} 
Skill demand forecasting evaluation typically employs three primary categories of metrics: classification metrics, ranking metrics, and regression metrics.
\begin{itemize}[left=0pt]
    \item \textbf{Classification Metrics}: \textsl{Chao et al.} cast the demand and supply forecasting task as a classification problem, to anticipate future skill demand trend~\cite{chao_cross-view_2024}. They evaluated their model using widely accepted classification metrics, such as ACC, F1, and AUC.
    \item \textbf{Ranking Metrics}: \textsl{Chen et al.} predicted the skill demand at the occupational level and framed it as a dynamic graph predicting task~\cite{xi_pretrain_2024}. Furthermore, they decomposed this task into a dynamic link prediction task and a dynamic edge regression task. Evaluation of the model's effectiveness in dynamic link prediction employed ranking metrics, such as NDCG and MRR.
    \item \textbf{Regression Metrics}: \textsl{Macedo}~\cite{de_macedo_practical_2022}, \textsl{Wolf}~\cite{wolf_2023_generating}, and \textsl{Chen et al.}~\cite{xi_pretrain_2024} quantify the skill demand and predict the value of each skill demand. Thus, they utilized common regression metrics, RMSE, and MAE to assess the accuracy of demand value forecasting.
\end{itemize}

\noindent \textbf{$\diamondsuit$ Takeaway.} 
\begin{itemize}[left=0pt]
\item \textbf{Advantages of AI technology}: 
\textsl{
(1) Thanks to the advancements in AI technology, particularly in time-series learning, significant strides have been made in time-series forecasting tasks. By leveraging time-series learning models like MLP and RNN, it is now possible to accurately capture the skill demand changes and forecast the future skill demand over extended periods, spanning months or even longer.
(2) Traditional statistical methods often struggle to capture the intricate interrelationships among various skills. In contrast, deep learning methods excel at integrating diverse skill features, enabling the establishment of comprehensive frameworks to comprehend demand series across multiple skills under varying circumstances.}
\item \textbf{Limitations and future directions}:
\textsl{
(1) The current research largely overlooks the influence of geographical factors on skill demand. Yet, distinct cities possess unique industrial structures, and the evolution of their industries varies significantly. Consequently, skill demand exhibits considerable variation across cities. It is imperative to devote greater attention to forecasting skill demand in different cities, recognizing the importance of local industrial dynamics in shaping skill requirements.
(2) The current skill demand forecasting systems suffer from a deficiency in constructing labeled data, and there is a notable absence of discourse regarding skill word extraction methods in articles investigating skill demand prediction. Consequently, the extracted data is prone to biases. Future endeavors should aim to more systematically integrate the tasks of skill word extraction and skill demand prediction. This integration can mitigate data biases and reduce prediction inaccuracies stemming from biased environmental data.
(3) Current skill demand prediction methods rely heavily on job postings, which limits their scope by focusing only on immediate employer needs. This approach overlooks emerging skills, trends in informal and freelance markets, and internal workforce reskilling. As a result, it provides an incomplete picture of the evolving labor market.}
\end{itemize}

\subsubsection{Skill Value Estimation}

In recent times, the evaluation of skills has garnered significant attention from researchers and companies alike. This task holds importance not only for companies seeking to identify and retain top talent but also for individuals aiming to proactively acquire essential skills for their desired career path.
Formally, the task of estimating the value of skills can be defined as follows:

\begin{definition}[Skills Value Estimation]
Given a set of job postings $\mathcal{P}\{(J_i,S_i,(D_i,T_i)^*)\rvert i=1,2,...\}$, where * indicates optional information. The goal of skills value estimation is to measure the value of each skill $S_i$.
\end{definition}

\textsl{Rahman} devised a method to estimate individual worker skills derived from the outcomes of team-based tasks~\cite{rahman_worker_2015}. They formulated skill aggregation functions to effectively estimate the skills of workers involved in such endeavors, solving these functions using an efficient heuristic solution based on hill climbing.
\textsl{Sun et al.} proposed an enhanced neural network with a cooperative structure, Salary-Skill Composition Network (SSCN), for separating the job skills and measuring their value from the massive job postings~\cite{sun2021market}. Figure~\ref{SSCN} shows the overview of the workflow. In detail, this method mainly contains two modules, one is a Context-aware Skill Valuation Network (CSVN) for dynamically modeling the skills, extracting the context-skill interaction, and estimating the context-aware skill value. Another is the Attentive Skill Domination Network (ASDN), which can extract an influence representation for each skill to model their influence on domination to each other from the skill graph. The value of job skills can help companies in formulating talent strategies.
\textsl{Stephany et al.} argued for the importance of complementarity in skill estimation~\cite{stephany_what_2024}. They assigned interpretable market values to individual skills, measuring their worth through a linear regression approach. Additionally, they extended the theory of skill complementarity, constructing a skill network based on the characteristics of each skill's closest neighbors. Finally, they computed the value of complements as the average premium of the three most adjacent skills.

\textbf{Evaluation.} 
Current works on skill valuation aim to quantify skill value either as a deterministic value or as a probability distribution. Therefore, \textsl{Rahman} compared and contrasted different algorithms using Average Absolute Error and Normalized Relative Error~\cite{rahman_worker_2015}. Specifically, Normalized Relative Error is computed as $\frac{\sqrt{\sum_t e_t \times e_t}}{\sqrt{\sum_t q^t \times q^t}}$, where $t$ denotes a task, $q^t$ represents the skill value of task $t$, and $e_t$ stands for the predicted error.
In another study by \textsl{Sun et al.}, performance was evaluated using root mean square error (RMSE) and mean absolute error (MAE), both popular metrics for measuring differences between observations and predictions~\cite{sun2021market}.
\textsl{Stephany et al.} introduced a new metric called skill premium to explore how variance in values can be described by proposed features such as supply, demand, and complementarity~\cite{stephany_what_2024}.

\noindent 
\textbf{$\diamondsuit$ Takeaway.} 
\begin{itemize}[left=0pt]
\item \textbf{Advantages of AI technology}: 
\textsl{
(1) AI offers efficient processing of extensive job advertisement data sourced from online recruitment platforms, thereby providing researchers with a substantial dataset for analyzing job skill worth. Additionally, AI algorithms streamline data processing and analysis, significantly enhancing efficiency and scalability compared to manual methods.
(2) In contrast to conventional survey methods, AI-powered analysis delivers real-time insights into the evolving landscape of job skill value, accurately reflecting current market dynamics and trends. By harnessing AI, predictive models can be developed to anticipate future trends in job skill value, empowering individuals and organizations to proactively adapt to shifting market conditions.
}
\item \textbf{Limitations and future directions}:
\textsl{
(1) Various studies have approached skill valuation from different perspectives. \textsl{Rahman} quantified the skill value of individual workers within a teamwork context~\cite{rahman_worker_2015}, \textsl{Sun et al.} employed salary prediction as a collaborative task to address the salary-skill value composition issue~\cite{sun2021market}, and \textsl{Stephany et al.} aimed to explore the significance of complementarity in skill estimation~\cite{stephany_what_2024}. Consequently, a unified and standard skill valuation system and corresponding estimation method represent valuable directions for future research.
{(2) In the context of skill value estimation, there are multiple forms of data, including numerical (e.g., salary data), textual (e.g., recruitment needs), and geospatial (e.g., company location). Current approaches treat these data types and attributes independently, which limits the effectiveness of holistic labor market analysis. This lack of multi-modal fusion approaches restricts the ability to derive deeper insights from the combined data, which is critical for more accurate and nuanced skill value predictions. }
}
\end{itemize}

\subsection{Brand Analysis}

Brands are one of the most precious assets for a company, {which highlights the talent attractiveness in the labor market and related market}, and the corporate image attributes from employees and public opinion. It is crucial for corporate to manage brands as a talent strategic tool to keep up with the continuously changing business world. How to formulate a strategy for improving {a brand is getting} increasing attention in the area of talent management. Traditionally, the approaches for brand analysis mainly depend on surveys and interviews with expert knowledge. For example, \textsl{Ambler et al.} interviewed respondents from some companies about the relevance of branding to HRM~\cite{ambler1996employer}. \textsl{Arasanmi et al.} used an online survey method to collect the data and analyzed the relationships between employer branding, job designs, and employee performance by statistical methods~\cite{arasanmi2019employer}. \textsl{Fatma et al.} collected data through a social survey and analyzed the impact of Corporate Social Responsibility (CSR) on corporate brand equity~\cite{fatma2015building}. 


Recently, with the development of the Internet and online professional social networks, a large amount of company review data and various public data related to companies, e.g., online reviews, news, Twitter, and so on, can be collected. These data can provide new perspectives and opportunities for more comprehensive company brand analysis. {However, traditional methods struggle to analyze large volumes of unstructured text data. The rapidly developing} AI technology provides suitable methods for this kind of data. In particular, the topic model method, which can cluster the latent semantic structure of the corpus in an unsupervised learning manner, is good at semantic analysis and text mining {for} brand analysis~\cite{moniz2014sentiment}. 
\subsubsection{Company Profiling}
Company profiling is a kind of analytical task to understand the fundamental characteristics of companies. AI-driven approaches provide an opportunity to profile companies from abundant and various online employment data.

\noindent 
\textbf{Employer Branding.} Employer Branding is to understand an employer’s unique characteristics to identify competitive edges.
In~\cite{lin2017collaborative}, \textsl{Lin et al.} proposed CPCTR, which is a Bayesian model that combines topic modeling with matrix factorization to obtain the company profiles from online job and company reviews. In detail, CPCTR groups reviews by their job positions and companies denote two words lists as $\{w^P_{n,j,e}\}^N_{n=1}$ and $\{w^C_{m,j,e}\}^M_{m=1}$ to represent the positive opinion and negative opinion for a specific job position $j$ and company $e$, then formulates a joint optimization framework for learning the latent patterns of companies with different jobs $v_{j,e}$, which leads to a more comprehensive interpretation of company profiling and provides a collaborative view of opinion modeling. Subsequently, in~\cite{lin2020enhancing}, they provided a Gaussian process–based extension, GPCTR, which can capture the complex correlation among heterogeneous information and improve the profiling performance.
\textsl{Bajpai et al.} provided a hybrid algorithm, which works as an ensemble of unsupervised and machine learning approaches, for company profiling from online company reviews data~\cite{bajpai2019aspect}. First, this work uses CNN and Doc2Vec to extract the important opinion aspects from reviews. Then they combined universal dependent modifiers and sentiment dictionaries to assign polarity to each aspect of each company. If it fails to assign a score to the aspect, the ELM model can be used to predict the polarity. Finally, each company can be embedded in an $n$-dimensional representation space where $n$ is the number of aspects.
\textsl{Reis} delved into the correlation between employer branding and talent management, showcasing employer branding's potential to attract and retain top-tier employees within organizations~\cite{reis_employer_2021}. 
\textsl{Spears et al.} investigated the long-term impact of public opinion on company earnings, utilizing data extracted from news and social media alongside earnings reports~\cite{spears2021impact}. They employed a Markov switching model to quantify the relationship between adverse publicity and company finances, offering insights to enhance brand impact. 
\textsl{Yuan et al.} introduced the Self-Supervised Prototype Representation Learning (SePaL) framework for dynamic corporate profiling~\cite{yuan_self-supervised_2021}. This involved inferring initial cluster distributions of noise-resistant event prototypes and utilizing self-supervision signals for representation learning, leading to improved predictions of stock price spikes and evaluations of corporate default risk. 
\textsl{Bose et al.} compiled a vast dataset of corporate reviews and applied an ensemble approach to sentiment analysis, facilitating better insights for customers in selecting businesses~\cite{bose_sentiment_2022}. 
\textsl{Savin et al.} proposed an expert-bias-free classification of {startups} across 38 topics, quantifying relevance for each~\cite{savin_topic-based_2023}. Their analysis of industry and topic distributions provides valuable insights for entrepreneurs.

\noindent 
\textbf{Corporate Social Responsibility Communication.} 
Corporate social responsibility (CSR) stands as a cornerstone in both industry practices and academic discourse. Companies are compelled to devise robust CSR communication strategies to fortify their legitimacy and reputation. Numerous studies have been dedicated to unraveling this topic.
\textsl{Chae et al.} scrutinized CSR through Twitter posts, employing the Structural Topic Model algorithm to unveil correlations between diverse responsibility topics and their sequential trends~\cite{chae2018corporate}. Their findings underscore the significance of CSR and corporate reputation in bolstering brand equity.
\textsl{Mazza} delved into the evolution of CSR communication in the post-COVID-19 era using topic modeling techniques~\cite{mazza_text_2022}.
\textsl{Pilgrim} conducted a comprehensive review of social media mining methodologies, identifying four key approaches—topic models, network analysis, sentiment analysis, and regression analysis—to elucidate relevant CSR topics~\cite{pilgrim_csr_2023}.
\textsl{Thakur} underscored the pivotal role of CSR discussions on social media platforms, offering practical recommendations for firms and CEOs based on critical insights derived from these discussions~\cite{thakur_impact_2023}.

\noindent 
\textbf{Evaluation.}
Generally, company profiling encompasses two distinct evaluation methods tailored to different analysis techniques. One approach involves quantitative analysis, utilizing regression metrics as evaluation criteria. The other revolves around qualitative type analysis, employing classification metrics for measurement.
\begin{itemize}[left=0pt]
    \item \textbf{Regression Metrics}: \textsl{Lin et al.} utilized two commonly employed metrics, namely RMSE and MAE, as measures for Topic Regression~\cite{lin2017collaborative, lin2020enhancing}.
    \item \textbf{Classification Metrics}: \textsl{Bajpai} employed Accuracy and Macro F1-score to assess the performance of aspect-level sentiment analysis~\cite{bajpai2019aspect}. \textsl{Yuan} utilized Precision, Recall, and F1-score, widely adopted metrics, for both stock price spike prediction and corporate default risk~\cite{yuan_self-supervised_2021}. 
\end{itemize}
\noindent \textbf{$\diamondsuit$ Takeaway.} 
\begin{itemize}[left=0pt]
\item \textbf{Advantages of AI technology}: 
\textsl{
(1) Data Accessibility: AI can scrape and analyze vast amounts of data from online sources, including employee reviews and ratings on platforms like Glassdoor, Indeed, or LinkedIn. This provides access to information about companies that may not be readily available through traditional financial reports.
(2) Real-time Insights: AI algorithms can process and analyze data in real-time, providing up-to-date insights into a company's reputation, culture, and employee satisfaction. This real-time aspect is particularly valuable in today's fast-paced business environment, where conditions and perceptions can change rapidly.
}
\item \textbf{Limitations and future directions}:
\textsl{
(1) Current research efforts predominantly focus on utilizing textual and numerical data, such as reviews and salary information, to assess employer branding. However, they often overlook the dynamic evolution of companies, leading to an inability to adapt to the fast-paced labor market changes. Looking forward, integrating considerations of the dynamic development of the labor market will facilitate the construction of a timely framework for company profiling. This approach will enable a more nuanced understanding of companies' positioning and performance in response to the evolving demands of the labor market. (2) Existing methodologies for company profiling that leverage social media information often overlook the reliability of these data sources. Given the prevalence of potential misinformation within these platforms, there is a significant risk to the accuracy and trustworthiness of the derived insights. The field of web information mining, which has long addressed challenges related to fake news, offers promising techniques that could enhance AI-based company profiling systems. Incorporating these advanced detection methods to filter and verify social media content represents a critical future direction for improving the robustness and reliability of company profiling.
}
\end{itemize}

\subsubsection{Employee Sentiment Analysis}
The employee sentiment analysis task is more focused on the analysis of the company's public reviews, especially the reviews of employees. 
Employee sentiment analysis holds significant importance in the operations of firms and organizations worldwide due to its impact on employee turnover and customer satisfaction. 
These indicate the feedback of the company's talent strategy, which is important for further iteration of the right talent strategy.
As a widely used text analysis model, the topic model is an appropriate method for the employee sentiment analysis task. 

For example, \textsl{Moniz et al.} proposed an aspect-sentiment model based on the LDA approach for analyzing company reviews~\cite{moniz2014sentiment}. This kind of LDA approach can identify salient aspects in company reviews and manually infer one latent topic that appears to have a relationship with the firm's vision. Then, they combined the satisfaction topic information of company reviews with existing methods for earnings prediction. According to the results, employee satisfaction is important for firm earnings.
\textsl{Ikoro et al.} proposed a lexicon-based sentiment analysis method for analyzing the public opinion of corporate brands~\cite{ikoro2018analyzing}. In detail, they combined two sentiment lexicons and extracted two levels of sentiment terms, and collected over 60,000 tweets split over nine companies from Twitter. Then, the LDA methods are deployed to discover the sentiment topics.
\textsl{Chae et al.} analyzed the CSR based on posts data from Twitter, then, they applied the Structural Topic Model algorithm to discover the correlation between different responsibility topics and the sequential trend of topics~\cite{chae2018corporate}. The results also show that the CSR and the corporate reputation of a firm are important to its brand equity.
In addition, some authors explored other machine learning methods for brand analysis. For example,
\textsl{Spears et al.} investigated the impact of public opinion on companies’ earnings over time~\cite{spears2021impact}. {Public opinions were extracted from news and social media, while earnings data were collected from financial reports. The authors then employed a Markov switching model to quantify the relationship between negative publicity and company financial performance, providing insights that can help organizations strengthen their brand impact.} 
\textsl{Gaye et al.} employed TF-IDF, bag of words, and global vectors to extract three features and proposed a hybrid/voting model called Regression Vector-Stochastic Gradient Descent Classifier (RV-SGDC) for sentiment classification~\cite{gaye_sentiment_2021}.
\textsl{Shi} investigated factors influencing hotel employee satisfaction and explored different sentiments expressed in online reviews by hotel type (premium versus economy) and employment status (current versus former)~\cite{shi_listening_2021}. Structural topic modeling (STM) and sentiment analysis were utilized to extract topics influencing employee satisfaction and examine sentiment differences across each topic.
\textsl{Ganga} examined employee review literature and provided insights, noting emerging leading topics for satisfaction such as work environment and work-life balance, which are significant factors across various industries~\cite{ganga_employees_2022}.
\textsl{Rehan} implemented a purely supervised machine learning approach with two modules to classify employees as satisfied/unsatisfied and proper/improper, respectively~\cite{rehan_employees_2022}.
\textsl{Mouli} endeavored to address this gap by applying sentiment analysis techniques to a large dataset compiled from Glassdoor, primarily exploring worker sentiments using Bidirectional Gated Recurrent Unit, which demonstrated superior performance across various metrics~\cite{mouli_unveiling_2023}.

\noindent 
\textbf{Evaluation.}
In the realm of employee sentiment analysis, two primary evaluation methods were employed. One approach centered on the quantitative analysis concerning firm earnings derived from sentiment analysis, utilizing regression metrics for measurement. The other approach focused on evaluating employee satisfaction or attitudes, utilizing classification metrics for assessment.
\begin{itemize}[left=0pt]
    \item \textbf{Regression Metrics.} \textsl{Moniz et al.} employed quantitative analysis to explore the relationship between employee satisfaction and firm earnings, utilizing ordinary least squares regression and RMSE as the measurement~\cite{moniz2014sentiment}.
    \item \textbf{Classification Metrics.} \textsl{Gaye et al.}~\cite{gaye_sentiment_2021}, \textsl{Shi et al.}~\cite{shi_listening_2021}, \textsl{Rehan et al.}~\cite{rehan_employees_2022}, and \textsl{Mouli et al.}~\cite{mouli_unveiling_2023} assessed the performance of their models using metrics such as accuracy, precision, recall, and F1 score.
\end{itemize}

\noindent \textbf{$\diamondsuit$ Takeaway.} 
\begin{itemize}[left=0pt]
\item \textbf{Advantages of AI technology}: 
\textsl{
The advent of AI technology, particularly advancements in topic modeling, has revolutionized opinion extraction from vast amounts of review data. Unlike traditional methods that were susceptible to human bias and relied on time-consuming expert labeling, AI-based methods offer the potential to streamline the process and eliminate subjective influences.}
\item \textbf{Limitations and future directions}:
\textsl{
(1) The current landscape of employee sentiment analysis relies on traditional natural language processing techniques like {topic modeling}. However, recent advancements in LLMs have propelled sentiment analysis forward significantly~\cite{krugmann_sentiment_2024}. Looking ahead, integrating LLMs into employee sentiment analysis represents a more efficient approach for future endeavors.
(2) Sentiment and emotion analysis systems used in employee feedback analysis can inadvertently perpetuate and amplify existing biases related to race, gender, and other social factors. For instance, these systems may assign lower sentiment scores to text produced by individuals from certain demographic groups, such as specific races or genders, leading to skewed and unfair insights. This issue is particularly pronounced in NLP systems, where biases often stem from the training data, corpora, lexicons, or word embeddings employed in model development~\cite{kiritchenko2018examining}. Without addressing these biases, sentiment analysis in the workplace may result in inaccurate reflections of employee emotions and reinforce inequalities, ultimately affecting decision-making processes in areas like performance evaluations, promotions, or employee engagement strategies.} 

\end{itemize}

\subsection{Summary}
In general, AI-related labor market analysis primarily focuses on four key aspects: talent flow analysis, job analysis, skill analysis, and brand analysis. The research data predominantly consists of various data sources, including OPNs, social media platforms, job postings, and job and company reviews.
The talent flow analysis mainly includes the talent flow prediction task and other various flow pattern analysis tasks. {The job analysis mainly focuses on the analysis of trends, such as new job trends, demand trends, and topic trends.} The skill analysis mainly explored {potential skill prediction}, skill demand forecasting, and skill valuation.
Moreover, the brand analysis aims to model the brand and culture of the company, including the CSR communication, and analyze the employee sentiment of the company. 
Existing AI-driven labor market research has initially demonstrated its advantages. Instead of the traditional questionnaire survey method, AI-driven techniques can mitigate human bias and obtain more objective conclusions. In addition, AI techniques can make extensive use of large-scale data to mine potential patterns in the labor market, capture dynamic changes therein, and accurately predict future trends. Furthermore, with the rapid development of NLP technology, AI can deeply mine the intrinsic correlation of occupations, skills, and talents from data such as job postings, explore the correlation between them, and obtain more accurate analysis conclusions.
Despite these remarkable results, AI technology still has many limitations: fine value prediction cannot be achieved on some dynamic prediction tasks, and a large amount of work can only predict the extent of trend changes. In addition, a large amount of work requires label extraction and expert identification based on raw data. These processes still bring human bias and deviations in data distribution, further affecting the accuracy and fairness of downstream tasks.
{Moreover, current analyses heavily depend on traditional data sources like job postings and surveys, which limit the depth of insights. For instance, in skill demand forecasting, these sources primarily capture immediate employer needs, neglecting emerging skills and trends in informal or freelance markets, as well as internal workforce reskilling. This results in an incomplete understanding of labor market dynamics.
Another challenge is the integration of diverse data types—textual, numerical, and geospatial—into a cohesive analysis. Existing approaches treat these data sources independently, hindering the potential for more comprehensive insights, particularly in skill value estimation.
Furthermore, bias and fairness in labor market algorithms are insufficiently addressed. AI models trained on biased data can reinforce inequalities, affecting decisions related to performance evaluations and promotions. Sentiment analysis, in particular, may amplify racial, gender, or social biases, skewing employee feedback analysis and leading to unfair outcomes.}
Lastly, this related research about the labor market is still at an early stage; many advanced and potential AI methods, such as LLM, can be combined with labor market analysis-related tasks and improve the intelligence of talent management.

\section{Prospects}\label{prospects}
{Despite recent advances in AI-based talent analytics for human resource management, several critical challenges remain. In this section, we highlight key research directions to address these challenges and drive further progress.}

\vspace{-3mm}
\subsection{Multimodal Talent Analytics}
Information about a phenomenon or a process in talent analytics-related scenarios usually comes in different modalities. For instance, we can obtain communication and project collaboration networks in employee collaboration analysis. Indeed, mining the multimodal data in talent analytics can help us enhance the effectiveness of different applications. For example, \textsl{Hemamou et al.} collected the multimodal data in the job interview process, including text, audio, and video, and proposed a hierarchical attention model to achieve the best performance in predicting the {hireability} of the candidates~\cite{hemamou2019hirenet}. 
Moreover, the utilization of multimodal data to train AI models, rather than solely relying on traditional interviews, offers a more comprehensive understanding of the participants' personality test results~\cite{koutsoumpis2024beyond}.
Recently, multimodal learning has been used to achieve multimodal data representation, translation, alignment, fusion, and co-learning in various domains, such as commercial, social, biomedical~\cite{lahat2015multimodal,baltruvsaitis2018multimodal}. 
We can foresee that more multimodal learning methods will gradually be extensively used in talent analytics.

\subsection{Talent Knowledge Management} 
{Although AI-based approaches have made significant advances in talent acquisition and development, the management of talent knowledge through AI technologies remains underexplored, despite being a key driver of the economics of ideas~\cite{wiig1997knowledge}.} There is an urgent need for additional AI-based technologies that focus on talent knowledge creation, sharing, utilization, and management. Such efforts are crucial for maximizing the potential of human resources and enhancing organizational productivity. Indeed, we can leverage knowledge graph-related technologies~\cite{ji2021survey} to construct the talent's knowledge base and achieve efficient knowledge management. Moreover, we can transform the scenarios in talent development, such as knowledge learning and collaboration, into different recommendation scenarios and utilize recommendation algorithms to solve these problems. Recently, \textsl{Wang et al.} developed a personalized online courses recommendation system based on the employees' current profiling~\cite{wang2020personalized,wang2021personalized}. However, there is still a lack of an algorithm that can recommend heterogeneous knowledge. The existing algorithms are only from the individual perspective and have not been analyzed from the organizational aspect, such as the organizational knowledge diversity or competitiveness. 

\subsection{Market-oriented Talent Analytics}
AI technology has been effectively applied in labor market analytics~\cite{zhang2020large}. However, those approaches mainly focus on the perspective of global market analysis and have not explored how the changing environment of the labor market affects internal talent management or organization management. Combining macro and micro data in talent analytics is a vital research direction. For instance, \textsl{Hang et al.} leveraged the job posting data to capture the potential popularity of employees in external markets~\cite{hang2022outside}. Moreover, the rapid development of AI technologies provides an excellent technical foundation for this direction. {We can utilize multi-task learning to learn both macro and micro talent analytics-related tasks jointly. Besides, we can use AI technologies to identify high-potential employees in enterprises~\cite{ye2019identifying,cheng2021effectiveness}.} In the labor market, accurate recruitment of digital talents is crucial for the digital transformation of enterprises. \textsl{Harrigan et al.}~\cite{harrigan2021march} found a specific relationship between enterprise talent investment and digitization by identifying digital skills-related talents within the organizations. \textsl{Babina et al.}~\cite{babina2024artificial} classified different categories of AI skill-related talents based on job postings on recruitment platforms and analyzed the impact of these talents on the development of the enterprise after entering the enterprise. \textsl{Kim et al.}~\cite{kim2023ai} proposed a dynamic co-occurrence method, which dynamically calculates the AI relevance of various types of skills in the labor market, so as to identify AI talents more accurately. Moreover, AI has a broad prospect in the identification of green talents and Environmental, Social and Governance (ESG)-related talents. AI and human resource management still need large-scale research in the future.

\subsection{Organizational Culture Management}
The culture of an organization plays a crucial role in sustaining its effectiveness and viability. Generally, the culture mainly contains three aspects: Mission, Vision, and Values (MVVs), which can help employees understand what is encouraged, discouraged, accepted, or rejected within an organization, and facilitate the organization to thrive with the shared purpose. 
Recent developments reveal that the availability of extensive datasets covering the entire lifecycle of talents and organizations offers opportunities to realize effective culture management. For example, the interconnection between culture and leadership is evident, with exceptional team leaders significantly shaping organizational culture~\cite{corporateculture}. Some researchers~\cite{lee2020determining} discussed how machine learning techniques can be used to inform predictive and causal models of leadership effects. Meanwhile, several studies analyze leadership styles with data mining algorithms, demonstrating that the different leadership styles significantly influence leadership outcomes~\cite{ahmad2018investigation}. {Moreover, some researchers try to} utilize text mining to analyze the organizational culture. For instance, \textsl{Schmiedel et al.} leveraged the online company reviews data and topic model to explore the employees’ perception of corporate culture~\cite{schmiedel2019topic}. 
As a famous saying goes, ``Culture eats strategy for breakfast'', employing AI technologies in organizational culture management will become one of the most critical research directions in the future, as it can help managers scientifically address cultural management.

\subsection{Ethical AI in Talent Analytics}
Admittedly, AI technologies are increasingly employed in talent analytics, significantly enhancing management efficiency and accuracy.
However, there are ongoing concerns about how to ensure that AI technologies adhere to well-defined ethical guidelines regarding fundamental values. Recently, some researchers have made efforts from two perspectives, i.e., fairness and explainability.

Although AI technologies have achieved various successes in talent analytics, there is growing concern that such approaches may bring issues of unfairness to people and organizations, as evidenced by some recent reports~\cite{amazonfairness}. For instance, Amazon scrapped its AI-based recruitment system 
due to its discriminatory outcomes against women~\cite{amazonfairness}.
Recent studies have delved into the fairness of AI technologies in talent management from different perspectives. For instance, \textsl{Qin et al.}~\cite{qin2020enhanced} verified that when incorporating sensitive features, such as gender, age, etc., into the person-job fit model, the model will easily learn the bias from the original data. Intuitively, we can solve this problem by removing sensitive features. However, a large amount of unstructured data already contains sensitive features, such as audio and video data in the interview process. The model can easily infer the potentially sensitive attributes of data, which may still cause the bias of AI algorithms. 
In their study, {Pena~et~al.} examined multimodal systems to predict recruitable candidates using both image and structured data from resumes~\cite{pena2020bias}.
Similarly, \textsl{Yan et al.} both leveraged data balancing and adversarial learning to mitigate bias in the multimodal personality assessment ~\cite{yan2020mitigating}. 

Recently, there has been an increasing concern among employees and managers regarding the black-box AI algorithms. 
Therefore, the research interest in increasing the transparency of AI-based automated decision-making in talent management is re-emerging. For instance, \textsl{Qin et al.} proposed to leverage the attention mechanisms to explain the matching degree between the content of job postings and resumes~\cite {qin2018enhancing}. \textsl{Zhang et al.} further introduced the hierarchical attention and collaborative attention mechanisms to increase the person-job fit model explainability both at the structured and unstructured information level~\cite{zhang2021explainable}. 
However, the current approaches only stay in the perspective of AI model design and fail to consider whether employees or managers can easily comprehend and grasp the explanatory conclusions provided by the model. Indeed, visual analytics is an inherent way to help people who are inexperienced in AI understand the data and model. Therefore, combining visual analysis and explainable AI and building an intelligent talent management system is a valuable research direction. 

\subsection{Generative AI in Talent Analytics}

With technological advances, AI has gradually developed a branch called generative AI, and the language in which AI interacts with humans has changed to ``natural language"~\cite{wu2023survey}. 
The advent of generative large language models {has} sparked widespread discussion and excitement in both academia and industry.
First of all, LLMs have a strong generation ability. {\textsl{Zinjad et al.} proposed a resume generation tool, which allows users to generate tailored personalized resumes by providing simple personal information and job information~\cite{zinjad2024resumeflow}.} Second, LLMs facilitate the innovation of traditional AI technologies. For example, in the context of recommender systems, LLMs utilize their high-quality representation of textual features and their extensive coverage of external knowledge to achieve high-quality recommendations~\cite{wu2023survey}.

Besides, in recent years, LLM-based agents are utilized to solve various tasks such as software development, social simulation, and policy simulation~\cite{guo2024large}. Research on intelligent simulation of management-related problems has also been emerging, such as the study of member autonomy in organizations using human-machine systems~\cite{ellinger2023skin}. Further, by integrating multiple LLM agents, model efficiency can be further improved and simulation effects can be optimized~\cite{liu2023dynamic}. {However, there is not much research on using agents based on LLMs for simulating organizational behavior like cooperative behaviors among employees.} We should use management theory to simulate the framework of an organization and the behavior of individuals within the organization, i.e., the willingness to put one's self-interest at risk based on the positive expectations of others~\cite{xie2024can}.

{Finally, to further enhance decision-making efficiency, many organizational managers have introduced AI technologies for assistance. In particular, the recent rise of LLMs offers new opportunities. These models demonstrate proactive interaction capabilities and can support not only individual decision-making and task execution but also organizational-level planning and scheduling~\cite{bouschery2023augmenting}.} For instance, \textsl{Zheng et al.} developed an LLM-based generative job recommendation system to provide individuals with a more personalized and comprehensive job seeking experience~\cite{zheng2023generative}. However, due to the algorithmic black box characteristics of generative AI, it is difficult for {humans} to intervene and predict the content generated by generative AI. Therefore, in talent management, how to dialectically view AI and how to effectively use AI to manage and avoid risks is a key topic worthy of future research.

\subsection{The Impact of Emergent Events on Talent Analytics}
{Emergent events such as pandemics have had a huge impact on businesses and organizations, in talent management, because of the pandemic, the daily tasks of the employees in many companies will change.} To make the company adapt to this fluctuation and to mitigate {the impact of these emergent events}, we can predict the performance of the employees by using AI methods~\cite{hasan2024employee}. From the perspective of an organization's leader, it is essential to understand the impact of the epidemic on customers so that an effective response can be taken. Therefore, completing sentiment analysis of customers {using} text is a very effective approach~\cite{chang2022predicting,shah2021mining}. 
{Besides organizational resilience}, the forms of talent analytics are facing a transformation due to {some emergent events}. In the recruitment process, since global cooperation and telecommuting have become frequent, we can use AI technologies such as natural language processing technology to screen job applicants in global real-time big data~\cite{strang2022erp}, to reduce the risk of disease transmission in densely populated scenarios, the use of chatbots to communicate with onboarding employees during the onboarding process can avoid the contact of the employees and simplify the process of onboarding employees~\cite{westberg2019applying}. Once employees are onboarded, they can also be trained with the help of a robot~\cite{vrontis2022artificial}.

\section{Conclusions}
{AI-driven talent analytics is emerging as a powerful frontier in today’s competitive and fast-evolving business environment. 
This survey provides a comprehensive overview of recent advancements in AI-driven talent analytics. We first established a detailed data taxonomy as a foundation for understanding talents, organizations, and management. We then reviewed research efforts across three key areas: talent management, organization management, and labor market analysis. Finally, we outlined open challenges and future directions. We hope this survey offers readers a clear understanding of this emerging field and inspires further exploration at the intersection of AI and talent analytics.}


 \clearpage
\appendices

\renewcommand{\thetable}{S\arabic{table}}
\renewcommand{\thefigure}{S\arabic{figure}}
\setcounter{table}{0}
\setcounter{figure}{0}

\section{Systematic Resources for Talent Analytics}
To help the readers learn more effectively, we highlight the systematic resources provided in this survey as follows,

\begin{itemize}
    \item Table~\ref{tab:dataintroduction} and \ref{tab:data} summarizes the data for talent analytics.
    \item Table~\ref{tab:talentworks} summarizes recent AI-based talent analytics efforts in the talent management scenario.
    \item Table~\ref{tab:orgworks} summarizes recent AI-based talent analytics efforts in the organization management scenario.
    \item Table~\ref{tab:cultureworks} summarizes recent AI-based talent analytics efforts in the labor market analysis scenario.
\end{itemize}

\section{Data Processing for Talent Analytics}

In Section 2, we have introduced the major categories of data involved in talent analytics. In the Appendix, we provide additional details on data collection, data preprocessing, data cleaning and debiasing techniques, as well as the limitations of the current datasets and processing methods.

\subsection{Data Collection}
The source of recruitment data can be categorized into two types: internal data and external data. Correspondingly, data collection methodologies align with these source categories.

\noindent {\textbf{Internal Data.}}
The Current business environment is typically dependent on data systems~\cite{kropsuvehkapera_product_2009}. Internal data are collected from the internal enterprise management systems, which are known as enterprise resource planning (ERP) system~\cite{shehab_enterprise_2004}, customer relationship management (CRM) system~\cite{frow_customer_2009}, and applicant tracking system (ATS) system~\cite{mukherjee2014role}.
ERP systems are comprehensive business management tools that integrate various functions such as finance, sales, materials management, HR, production planning, and supply chain. CRM systems facilitate customer interaction and communication, encompassing customer information management, sales opportunities, and customer service.
An ATS is computer software that human resource departments use to process the overwhelming number of applications they receive for job openings.
These systems orderly store recruitment, employee, and organizational data, facilitating efficient collection and processing.

\noindent {\textbf{External Data.}}
As online services rapidly evolve, a growing number of individuals are turning to social media and job search websites to exchange job-related information and explore employment opportunities. These interactive platforms host a vast array of talent information due to their extensive user base. Additionally, third-party business investigation platforms provide detailed insights into companies and the relationships of their board members.
Data acquisition methods on these platforms vary. Third-party data collection websites typically aggregate data from their participating members. Meanwhile, web crawlers~\cite{kong2012case} can extract rich information from website pages, provided legal and regulatory compliance is ensured. Moreover, many job search websites retain substantial amounts of user data that isn't publicly disclosed. Typically, this data can be utilized for scientific research purposes following encryption and other privacy safeguards.

\subsection{Data Preprocessing}
After collecting a large amount of recruitment data, it is essential to preprocess the data for downstream applications, especially removing noisy, redundant, irrelevant, and potentially toxic data. In this part, we review the detailed data preprocessing strategies to improve the quality of the collected data according to various data types.

\begin{table*}[t]
\centering
\caption{The table of collected papers related to talent analytics-related data.}
\label{tab:dataintroduction}
\begin{tabular}{ccc}
\toprule
\textbf{Categories}      & \textbf{Data}        & \textbf{Reference} \\
\midrule
\midrule
\multirow{3}{*}{Internal: Recruitment} 
& Resume            & ~\cite{yao2021interactive, pena2020bias, zhang2018resumevis, meng2019hierarchical, zhang2021attentive, wang2021variable}  \\
& Job Posting       & ~\cite{qin2018enhancing, bian2019domain, zhu2018person, zhang2021talent, xu2018measuring, wu2019trend, sun2021market}  \\
& Interview-related & ~\cite{shen2018joint, shen2021joint, chen2016automated, chen2017automated, hemamou2019hirenet, chen2020hierarchical} \\
\midrule
\multirow{2}*{Internal: Employee} 
& Employee Profiles & ~\cite{ye2019identifying, li2017prospecting, sun2019impact, teng2019exploiting, teng2021exploiting, hang2022outside, zhang2021attentive} \\
& Training Records & ~\cite{wang2020personalized, wang2021personalized, srivastava2018s, li2020data,wu2005simplifying,liu2009application,hong2019evaluation} \\
\midrule
\multirow{2}*{Internal: Organization}
& Reporting Lines & ~\cite{sun2019impact, sun2021modeling} \\
& In-firm Social Network & ~\cite{klunder2020identifying, cao2021my, ye2019identifying, teng2021exploiting} \\
\midrule
\multirow{2}*{External}
& Social Media & ~\cite{ikoro2018analyzing, spears2021impact} \\
& Job Search Websites & ~\cite{lin2017collaborative, lin2020enhancing, bajpai2019aspect, schmiedel2019topic, park2019global}{\cite{xi2024job}} \\
\bottomrule
\end{tabular}
\end{table*}

\setcellgapes{3pt}
\begin{table*}[t]
\raggedright\makegapedcells
\centering
\caption{The table of collected talent analytics-related open datasets.}
\label{tab:data}
\begin{tabular}{llll}
\toprule
\textbf{Categories}&\textbf{Dataset}&\textbf{Link}&\textbf{Note}\\
\midrule
\multirow{12}{*}{Internal:Recruitment}&Kaggle-Entity\_Recognition\_Resumes&\makecell[l]{https://www.kaggle.com/datasets/\\ dataturks/resume-entities-for-ner}&\makecell[l]{220 resumes; 10 categories;\\ Resume Understanding}\\
&LinkedIn-Job-Scraper&\makecell[l]{https://www.kaggle.com/datasets\\ /arshkon/linkedin-job-postings/data}&\makecell[l]{33,000+ job postings;\\ 27 valuable attributes including\\ the title, job description,\\ salary, location, etc}\\
&Job Dataset&\makecell[l]{https://www.kaggle.com/datasets/\\ ravindrasinghrana/job-description-dataset}&\makecell[l]{synthetic job postings;\\23 attributes including job title,\\ job salary, job skill, etc}\\
&Linkedin Jobs \& Skills (2024)&\makecell[l]{https://www.kaggle.com/datasets/asaniczka/\\1-3m-linkedin-jobs-and-skills-2024}&\makecell[l]{12,96,381 job postings;\\ skills mapping,\\ job recommendation systems}\\
\hline
\multirow{4}{*}{Internal:Employee}&HR Analytics&\makecell[l]{https://www.kaggle.com/datasets\\/colara/hr-analytics}&\makecell[l]{14999 employees;\\10 attributes,including\\ satisfaction\_level, salary, etc;\\turnover prediction}\\
&Human Resources Data Set&\makecell[l]{https://www.kaggle.com/datasets/\\ rhuebner/human-resources-data-set/data}&\makecell[l]{311 records; 36 attributes,\\including name, salary, etc}\\
\bottomrule
\end{tabular}
\end{table*}

\noindent \textbf{Structured Data.}
Structured data, in simple terms, is a database such as ERP system~\cite{elragal2014erp}, that has a standardized format for efficient access by software and humans alike.  
Besides, the information of employees and companies is orderly collected into the database. On some data collection websites, a lot of user information is also strictly stored through databases, such as records of interactions between users and the platform, including clicking and browsing.
Since these data have been stored in a standardized format, filtering and integration of corresponding data can generally be achieved by connecting different tables and setting data filtering conditions on the tables~\cite{bakaev2015intelligent}.

\noindent \textbf{Semi-Structured Data.}
Semi-structured data, or partially structured data, diverges from the conventional tabular format characteristic of relational databases or other tabular data forms~\cite{theobald2008topx}. Instead, it incorporates tags and metadata to delineate semantic elements and establish hierarchical relationships among records and fields. This type of data is prevalent in the labor market~\cite{signore_2023d_igital}, encompassing employee resumes~\cite{zhang_2018_resumevis}, individual interviews~\cite{bovstjanvcivc_2018_role}, job postings~\cite{cohen2001structured}, and web pages~\cite{de2018human}. Due to the absence of standardized formats and the diversity of semi-structured data types, it necessitates thorough exploration of commonalities, extraction of multidimensional information, and comprehensive filtering to transform it into structured data.
For example, \textsl{Sun et al.} collected job postings from an online recruitment website~\cite{sun2021market}.
On this website, each job opening is displayed in HTML, which contains information of salary range, company, location, time, and job description text. They parsed the HTML and obtained structured job posting information.

\noindent \textbf{Unstructured Data.}
Unstructured data~\cite{vijayarani2015preprocessing} refers to information that does not conform to the conventional row-column structure found in traditional databases. Within the labor market context, data primarily manifests as text, comprising a blend of structured and unstructured fields. Structured fields denote specific categories like job titles (e.g., occupation), location, etc., while unstructured fields provide a broader description of vacancy content.
Approximately 80\% of data held by firms today is unstructured~\cite{balducci2018unstructured}, expanding at a rate fifteen times faster than structured data. While some approaches skip the processing of natural text, opting instead for direct classification tasks on such text, as observed in the classification of web job vacancies~\cite{boselli2018wolmis}, others require extraction and subsequent processing of information from unstructured data into structured data.
For instance, extracting skill information from job descriptions~\cite{xi_pretrain_2024} involves the identification of skill words through regular expression matching. These identified words are then subjected to expert evaluation for further refinement, resulting in a set of relevant skill words for each job description. Subsequent statistical analysis across all jobs yields a comprehensive frequency distribution of skill words as structured data, facilitating downstream tasks such as skill demand forecasting.

\subsection{Data Cleaning and Debias}
After initial data preprocessing, standard structured data is obtained. However, given the potential for noise introduced during the data acquisition process, coupled with non-uniform acquisition methods, the provided data may not be of high quality. Therefore, a cleanup and debiasing of the data becomes imperative. In this part, we first introduce several data quality issues commonly seen in AI-driven talent analytics. Subsequently, several clean and debias methods are introduced to solve these issues.

\noindent \textbf{Data Quality Issues.} There are various data quality issues: missing data, duplicated data, extraneous data, and inconsistent data~\cite{corrales2018address}. These issues will introduce biases into analyses and lead to inaccurate conclusions. We introduce these issues as follows:
\begin{itemize}
\item \textbf{Missing Data:}
Missing data occurs when essential data is absent from a dataset, which can result from factors like data corruption, or failure to capture specific information. Within the talent market, missing values often stem from inconsistent information sources. For instance, job postings that should contain recruitment requirements and descriptions may be missing key fields~\cite{huang2009study}. Resumes also frequently omit crucial information such as email addresses and physical addresses~\cite{shao2023exploring}.
\item \textbf{Duplicated Data:}
Duplicated data refers to the presence of identical or nearly identical records within a dataset, which can arise from data entry errors, erroneous dataset merging, or technical malfunctions. Duplicated data can distort statistical analyses and exaggerate the significance of certain data points. In the talent market, this issue can arise due to data sources providing overly homogeneous information. Repetitive job postings may be erroneously interpreted as multiple postings for the same position in certain analysis and prediction scenarios~\cite{sun2021market}.
\item \textbf{Extraneous Data:}
Extraneous data comprises irrelevant or unnecessary information within a dataset, often included mistakenly due to human errors or incorrect data integration processes. Such data can complicate analyses, waste computational resources, and yield inaccurate results. This data often requires further filtering to retain only relevant fields and eliminate irrelevant information, as noted in~\cite{balaji2019airesume, berkesewicz2024text}. Redundancies in job offers can hinder the accuracy of downstream classification tasks~\cite{berkesewicz2024text}.
\item \textbf{Inconsistent Data:}
Inconsistent data encompasses conflicting or contradictory information within a dataset, stemming from sources like data entry errors, incompatible formats, or changes in data collection methods over time~\cite{huang2009study, balaji2019airesume, gaikwadeffective}. Such inconsistencies impede meaningful insights and necessitate thorough validation and cleansing to ensure data integrity and accuracy. Common inconsistencies include misspellings and variations in job titles, which require resolution to maintain consistency across professional documents~\cite{balaji2019airesume, gaikwadeffective}. Besides, erroneous labeling of job postings as job titles is also a common problem~\cite{gaikwadeffective}.
\end{itemize}

\begin{table*}[]
\centering
\caption{The table of collected papers related to talent management.  }
\label{tab:talentworks}
\resizebox{\linewidth}{!}{
\begin{tabular}{cccc}\toprule
 \textbf{Task} & \textbf{Method}  & \textbf{Data} & \textbf{Reference}  \\
\midrule
\midrule
\multicolumn{4}{c}{\textbf{Talent Recruitment}}\\\midrule
 Job Posting Generation & RNN & Job posting & \cite{liu2020hiring,qin2022towards}  \\
  {Job Posting Generation} & {LLMs} & {Job posting} & {\cite{lorincz2022transfer,borchers2022looking}}  \\
  {Resume Understanding} & {Rule-base method} & {Resume} & {\cite{kopparapu2010automatic}}  \\
  {Resume Understanding} & {HMM} & {Resume} & {\cite{yu2005resume}}  \\
  {Resume Understanding} & {SVM} & {Resume} & {\cite{yu2005resume,chuang2009resume}}  \\
  {Resume Understanding} & {CRF} & {Resume} & {\cite{pawar2012automatic,chen2016information}}  \\
{Resume Understanding} & {LSTM,CNN} & {Resume} & {\cite{ayishathahira2018combination}}  \\
{Resume Understanding} & {LSTM,CRF} & {Resume} & {\cite{pham2018study,pham2018study,li2021method,luo2022novel}}  \\
{Resume Understanding} & {RoBERTa,GCN} & {Resume} & {\cite{wei2020robust}}  \\
{Resume Understanding} & {Multimodal pre-trained model} & {Resume} & {\cite{yao2023resume,jiang2024efficient}}  \\
{Talent Searching} & {Keywords matching} & {Query, Resume} & {\cite{manad2018enhancing}}  \\
{Talent Searching} & {Keywords matching, knowledge graph} & {Query, Resume} & {\cite{wang2021analysing}}  \\
{Talent Searching} & {Traditional classifiers} & {Query, Resume} & {\cite{apatean2017machine,ozcaglar2019entity}}  \\
{Talent Searching} & {Topic model, multi-armed bandit algorithm} & {Query, Resume} & {\cite{geyik2018session}}  \\
{Talent Searching} & {Learning-to-rank algorithm} & {Query, Resume} & {\cite{ha2016search,ha2017query}}  \\
{Talent Searching} & {DNN, learning-to-rank algorithm} & {Query, Resume} & {\cite{ramanath2018towards,yang2021cascaded}}  \\
 Person-Job Fitting   & Latent variable model   & Job posting, resume & \cite{malinowski2006matching}\\
 Person-Job Fitting   & CNN & Job posting, resume & {\cite{zhu2018person,zhao2021embedding,he2021finn,zhenhong2021person}}\\
 Person-Job Fitting   & RNN & Job posting, resume & \cite{qin2018enhancing,qin2020enhanced,yan2019interview}\\
  Person-Job Fitting   & CNN, RNN & Job posting, resume & {\cite{bian2019domain,luo2019resumegan,jiang2020learning}}\\
 Person-Job Fitting   & BERT & Job posting, resume & \begin{tabular}[c]{@{}c@{}} {\cite{lavi2021consultantbert,abdollahnejad2021deep,liu2022job,shao2023exploring}}, \\ {\cite{kaya2023exploration,fang2023KDD,guan2024jobformer,gong2024your}}\end{tabular}\\
 Person-Job Fitting   & GNN, BERT & Job posting, resume & {\cite{bian2020learning, yao2022knowledge, wang2022person, yang2022modeling, chen2024professional}}  \\
 Person-Job Fitting   & Attention mechanism   & Job posting, resume & {\cite{he2021self,fu2021beyond,zhang2021explainable,hou2022leveraging}}\\
 {Person-Job Fitting}   & {Reinforcement learning}  & {Job posting, resume} & {\cite{fu2022market}}\\
  {Person-Job Fitting}   & {Federated learning} & {Job posting, resume} & {\cite{zhang2023fedpjf}} \\
  {Person-Job Fitting}   & {LLMs} & {Job posting, resume} & {\cite{zheng2023generative,ghosh2023jobrecogpt,wu2024exploring,du2024enhancing,yu2024confit}} \\
 Person-Job Fitting   & Ranking & \begin{tabular}[c]{@{}c@{}}Job posting, resume, \\ social media\end{tabular}    & \cite{faliagka2012integrated,faliagka2014line}\\
 Person-Job Fitting   & K-means & Social media& \cite{giri2016vritthi}\\
 Person-Job Fitting   & Traditional classifiers  & Social media& \cite{menon2016novel}\\
 Person-Job Fitting   & Gamma-Poisson model & User behavior & \cite{borisyuk2017lijar} \\
 \midrule
 \midrule
 \multicolumn{4}{c}{\textbf{Talent Assessment}}\\\midrule
 Interview Question Recommendation & Topic model  &\begin{tabular}[c]{@{}c@{}}Job posting, resume, \\ assessment report\end{tabular}  & \cite{shen2018joint,shen2021joint}\\
 Interview Question Recommendation & Knowledge graph & \begin{tabular}[c]{@{}c@{}}Job posting, resume, \\ search engine log\end{tabular}   & \cite{qin2019duerquiz}\\
  {Interview Question Recommendation} & \begin{tabular}[c]{@{}c@{}}{Knowledge graph,} \\{integer linear programming}\end{tabular} & {Question bank} & {\cite{datta2021generating}}\\
 Interview Question Recommendation & BERT & Job posting & \cite{shi2020learning}\\
{Interview Question Recommendation} & GCN & KSC, search engine log & \cite{qin2023automatic}\\ 
 Assessment Scoring & Regression models & Interview videos & {\cite{nguyen2014hire, naim2015automated, naim2016automated}} \\
 Assessment Scoring & Doc2Vec & Interview videos & \cite{chen2016automated,chen2017automated}\\
 Assessment Scoring & GNN & Interview records & \cite{chen2020hierarchical}  \\
 Assessment Scoring & Attention mechanism & Interview videos & {\cite{hemamou2019hirenet, hemamou2019slices, hemamou2021multimodal}} \\
  Assessment Scoring & {Transformer} & Interview videos & {\cite{singhania2020grading,verrap2022answering}} \\
  {Assessment Scoring} & {Adversarial learning} & {Interview videos} & {\cite{hemamou2021don}} \\
 Assessment Scoring & traditional classifiers & Employee profiles & \cite{liu2009application, hong2019evaluation, li2020data} \\
 \midrule
 \midrule
 \multicolumn{4}{c}{\textbf{Career Development}}\\\midrule
 Course Recommendation & Collaborative filtering & Trainees’ profiles  & \cite{wang2020personalized,wang2021personalized} \\
 Course Recommendation & Markov decision process& Learning records& \cite{srivastava2018s}\\
  {Course Recommendation} & {Neural Network} & {Learning records} & \cite{patki2021personalised}\\
 {Course Recommendation} & {Reinforcement Learning} & {Learning records} & \cite{zhengzhi2023recsys}\\
 {Course Recommendation} & {KG-based Transformer} & {Learning records, Knowledge Graph} & \cite{yangyang2023TOIS}\\
 Promotion Prediction & Traditional classifiers  & Social network  & \cite{yuan2016promotion} \\
 Promotion Prediction  & Traditional classifiers    & \begin{tabular}[c]{@{}c@{}}Personal profile, \\job posting log\end{tabular}  & \cite{long2018prediction}\\
 {Promotion Prediction} & {Multiple classification}  & {Employee's Detail Record}  & {\cite{liu2021data}} \\
 Promotion Prediction & Survival analysis   & Personal profile, career paths& \cite{li2017prospecting} \\
 Turnover Prediction & Traditional classifiers & HR dataset & \cite{sisodia2017evaluation, nagadevara2008establishing, alao2013analyzing} \\
 Turnover Prediction & GNN, RNN & profile, turnover records & \cite{teng2019exploiting} \\
 Turnover Prediction & neural network & profile, turnover records & \cite{teng2021exploiting} \\
 Turnover Prediction & Traditional classifiers & HR Information Systems & \cite{ajit2016prediction} \\
 Turnover Prediction & GNN, RNN, survival analysis & \begin{tabular}[c]{@{}c@{}}job description, organizational tree, \\ profile, turnover records\end{tabular} & \cite{hang2022outside} \\
 Job Satisfaction & Traditional classifiers  & \begin{tabular}[c]{@{}c@{}}Personal profile, \\ job profile \end{tabular} & \cite{arambepola2021makes}\\
 Job Satisfaction & Traditional classifiers  & Social media& \cite{saha2021social}\\
 Career Mobility Prediction & RNN & Career paths & \cite{li2017nemo,he2021your}\\
 Career Mobility Prediction & Attention mechanism & Career paths, employee profile & \cite{meng2019hierarchical} \\
  {Career Mobility Prediction} & {Transformer} & {Career paths, employee profile} & {\cite{yamashita2022looking}} \\
 Career Mobility Prediction & Collaborative filtering & Career paths, employee profile & \cite{wang2021variable} \\
 Career Mobility Prediction & GNN & Career paths, employee profile & \cite{zhang2021attentive} \\
  {Career Mobility Prediction} & {Representation learning} & {Career paths, employee profile} & {\cite{decorte2023career,zha2024towards}} \\
  {Career Mobility Prediction} & {Reinforcement learning} & {Career paths, employee profile} & {\cite{guo2022intelligent,guo2023preference,avlonitis2023career}} \\
\bottomrule
\end{tabular}}
\begin{tablenotes}
\item  Traditional classifiers represent one or multiple traditional classifiers, including SVM, Logistic regression, GBDT, XGBoost, Decision tree, Adaboost, KNN and so on.
\end{tablenotes}
\end{table*}

\noindent \textbf{Cleaning and Debias Methods.} Data cleaning and debiasing is an iterative process tailored to the requirements and semantics of specific analysis tasks~\cite{krishnan2016towards}. This process consists of transforming raw data into consistent data that can be analyzed. Herein, we delineate several frequently employed methodologies in talent analytics for data cleaning and debiasing.
\begin{itemize}
\item \textbf{Data Selection:}
Data selection involves the meticulous identification and extraction of pertinent data subsets from a broader dataset, guided by specific criteria or requisites~\cite{liu1998feature}. This approach serves to streamline the data analysis process by honing in on the most relevant information, thereby mitigating noise and extraneous data. For instance, \textsl{Zhang et al.}~\cite{zhang2021talent} opted to discard company-position pairs with a monthly averaged talent demand of less than 2, considering scenarios where certain companies may not extensively recruit for particular research positions, resulting in consistently low demand. \textsl{Shao et al.} conducted data cleaning by eliminating job posts and resumes with incomplete attributes~\cite{shao2023exploring}. \textsl{Balaji et al.} focused on extracting useful fields from job titles to curate the most pertinent and distinctive set of work activities corresponding to selected company job titles~\cite{balaji2019airesume}.
\item \textbf{Data Filtering:}
Data filtering encompasses the systematic elimination or exclusion of undesirable or extraneous data from a dataset, thereby diminishing noise and refining its quality. In typical text cleaning tasks, the compilation of a stop word list proves indispensable for sieving out irrelevant information~\cite{guo2016resumatcher}. Moreover, the adoption of regular expressions to delineate fields and detect anomalies is widely embraced in personalized resume-job matching systems~\cite{guo2016resumatcher} and job search engines~\cite{muthyala2017data}.
\item \textbf{Data Clustering:}
Data clustering involves grouping similar data points together based on certain characteristics or features. \textsl{Wakchaure et al.}~\cite{wakchaure2008technique} leverages the Levenshtein edit distance, a metric gauging string similarity, to designate matches exceeding 0.85 as identical individuals. \textsl{Gaikwad et al.}~\cite{gaikwadeffective} employs Levenshtein edit distance for duplicate detection within XML documents. \textsl{Sun et al.}~\cite{sun2021market} utilize text embedding and measure edit distance to approximate similar job descriptions.\textsl{Zhang et al.}~\cite{zhang2021talent} employ a classification approach to categorize original job titles into 16 distinct categories, thereby mitigating noise and standardizing job titles.
\item \textbf{Data Synthesis:}
Data synthesis encompasses the generation of fresh data points to augment extant datasets, proving instrumental in addressing gaps or amplifying the efficacy of analysis and model functionalities~\cite{huang2009study}. For instance, \textsl{Magron et al.}~\cite{magron2024jobskape} devised synthetic job postings to refine skill alignment, while \textsl{Skondras et al.}~\cite{skondras2023generating} utilized Large Language Models to craft synthetic resume data, thereby bolstering job description classification.
\item \textbf{Data Normalization:}
Data normalization is a crucial procedure aimed at standardizing the scale or distribution of data values within a dataset, thereby facilitating precise comparisons and analyses while ensuring uniformity across variables. It finds widespread application in various domains, particularly in time series forecasting. For instance, \textsl{Liu et al.} conducted normalization operations, which involve subtracting the minimum value of each node and dividing it by the difference between the maximum and minimum values of the node~\cite{liu_talent_2023}. Similarly, \textsl{Zhang et al.} normalized the number of job transitions along the company axis to achieve consistency~\cite{zhang2019large}.
\end{itemize}

\begin{figure*}[t!]
\centering
\includegraphics[scale=0.25]{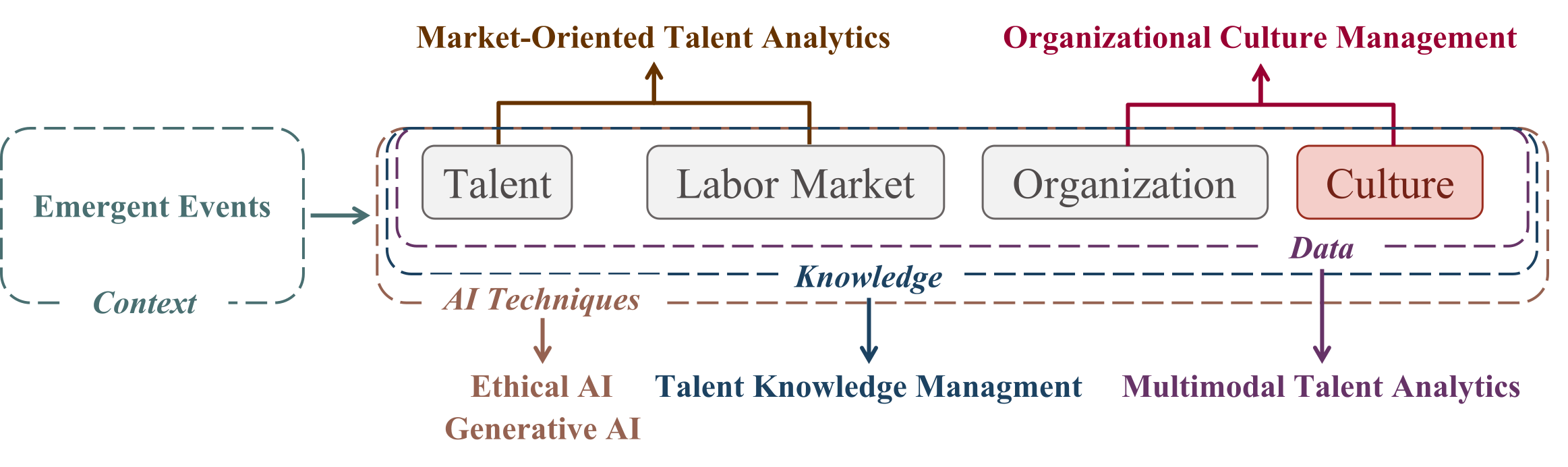}
\caption{The potential research directions in talent analytics.}
\label{fig:culture} 
\end{figure*}

\subsection{Limitations}
Existing data and processing methods in the field of talent analytics continue to face many limitations and challenges, which constrain the development of AI-based approaches. We summarize these issues as follows.
\begin{itemize}
    \item \textbf{Lack of benchmark datasets:} Most current research in talent analytics heavily depends on proprietary data related to job applicants, employees, and organizations. {As previously mentioned, internal data includes sensitive private information, such as the profiles of both job seekers and employees.} The sensitive nature of this data often prevents it from being publicly accessible, severely restricting the development of publicly available benchmark datasets. This limitation impedes the standardization of problem definitions and comparative analyses of methodologies, thus slowing down technological advancements in the field. While some job-related data has been made open-source, it remains limited in scale and temporal coverage. {Moreover, while existing methods utilize these data, they lack sufficient data analysis to provide researchers with meaningful insights.} Consequently, there is an urgent need to create comprehensive open-source benchmark datasets for pivotal tasks within this field. We present several open datasets, as shown in the Appendix, primarily related to resume parsing and employee turnover prediction. These datasets can help researchers in developing standardized datasets for uniform comparisons. Additionally, the data anonymization and de-identification methods require more extensive consideration to facilitate the construction of more open-source datasets for various scenarios. {Indeed, such technologies have been widely researched in the medical field~\cite{friedrich-etal-2019-adversarial,trienes2020comparing,chevrier2019use}. In the context of talent analytics, \textsl{Jensen et al.} recently enhanced de-identification performance by integrating NER techniques with BERT for job postings~\cite{jensen-etal-2021-de}. }
    {However, current techniques often compromise data utility, especially for fairness and diversity analysis. Developing more advanced methods will allow the creation of comprehensive, privacy-preserving datasets that retain key demographic variables, supporting both benchmarking and fairness research in talent analytics.}

    \item \textbf{Diversity deficiencies in datasets:} The datasets commonly used in AI-based talent analytics often lack adequate consideration of diverse populations, especially underrepresented minorities. This deficiency hampers efforts to evaluate and ensure algorithmic fairness. Models developed from such datasets may exhibit biases, leading to inequitable outcomes among various demographic groups. {Furthermore, the absence of publicly available gender and racial data restricts the ability to conduct comprehensive fairness evaluations, making it difficult to detect and mitigate biases effectively.} Therefore, it is essential to integrate more diverse datasets to promote fairness and enhance the overall performance of AI algorithms in talent analytics.
    \item \textbf{Challenges with subjective data:} In many talent analytics scenarios, such as employee promotion decisions, outcomes are frequently based on the subjective judgment of supervisors. This practice can introduce inherent biases and noise into the real datasets collected from these decisions. Traditional data preprocessing techniques often fail to detect and address these anomalies effectively. Therefore, there is a pressing need for specialized noise detection algorithms or the development of robust AI-based talent analytics models that can handle such irregularities. These approaches are essential to ensure the accuracy and fairness of the analytics outcomes.
\end{itemize}

\begin{table*}[t!]
\centering 
\caption{The table of collected papers related to organizational management.}
\begin{tabular}{c  c  c  c}
\toprule
\textbf{Task} & \textbf{Method} & \textbf{Data} & \textbf{Reference} \\ 
\midrule
\midrule
\multicolumn{4}{c}{\textbf{Organizational Network Analysis}}\\
\midrule
Organizational Network Modeling	&Representation learning &	Organizational network dataset &	~\cite{ye2022mane}\\
Organizational Network Modeling &	GCN, RNN &	Working related records&	~\cite{teng2021exploiting}\\
Organizational Network Modeling &	Community Search &	Public data, in-firm data &	~\cite{dong2021butterfly}\\
High-potential Talent Identification &	GNN, LSTM &	Organizational network dataset &	~\cite{ye2019identifying}\\
\midrule
\midrule
\multicolumn{4}{c}{\textbf{Organizational Stability}}\\
\midrule
Team formation & Heuristic algorithm & DBLP, IMDb & \cite{kargar2012efficient} \\
Team formation & Greedy algorithm  & DBLP & \cite{zihayat2016authority} \\
Team formation & 
Variational Bayes Network
& DBLP & \cite{hamidi2020learning} \\
Team formation & Hybrid approach & Crowdsourcing
data & \cite{liu2015efficient} \\   
Team formation &  Greedy algorithm & Freelancer dataset & \cite{barnabo2019algorithms} \\
Team formation & 
Adaptive algorithm
& Crowdsourcing
data & \cite{anagnostopoulos2018algorithms} \\
Team optimization & Graph kernel & DBLP, Movie and NBA & \cite{li2015replacing,li2016enhancing}\\
Team optimization & Deep learning & GitHub, DBLP & \cite{zhao2018team} \\
Team optimization & Reinforcement learning & Movie, DBLP &  \cite{zhou2019towards}\\
Person-Organization Fit &	CNN, LSTM &	In-firm data &	~\cite{sun2019impact,sun2021modeling}\\
{Person-Organization Fit} &	{KNN} &	{Talent data} &	{~\cite{artar2024improving}}\\
\midrule
\midrule
\multicolumn{4}{c}{\textbf{Organizational Incentive Analysis}}\\
\midrule
Job title benchmarking & Representation Learning & OPNs & ~\cite{zhang2019job2vec}\\ 
{Job title benchmarking} &	{Semantic relatedness modeling} &	{CEDEFOP classify Online Job Vacancies} &	{~\cite{malandri2021meet}}\\
{Job title benchmarking} &	{Graph learning} &	{Working experiences and resume data} &	{~\cite{zhu2022towards}}\\
{Job title benchmarking} &	{BiLSTM, Unsupervised representation learning} &	{A labeled job similarity dataset} &	{~\cite{zbib2022learning}}\\
{Job title benchmarking} &	{Transformer, RNN} &	{Employee career paths in IT field} &	{~\cite{liu2022job}}\\
{Job title benchmarking} &	{Bert, Graph embedding} &	{Resume data} &	{~\cite{yamashita2023james}}\\
Job salary benchmarking &	Matrix completion	& Job posting &	~\cite{meng2018intelligent, meng2022fine} \\
{Job salary benchmarking} &	{Matrix equation}	& {Job post and review data} &	{~\cite{hung2021aggregating}} \\
\bottomrule
\end{tabular}
\label{tab:orgworks}
\end{table*}

\begin{table*}[t!]
\centering 
\caption{The table of collected papers related to labor market analysis.}
\begin{tabular}{c  c  c  c}
\toprule
\textbf{Task} & \textbf{Method} & \textbf{Data} & \textbf{Reference}\\ 
\midrule
\midrule
\multicolumn{4}{c}{\textbf{Talent Flow Analysis}}\\\hline
Flow prediction&RNN&OPNs&~\cite{xu2018dynamic}\\
Flow prediction&Tensor Factorization&OPNs&~\cite{zhang2019large}\\
Flow prediction& Latent variable model&OPNs&~\cite{zhang2020large}\\
{Flow prediction}&{GAT}&{Questionnaires}&{~\cite{liu_talent_2023}}\\
{Flow prediction}&{Graph Clustering, HITS}&{OPNs}&{~\cite{sun2024large}}\\

Flow pattern analysis&Learning algorithms&OPNs&~\cite{cheng2015mining}\\
Flow pattern analysis&Optimization algorithm&OPNs&~\cite{cheng2013jobminer}\\
Flow pattern analysis &PageRank&OPNs&~\cite{oentaryo2017analyzing,oentaryo2018talent}\\
Flow pattern analysis&Clustering model&OPNs&~\cite{xu2016talent}\\
{Flow pattern analysis}&{Clustering model, PageRank}&{OPNs}&{~\cite{li_measuring_2024}}\\
\midrule
\midrule
\multicolumn{4}{c}{\textbf{Job Analysis}}\\\midrule
Demand trend analysis&Latent Semantic Indexing&Job postings&~\cite{karakatsanis2017data}\\
Demand trend analysis&N-gram, SVM&Job postings&~\cite{lovaglio2020comparing}\\
Demand trend analysis&Attentive neural network&Job postings&~\cite{zhang2021talent}\\
{Demand Trend analysis}&{Dynamic Graph}&{Job postings}&{~\cite{guo_talent_2022}}\\
Topic Trend analysis&Latent variable model&Job postings&~\cite{zhu2016recruitment}\\
Topic Trend analysis&Bi-gram model&Job postings&~\cite{marrara2017language}\\
{Topic Trend analysis}&{Word2Vec, RNN}&{Job postings}&{~\cite{azzahra_text_2021}}\\
{Topic Trend analysis}&{NER}&{Job postings}&{~\cite{mahdavimoghaddam_utility_2022}}\\
\midrule
\midrule
\multicolumn{4}{c}{\textbf{Skill Analysis}}\\\midrule
Potential Skill Prediction&language models,SVM&Job postings&~\cite{colombo2018applying}\\
Potential Skill Prediction&Apriori algorithm&Job postings&~\cite{akhriza2019constructing}\\
Potential Skill Prediction&FP-growth algorithm&Job postings&~\cite{patacsil2021analyzing}\\
Potential Skill Prediction&KNN algorithm&Job postings&~\cite{wowczko2015skills}\\
Potential Skill Prediction&Tensor factorization&Job postings&~\cite{wu2019trend}\\
Potential Skill Prediction&Topic model&Job postings&~\cite{xu2018measuring}\\
{Potential Skill Prediction}&{GNN}&Job postings&{~\cite{liu_2021_learning}}\\
{Potential Skill Prediction}&{SDCA logistic regression}&Job postings&{~\cite{walek_2021_data}}\\
{Potential Skill Prediction}&{Bert}&Job postings&{~\cite{leon_2024_hierarchical}}\\

{Skill demand forecasting}&{Logistic Regression, Granger Causality}&Job postings&{~\cite{mahdavimoghaddam_2021_congruence}}\\
{Skill demand forecasting}&{Bert, Logistic Regression, Random forest, SVC}&Job postings&{~\cite{mahdavimoghaddam_2022_exploring}}\\
{Skill demand forecasting}&{RNN}&Job postings&{~\cite{de_macedo_practical_2022}}\\
{Skill demand forecasting}&{LSTM, TimeGAN}&Job postings&{~\cite{wolf_2023_generating}}\\
{Skill demand forecasting}&{Hypernetwork, RNN}&Job postings&{~\cite{chao_cross-view_2024}}\\
{Skill demand forecasting}&{Pretraining Graph Autoencoder}&Job postings&{~\cite{xi_pretrain_2024}}\\
{Skill Valuation}&{Hill climbing}&{Dblp}&~\cite{rahman_worker_2015}\\
Skill Valuation&Neural network&Job postings&~\cite{sun2021market}\\
{Skill Valuation}&{Linear regression}&{Job postings}&~\cite{stephany_what_2024}\\
\midrule
\midrule
\multicolumn{4}{c}{\textbf{Brand Analysis}}\\\midrule

{Employer branding} & {Topic Regression} & Job reviews& ~\cite{lin2017collaborative,lin2020enhancing}\\ 
{Employer branding}&ELM classifier&Company reviews&~\cite{bajpai2019aspect}\\
{Employer branding} & {Markov switching model} & {News, Social media}& {~\cite{spears2021impact}}\\ 
{Employer branding} & {Prototype representation} & {Corporate events}& ~\cite{yuan_self-supervised_2021}\\ 
{Employer branding} & {Extreme learning machine} & {Job reviews}& {~\cite{bose_sentiment_2022}}\\ 
{Employer branding} & {Topic model} & {Company descriptions}& {~\cite{savin_topic-based_2023}}\\ 
{CSR communication} & {Topic model} & {Social media}& {~\cite{chae2018corporate, mazza_text_2022, pilgrim_csr_2023, thakur_impact_2023}}\\ 
{Employee sentiment analysis}&{Topic Model}&{Company reviews}&{~\cite{moniz2014sentiment}}\\
{Employee sentiment analysis}&Topic model&Social media&~\cite{ikoro2018analyzing}\\
{Employee sentiment analysis}&{Topic model}&{Job reviews}&{~\cite{shi_listening_2021,ganga_employees_2022}}\\
{Employee sentiment analysis}&{TF-IDF, Bag of words, SVM}&{Job reviews}&~\cite{gaye_sentiment_2021, rehan_employees_2022}\\
{Employee sentiment analysis}&{CNN, RNN}&{Job reviews}&{~\cite{mouli_unveiling_2023}}\\
\bottomrule
\end{tabular}
\label{tab:cultureworks}
\end{table*}

\section{Prospects}\label{prospects}
In this appendix, we provided a broader and specific
 recent efforts in AI-based talent analytics in human resource management from three different aspects: talent management, organization management, and labor market analytics, which can provide a more detailed explanation and discussion for the prospects section in the main text.

\subsection{Multimodal Talent Analytics}
Information about a phenomenon or a process in talent analytics-related scenarios usually comes in different modalities. For instance, we can obtain communication and project collaboration networks in employee collaboration analysis. Indeed, mining the multimodal data in talent analytics can help us enhance the effectiveness of different applications. For example, \textsl{Hemamou et al.} collected the multimodal data in the job interview process, including text, audio, and video, and proposed a hierarchical attention model to achieve the best performance in predicting the hirability of the candidates~\cite{hemamou2019hirenet}. 
Moreover, the utilization of multimodal data to train AI models, rather than solely relying on traditional interviews, offers a more comprehensive understanding of the participants' personality test results~\cite{koutsoumpis2024beyond}.
Recently, multimodal learning has been used to achieve multimodal data representation, translation, alignment, fusion, and co-learning in various domains, such as commercial, social, biomedical~\cite{lahat2015multimodal,baltruvsaitis2018multimodal}. 
We can foresee that more multimodal learning methods will gradually be extensively used in talent analytics.

\subsection{Talent Knowledge Management} 
Though AI-based approaches have achieved great success in acquiring talents and developing them, relatively few works explore managing those talents' knowledge with AI technologies, which is the primary driving force to the economics of ideas~\cite {wiig1997knowledge}. There is an urgent need for additional AI-based technologies that focus on talent knowledge creation, sharing, utilization, and management. Such efforts are crucial for maximizing the potential of human resources and enhancing organizational productivity. Indeed, we can leverage knowledge graph-related technologies~\cite{ji2021survey} to construct the talent's knowledge base and achieve efficient knowledge management. Moreover, we can transform the scenarios in talent development, such as knowledge learning and collaboration, into different recommendation scenarios and utilize recommendation algorithms to solve these problems. Recently, \textsl{Wang et al.} developed a personalized online courses recommendation system based on the employees' current profiling~\cite{wang2020personalized,wang2021personalized}. However, there is still a lack of an algorithm that can recommend heterogeneous knowledge. The existing algorithms are only from the individual perspective and have not been analyzed from the organizational aspect, such as the organizational knowledge diversity or competitiveness. 
Furthermore, although AI-based technologies have greatly improved the efficiency of knowledge management, the social and ethical implications (algorithmic discrimination, etc.) that they bring with them are also worthy of continued research~\cite{wang2022artificial}.

\subsection{Market-oriented Talent Analytics}
AI technology has been effectively applied in labor market analytics~\cite{zhang2020large}. However, those approaches mainly focus on the perspective of global market analysis and have not explored how the changing environment of the labor market affects internal talent management or organization management. In fact, combining macro and micro data in talent analytics is a vital research direction~\cite{huang2016participation}. With the recent accumulation of internal and external data, there exists an exceptional opportunity to implement market-oriented talent analytics. For instance, \textsl{Hang et al.} leveraged the job posting data to capture the potential popularity of employees in external markets specific to skills and further achieve more accurate employee turnover prediction based on the market trend~\cite{hang2022outside}. Moreover, the rapid development of AI technologies provides an excellent technical foundation for this direction. We can utilize multi-task learning~\cite{zhang2021survey} to jointly learn both macro and micro talent analytics-related tasks. Heterogeneous graph learning~\cite{zhang2019heterogeneous,hang2022outside} can help us effectively model the correlation of macro and micro data. 

Besides, we can use AI technologies to identify various types of talents in the market. Identifying high-potential employees has always been an important issue in enterprises. \textsl{Ye et al.}~\cite{ye2019identifying} proposed a neural network-based dynamic social profiling approach for quantitatively identifying high-potential talents. \textsl{Cheng et al.}~\cite{cheng2021effectiveness} conducted a quasi field experiment study and found that AI outperforms humans in the identification of high-potential talents. In the labor market, accurate recruitment of digital talents is crucial for the digital transformation of enterprises. \textsl{Harrigan et al.}~\cite{harrigan2021march} found a specific relationship between enterprise talent investment and digitization by identifying digital skills-related talents within the organizations.  By collecting data on workers, \textsl{Goos et al.}~\cite{goos2021routine} found that people in regular jobs had a harder time finding jobs in new factories compared to digital talents. \textsl{Babina et al.}~\cite{babina2024artificial} classified different categories of AI skill-related talents based on job postings on recruitment platforms and analyzed the impact of these talents on the development of the enterprise after entering the enterprise. \textsl{Kim et al.}~\cite{kim2023ai} proposed a dynamic co-occurrence method, which dynamically calculates the AI relevance of various types of skills in the labor market, so as to identify AI talents more accurately. It can be seen that AI has a great impact on human resource management, especially in the current situation where enterprises are paying more and more attention to environmental protection and social responsibility, AI has a broad prospect in the identification of green talents and Environmental, Social and Governance (ESG)-related talents. AI and human resource management still need large-scale research in the future.

\subsection{Organizational Culture Management}
In our survey, we reviewed recent advancements in AI techniques for talent analytics in HRM from three perspectives, including talent management, organization management, and labor market analysis. 
The culture of an organization plays a crucial role in sustaining its effectiveness and viability~\cite{corporateculture}. Generally, the culture mainly contains three aspects: Mission, Vision, and Values (MVVs), which can help employees understand what is encouraged, discouraged, accepted, or rejected within an organization, and facilitate the organization to thrive with the shared purpose. 
Recent developments reveal that the availability of extensive datasets covering the entire lifecycle of talents and organizations offers opportunities to realize effective culture management. For example, the interconnection between culture and leadership is evident, with exceptional team leaders significantly shaping organizational culture~\cite{corporateculture}. Some researchers~\cite{lee2020determining} discussed how ML techniques can be used to inform predictive and causal models of leadership effects. Accordingly, they further provided a step-by-step guide on designing studies that combine field experiments to establish causal relationships with maximal predictive power. Meanwhile, several studies analyze leadership styles with data mining algorithms, demonstrating that the different leadership styles significantly influence leadership outcomes~\cite{ahmad2018investigation}. Moreover, some researchers try to utilize text mining to analyze the organizational culture. For instance, \textsl{Schmiedel et al.} leveraged the online company reviews data and topic model to explore the employees’ perception of corporate culture~\cite{schmiedel2019topic}. \textsl{Li et al.}~\cite{li2021role} applied a topic model to obtain firm-level measures of exposure and response related to emergent events for many U.S. firms. 
As a famous saying goes, ``Culture eats strategy for breakfast'', employing AI technologies in organizational culture management will become one of the most critical research directions in the future, as it can help managers scientifically address cultural management.

\subsection{Ethical AI in Talent Analytics}
Admittedly, AI technologies are increasingly employed in talent analytics, significantly enhancing management efficiency and accuracy.
However, there are ongoing concerns about how to ensure that AI technologies adhere to well-defined ethical guidelines regarding fundamental values. Recently, some researchers have made efforts from two perspectives, i.e., fairness and explainability.

\subsubsection{Fairness} 
The importance of fairness in talent analytics cannot be overstated, given its profound impact on employee well-being and alignment with organizational values.
Although AI technologies have achieved various successes in talent analytics, there is growing concern that such approaches may bring issues of unfairness to people and organizations, as evidenced by some recent reports~\cite{lee2018understanding,amazonfairness,bogen2019all}. For instance, Amazon scrapped its AI-based recruitment system 
due to its discriminatory outcomes against women~\cite{amazonfairness}.

Recent studies have delved into the fairness of AI technologies in talent management from different perspectives. For instance, \textsl{Qin et al.}~\cite{qin2020enhanced} verified that when involving sensitive features, such as gender, age, etc., into the person-job fit model, the model without special design will easily learn the bias from the original data. Intuitively, we can solve this problem by removing sensitive features, also regarded as one of the pre-processing methods for imposing fairness~\cite{d2017conscientious}. However, a large amount of unstructured data already contains sensitive features, such as audio and video data in the interview process. The model can easily infer the potentially sensitive attributes of data, which may still cause the bias of AI algorithms~\cite{de2019bias}. 
In their study, {Pena~et~al.} examined multimodal systems to predict recruitable candidates using both image and structured data from resumes~\cite{pena2020bias}.
The authors first demonstrated that the deep learning model could reproduce the biases from the training data, even without the sensitive features. 
To address this, they integrated an adversarial regularizer to remove sensitive information from unstructured data and promote algorithmic fairness~\cite{morales2020sensitivenets}. Similarly, \textsl{Yan et al.} both leveraged data balancing and adversarial learning to mitigate bias in the multimodal personality assessment ~\cite{yan2020mitigating}. Moreover, there exist several open-source tools, such as AIF360~\cite{bellamy2018ai}, FairML~\cite{adebayo2016fairml}, Themis-ML~\cite{bantilan2018themis}, that can facilitate systematic bias checks and embed fairness in the AI algorithms~\cite{mujtaba2019ethical}. In addition, AI technologies can also help to reduce human bias in different talent management scenarios. For instance, AI technologies have been applied to detect the potentially problematic words in the job posting that lead to bias or even legal risks and further assist employers in writing inclusive job descriptions~\cite{hirenotneutral}. 
Consequently, ensuring the fairness of AI algorithms is becoming an increasingly significant area of research within the field of talent analytics.

\subsubsection{Explainability} 
Recently, there has been an increasing concern among employees and managers regarding the decisions made by black-box AI algorithms. Questions arise regarding the basis of these decisions, understanding the factors behind algorithmic success or failure, and determining how to rectify any errors that occur.
Therefore, the research interest in increasing the transparency of AI-based automated decision-making in talent management are re-emerging~\cite{hunkenschroer2022ethics,zhang2020developing}. For instance, \textsl{Qin et al.} proposed to leverage the attention mechanisms to explain the matching degree between the content of job postings and resumes~\cite {qin2018enhancing}. \textsl{Zhang et al.} further introduced the hierarchical attention and collaborative attention mechanisms to increase the person-job fit model explainability both at the structured and unstructured information level~\cite{zhang2021explainable}. \textsl{Upadhyay et al.} leveraged the knowledge graph and name entity recognition technologies to generate the understandable textual job recommendation explanation~\cite{upadhyay2021explainable}. In ~\cite{kaya2017multi}, \textsl{Kaya et al.} focused on constructing an end-to-end system for explainable automatic job candidate screening from video resumes. The authors extracted the audio, face, and scene features and leveraged the decision trees to both predict whether the candidates will be invited to the interview and explain the decisions by using binarization with a threshold. \textsl{Liem et al.} further handled the job candidate screening problem from an interdisciplinary viewpoint of psychologists and machine learning scientists~\cite{liem2018psychology}. 
Moreover, \textsl{Juvitayapun et al.} utilized the tree-based model to calculate the importance of different features, enhancing the explainability of AI-based turnover prediction~\cite{juvitayapun2021employee}.

However, the current approaches only stay in the perspective of AI model design and fail to consider whether employees or managers can easily comprehend and grasp the explanatory conclusions provided by the model. Indeed, visual analytics is an inherent way to help people who are inexperienced in AI understand the data and model~\cite{choo2018visual,alicioglu2022survey}. Therefore, combining visual analysis and explainable AI and building an intelligent talent management system is a valuable research direction. 
Additionally, leveraging the extensive interaction data generated by users can facilitate iterative model improvements from various angles, including correcting errors in automated decision-making and enhancing the efficiency of visual information presentation.

\subsection{Generative AI in Talent Analytics}

With technological advances, artificial intelligence has gradually developed a branch called generative AI, and the language in which AI interacts with humans has changed from ``machine language" to ``natural language"~\cite{wu2023survey}. 
By employing suitable templates, many tasks can be addressed through generative AI. Specifically, the knowledge emergence facilitated by large-scale parameters enables generative AI to tackle a series of complex tasks~\cite{wang2024survey}. The advent of generative LLMs has sparked widespread discussion and excitement in both academia and industry.
First of all, LLMs have a strong generative ability, which enables them to easily understand various contexts and generate appropriate content according to the prompts. The following work is a good illustration of this aspect of the capabilities of LLMs: \textsl{Zinjad et al.} proposed a resume generation tool based on natural language understanding and information extraction capabilities of LLMs, which allows users to generate tailored personalized resumes by providing simple personal information and job information~\cite{zinjad2024resumeflow}. \textsl{Ayoobi et al.} proposed a method for recognizing LLM-generated resumes in online recruitment platforms, which is beneficial in solving the challenge of identifying fake and LLM-created resumes that are difficult to identify~\cite{ayoobi2023looming}. \textsl{Magron et al.} constructed a job posting dataset for skill matching that contains more implicit skills, longer sentences, and is closer to data from real job platforms. They conduct skill-matching experiments using LLMs and show that the method performs well in real-world data evaluation~\cite{magron2024jobskape}. Second, LLMs facilitate the innovation of traditional AI technologies. For example, in the context of recommender systems, LLMs utilize their high-quality representation of textual features and their extensive coverage of external knowledge to achieve high-quality recommendations~\cite{wu2023survey}. \textsl{Wu et al.} revealed the capability of LLMs to mine graph information by proposing a meta-path prompt constructor to help LLMs understand the semantics of behavioral graphs, and the framework helps provide personalized job recommendations for job seekers~\cite{wu2024exploring}. \textsl{Du et al.} improved the traditional LLM-based job recommendation method and verified its effectiveness by mining the explicit and implicit features of users in an online recruitment platform, and aligning the unmatched low-quality and high-quality generated resumes via a generative adversarial network~\cite{du2024enhancing}. \textsl{Zheng et al.} guided the LLM-based generative job recommendation system based on a Supervised Fine-Tuning strategy, generate suitable job descriptions according to the resumes of job seekers, and provide individuals with a more personalized and comprehensive job-seeking experience~\cite{zheng2023generative}. \textsl{Abu-Rasheed et al.} developed a group chat method based on knowledge graphs to enhance the interpretability of students in conversations with chatbots in the course recommendation task~\cite{abu2024supporting}.

Besides, in recent years, LLM-based agents are utilized to solve various tasks such as software development, social simulation, and policy simulation~\cite{guo2024large}. Research on intelligent simulation of management-related problems using digital technologies has also been emerging, such as the study of member autonomy in organizations using human-machine systems based on blockchain technology~\cite{ellinger2023skin}; how humans can use AI assistants to work effectively with large language models~\cite{raiaan2024review}; and how autonomous agents can utilize the generic capabilities of the underlying model for reasoning, decision-making, and environment interaction~\cite{yang2024towards}. Further, by integrating multiple LLM agents, model efficiency can be further improved and simulation effects can be optimized~\cite{liu2023dynamic}. However, there is not much research on using agents based on large language models for simulating organizational behavior. We should use management theory to simulate the framework of an organization and the behavior of individuals within the organization, which can focus on the trust behavior of inter-organizational personnel in interpersonal interactions, i.e., the willingness to put one's self-interest at risk based on the positive expectations of others~\cite{xie2024can}. An LLM-based agent can be used to generate conversational data for studying the behaviors and capabilities of chat agents~\cite{li2024camel}. They query APIs that read and write to web pages, generate content that shapes human behavior, and run system commands as autonomous agents~\cite{pan2024feedback}. We can also analyze LLMs' network formation behavior to examine whether the dynamics of multiple LLMs are similar to or different from human social dynamics~\cite{papachristou2024network}. In addition, LLM capabilities can be used to create professional agents with controlled, specialized, and interactive, professional-level capabilities to reshape professional services through evolving expertise~\cite{chu2024professional}. There is an empirical result that shows the agent named QuantAgent is useful in financial research, as it has the capability to uncover viable financial signals and enhance the accuracy of financial forecasts~\cite{wang2024quantagent}. In a study by \textsl{Yang et al.}~\cite{yang2023impacts}, an agent-based model was developed to simulate the decision-making behavior and interactive behavior of enterprises and employees based on the HRM characteristics of growing enterprises. The study found that firms should pay extra attention to recruitment programs, changes in workers' pay gaps are influenced by industry growth and their own capabilities, and pay cap policies have a positive impact on the development of growth firms in the start-up stage. However, there is still a lack of research on Agent-based simulation of cooperative behaviors among employees with different competencies within an organization based on a large language model.

Finally, in the organization, managers use the knowledge inside and outside the organization to make procedural and non-procedural decisions to improve the scientificity and accuracy of decision making~\cite{kuechler2006so}. To further enhance decision-making efficiency, many organizational managers have introduced AI technologies for assistance. In particular, the recent rise of generative large language models (LLMs) offers new opportunities. These models demonstrate proactive interaction capabilities and can support not only individual decision-making and task execution but also organizational-level planning and scheduling~\cite{bouschery2023augmenting}. For instance, \textsl{Zheng et al.} developed an LLM-based generative job recommendation system to provide individuals with a more personalized and comprehensive job seeking experience~\cite{zheng2023generative}. By imposing prompt-based organizational structures on LLM agents, \textsl{Guo et al.}~\cite{guo2024embodied} found that LLMs exhibit leadership and spontaneous cooperative behaviors in their organizations. In the decision-making process, it seems that human dominance has become weaker and AI's decision-making ability has become stronger, which makes it easier to correct the decision-making bias brought about by human intuition~\cite{susarla2023janus}. However, due to the algorithmic black box characteristics of generative AI, it is difficult for humans to intervene and predict the content generated by generative AI~\cite{deng2022benefits}.  Therefore, in the decision-making process, generative AI may pose the risk of false information and misleading decisions~\cite{shulner2022fairness,wang2023survey}. Therefore, in talent management, how to dialectically view AI and how to effectively use AI to manage and avoid risks is a key topic worthy of future research. However, the applications of existing LLMs in talent analysis also have limitations, such as the accuracy of generated information, interpretability, interaction forms, etc., which will affect the application of large models in business scenarios.

\subsection{The Impact of Emergent Events on Talent Analytics}
Emergent events such as pandemics have had a huge impact on businesses and organizations, and there are many management scenarios where we can apply AI methods to effectively deal with these dilemmas. In talent management, because of the pandemic, the daily tasks of the employees in many companies will change. In order to make the company adapt to this fluctuation and to mitigate the impact of these emergent events, we can predict the performance of the employees by using AI methods~\cite{hasan2024employee}. Additionally, the epidemic has taken a toll on numerous markets and industries, causing employees in companies to face high psychological stress, and some may even leave the organization. We can complete the simulation of employees' stress through AI technologies~\cite{garlapati2020predicting} and understand the factors that influence the employees' intention to leave the organization~\cite{mozaffari2023employee,tharani2020predicting}. From the perspective of an organization's leader, it is essential to understand the impact of the epidemic on customers so that an effective response can be taken. Therefore, completing sentiment analysis of customers using text is a very effective approach~\cite{chang2022predicting,shah2021mining}. Lastly, in the labor market, AI methods can effectively use large-scale data to provide researchers with research perspectives and technical support to study the impact of emergent events on the labor market~\cite{celbics2023impacts,alaql2023multi}.

Besides organizational resilience, the forms of talent analytics are facing a transformation due to some emergent events. Because issues such as the need to limit social distances that arose during the epidemic triggered us to think about increasing human-computer interaction in the workplace scenario~\cite{mer2023navigating}. In the recruitment process, since global cooperation and telecommuting have become frequent, we can use AI technologies such as natural language processing technology to screen job applicants in global real-time big data~\cite{strang2022erp}, to reduce the risk of disease transmission in densely populated scenarios, the use of chatbots to communicate with onboarding employees during the onboarding process can avoid the contact of the employees and simplify the process of onboarding employees~\cite{westberg2019applying}. Once employees are onboarded, they can also be trained with the help of a robot~\cite{vrontis2022artificial}. \textsl{Xu et al.} found that applying AI-enhanced VR simulators to employee training can reduce human resource costs and increase employee satisfaction, thereby enhancing the competitiveness of the organization~\cite{xu2020influence}. Finally, for managers of organizations, as the temporary workforce of organizations is increasing after the epidemic era, combining temporary workforce management with AI can reduce risks and costs and improve work efficiency and quality~\cite{matonya2020innovation}.


{
 \scriptsize
 \bibliographystyle{IEEEtran}
 \bibliography{reference}
}

\ifCLASSOPTIONcaptionsoff
  \newpage
\fi

\end{document}